\documentclass[iop]{emulateapj}
\usepackage[]{natbib}
\usepackage{graphicx}
\usepackage[]{hyperref}
\usepackage[figure]{hypcap}
\usepackage{afterpage}

\usepackage[capitalize]{cleveref}

\crefname{figure}{Figure}{Figures}
\Crefname{figure}{Figure}{Figures}
\crefname{table}{Table}{Tables}
\Crefname{table}{Table}{Tables}
\crefname{section}{Section}{Sections}
\Crefname{section}{Section}{Sections}
\crefname{chapter}{Chapter}{Chapters}
\Crefname{chapter}{Chapter}{Chapters}
\crefname{equation}{Equation}{Equations}
\crefname{equation}{Equation}{Equations}

\newcommand{\FIRST}{{FIRST}}
\newcommand{\SDSS}{{SDSS}}

\slugcomment{Accepted for publication in The Astrophysical Journal}
\shorttitle{SUBSTRUCTURE IN CLUSTERS WITH RADIO SOURCES}
\shortauthors{WING \& BLANTON}

\begin{document}

\title{An Examination of the Optical Substructure of Galaxy Clusters Hosting Radio Sources}

\author{Joshua D.\ Wing and
		Elizabeth L.\ Blanton}		
\affil{Astronomy Department and Institute for Astrophysical Research,
Boston University, Boston, MA 02215}
\email{jwing@bu.edu}

\begin{abstract}
Using radio sources from the Faint Images of the Radio Sky at Twenty-cm (FIRST) survey, and optical counterparts in the Sloan Digital Sky Survey (SDSS), we have identified a large number of galaxy clusters.  The radio sources within these clusters are driven by active galactic nuclei, and our cluster samples include clusters with bent, and straight, double-lobed radio sources.  We also included a single-radio-component comparison sample.  We examine these galaxy clusters for evidence of optical substructure, testing the possibility that bent double-lobed radio sources are formed as a result of large-scale cluster mergers.  We use a suite of substructure analysis tools to determine the location and extent of substructure visible in the optical distribution of cluster galaxies, and compare the rates of substructure in clusters with different types of radio sources.  We found no preference for significant substructure in clusters hosting bent double-lobed radio sources compared to those with other types of radio sources.
\end{abstract}

\section{Introduction} \label{introduction} \index{Introduction}
In the hierarchical formation model, galaxy clusters are the last structures in the universe to collapse and equilibrate \citep{springel2006}.  They form via mass accretion over a wide range of masses.  On the smallest mass accretion scales, surrounding filaments continuously fall towards the centers of mass of clusters.  This mass infall progresses all the way to the largest mass accretion scales where entire galaxy clusters merge, thereby creating dramatic changes in the structure of clusters.  These mergers are the most energetic events in the universe \citep{markevitch1998}.

Such large-scale cluster-cluster mergers have a significant impact on the environment of clusters, resulting in merging of the intracluster medium (ICM).  The cluster galaxies, instead of being distributed uniformly in a Gaussian distribution, are found in clumps and sub-clumps throughout a recently merged cluster.  Clusters exhibiting evidence of a recent large-scale merger have a larger fraction of galaxies with radio-loud active galactic nuclei (AGN) \citep{venturi2000,venturi2001,johnston-hollitt2008}.

Previous studies \citep{zhao1989,hill1991,allington-smith1993,blanton2000,blanton2000a,blanton2001,blanton2003,mao2010,mao2011,wing2011} have shown that bent double-lobed radio sources are often found in galaxy cluster environments, as much as $70\%$ of the time.  These include wide angle tail (WAT) and narrow angle tail (NAT) radio sources found in cluster centers, and cluster outskirts, respectively.  Bent, double-lobed radio sources are most likely bent by ram pressure resulting from the relative velocity between the radio host galaxy and the ICM.  \citet{eilek1984} calculated the velocity difference and densities needed to create observed swept-back lobes of WATs and found that the peculiar velocities of central giant ellipticals (often hosts to these bent radio sources) were insufficient to bend the lobes.  \citet{burns1990} calculated that the resulting motion of the ICM from a recent large-scale cluster-cluster merger was enough to create the velocity differences needed to bend radio lobes.  \citet{hardcastle2005} found that it is possible to bend radio lobes to the extent observed, assuming a high flow velocity and low density within the radio lobes, with radio-host-galaxy velocities as low as $100$-$300$ km s$^{-1}$, typical of galaxies located near the center of a galaxy cluster.  Bent radio sources may also be found in clusters that are relaxed (exhibiting little substructure) on large scales, e.g.\ Abell 2029 \citep{clarke2004}.  In these cases, the bending of the lobes may be related to gas motions induced by ``sloshing'' \citep{ascasibar2006,mendygral2012} of the central ICM.  This sloshing can persist for billions of years after an off-axis merger of a group or sub-cluster with a main cluster.

Here, we examine the optical substructure in a sample of clusters containing bent, double-lobed radio sources to determine whether the clusters exhibit significant substructure, as would be expected if the lobes were bent through large-scale cluster-cluster mergers.  If the lobes are bent through sloshing, we expect to see less substructure.  We examine the cluster environments of straight-lobed and single-component radio sources as a comparison.

Substructure can be identified through optical data.  Some earlier studies \citep{carter1980,rhee1987,rhee1991} examined the elongation of clusters using the positions of optically detected galaxies within the clusters.  The elongation represents a deviation from a spherical cluster.  Other studies examined the surface density contours of galaxies within clusters by looking for clumps of galaxies \citep{geller1982}.  Obtaining spectra for galaxies within clusters allows for a more detailed look at the three-dimensional distribution of the galaxies within them \citep{beers1982,baier1984,fitchett1987}.  Recent advances in multi-object spectroscopy now allow the study of clusters with hundreds of spectroscopically confirmed members.  As a result, robust statistical estimators for the presence of substructure within clusters have been developed \citep{dressler1988,west1988,west1990,beers1990,bird1993,colless1996,pinkney1996,kriessler1997,burgett2004,flin2006,ramella2007,owers2009,owers2011,einasto2012,hou2012}.  Using these estimators, it is possible to measure the significance of substructure within galaxy clusters.

Using radio sources from the \citet{wing2011} sample, we examine the optical environments surrounding those sources in rich cluster environments using the Sloan Digital Sky Survey \citep[\SDSS,][]{york2000}.  Specifically, we address the question of whether optical substructure within these clusters is related to the radio source morphology.  In \S~\ref{the_clusters}, we define the samples we used and the method for obtaining positions and redshifts used in the analysis of the substructure as well as our methods for determining and rejecting cluster interlopers.  In \S~\ref{substructure_analysis}, we discuss our different substructure analysis tools.  In \S~\ref{results}, we present our results, and in \S~\ref{conclusions}, we present our conclusions and a summary of our work.

\section{The Clusters} \label{the_clusters} \index{The Clusters}
In \citet{wing2011}, we compiled a sample of bent and straight double-lobed radio sources, as well as a single-component comparison sample from the Faint Images of the Radio Sky at Twenty-cm \citep[\FIRST\/ ][]{becker1995} survey, associated with galaxy clusters within the footprint of the Sloan Digital Sky Survey (\SDSS).  We showed that bent double-lobed radio sources are excellent tracers of galaxy clusters, as they are associated with rich clusters $>60\%$ of the time.  We use the four different samples of radio sources from \citet{wing2011} that have optical hosts in the \SDSS.  These include a visually-selected bent, double-lobed sample (the ``visual-bent'' sample) a bent, double-lobed sample selected using a computer algorithm (the ``auto-bent'' sample) a straight-lobed sample, and a single-component sample.  In order to use \SDSS\/ photometry and spectroscopy to measure substructure, we are limited to only the most nearby objects in our samples.

After rejecting interlopers within a cluster, and rejecting clusters with fewer than a minimum of $30$ spectroscopically confirmed cluster members (details of which are outlined below), we were left with $9$ and $11$ clusters in the visual-bent sample, using our fixed gap interloper rejection method and our shifting gapper interloper rejection method (see \S~\ref{cluster_center} and \S~\ref{interloper_rejection}), respectively.  The auto-bent sample is left with $7$ and $8$ clusters after use of the fixed gap and shifting gapper methods, respectively.  The straight sample contains $9$ and $13$ clusters, and the single-component sample has $5$ and $7$ clusters with the fixed gap and shifting gapper methods, respectively.  In all, we examine the substructure environments of $30$ clusters using the fixed gap interloper rejection method, and $39$ clusters using the shifting gapper interloper rejection method.

\subsection{Finding the Cluster Center} \label{cluster_center} \index{The Clusters!Finding the Cluster Center}
We started with these four samples of radio sources and searched Data Release 8 (DR8) \citep{aihara2011} of the \SDSS\/ for every optical source with a spectroscopically measured redshift within a $6$ Mpc projected radius of the radio source and within a recessional velocity of $\pm10,000$ km s$^{-1}$ of the recessional velocity of the radio source.  We only considered sources with spectroscopically measured redshifts within this volume.  While this redshift determination is not flawless, and photometric redshifts have improved over time, we get more reliable results when we only include potential cluster members with spectroscopically confirmed redshifts.  Of course, the limitations of \SDSS\/ spectroscopic observations (fiber collisions force minimum separations between galaxies of $\sim1\arcmin$) mean that the galaxies within the cluster are not likely to be fully sampled.  This is more of an issue for sources at higher redshifts where the angular distance between cluster galaxies will be small.

\begin{figure}
\centering
\capstart
\includegraphics[scale=0.4]{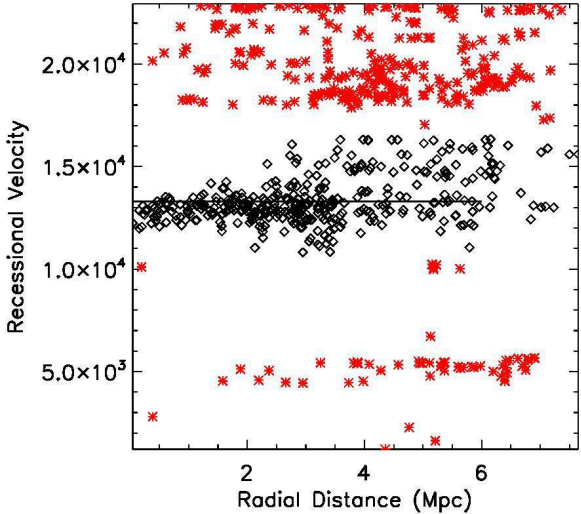}
\caption{The distribution of recessional velocities for sources within $6.0$ Mpc of the radio source (at the redshift of the radio source).  The (red) asterisks represent sources that were removed as outliers prior to finding the center of the cluster.  Any source with a gap of greater than $500$ km s$^{-1}$ in recessional velocity with the next closest galaxy was considered an outlier and removed.  The solid line represents the bi-weight mean of the recessional velocity of the potential cluster members after rejecting interlopers.  This is how we find the center of the cluster before again searching with \SDSS\/ to find cluster members.  (A color version of this figure is available online.)} \label{clustercenter}
\end{figure}

We start with this large search radius and velocity range since the radio sources may sometimes be located near the cluster outskirts.  Using only these spectroscopically confirmed sources within the volume specified above, we removed any obvious foreground and background galaxies by sorting all galaxies by their recessional velocities and identifying any adjacent galaxies (in recessional velocity space) with gaps greater than $500$ km s$^{-1}$.  \citet{de-propris2002} argue that galaxy clusters correspond to well-defined peaks with respect to recessional velocity and that gaps between successive galaxies of more than $1000$ km s$^{-1}$ indicate foreground and background galaxies.  This is also the value used by \citet{aguerri2010} for analyzing substructure in \SDSS\/ clusters.  We find that, for our data, a gap of $500$ km s$^{-1}$ is more successful at removing foreground and background galaxies.  We iteratively remove adjacent galaxies with gaps of greater than $500$ km s$^{-1}$ until the number of galaxies in the cluster remains constant.  An example of this is seen in \cref{clustercenter}.
If there are a minimum of $10$ galaxies remaining in the cluster at this point, we then determine the bi-weight mean \citep{beers1990} recessional velocity and positional center of the remaining galaxies.

\begin{figure}
\centering
\capstart
\includegraphics[scale=0.4]{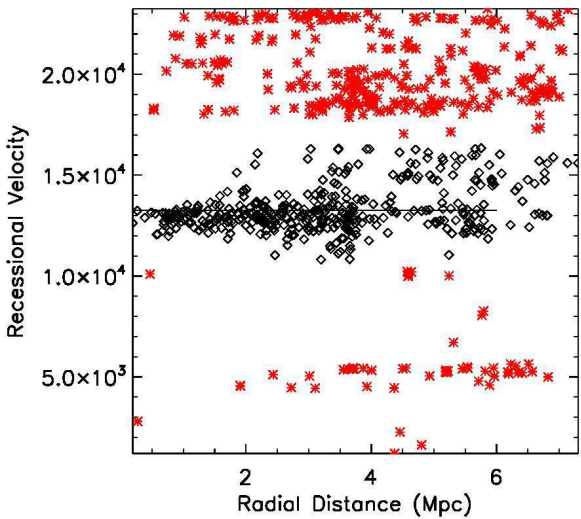}
\caption{The distribution of recessional velocities for sources within $6.0$ Mpc of the center of the galaxy cluster (at the redshift of the cluster).  The (red) asterisks represent sources that were removed as outliers prior to finding the center of the cluster.  Any source with a gap of greater than $500$ km s$^{-1}$ in recessional velocity with the next closest galaxy was considered an outlier and removed.  The solid line represents the bi-weight mean of the recessional velocity of the potential cluster members after rejecting interlopers.  Once we determine the center of the cluster, we are able to perform a more accurate interloper rejection.  (A color version of this figure is available online.)} \label{clustercenter_center}
\end{figure}

We assume that this center is more likely to be the center of the cluster than the position of the radio source, and we use this center as the new position around which to search in \SDSS\/ for all sources within $6$ Mpc and $\pm10,000$ km s$^{-1}$.  Using all \SDSS\/ spectroscopically confirmed galaxies within this new volume, we again remove foreground and background galaxies through the same method described above.  This process is illustrated in \cref{clustercenter_center}.
We then find the bi-weight mean center of the cluster, and we require that the radio source be located within a $3$ Mpc radius from the new center of the cluster, at the redshift of the center of the cluster.  We also require that the recessional velocity of the new center be within $\pm5000$ km s$^{-1}$ of the recessional velocity of the radio source.  We further require that there be a minimum of $30$ spectroscopically confirmed galaxies contained within a radius of $3$ Mpc and recessional velocity of $\pm5000$ km s$^{-1}$ of the center of the cluster.  The requirement of at least $30$ spectroscopically confirmed galaxies within the cluster is a requirement for obtaining statistically significant results with our substructure tests.

\subsection{Interloper Rejection} \label{interloper_rejection} 
\begin{figure*}
\begin{center}
\capstart
\includegraphics[scale=0.4]{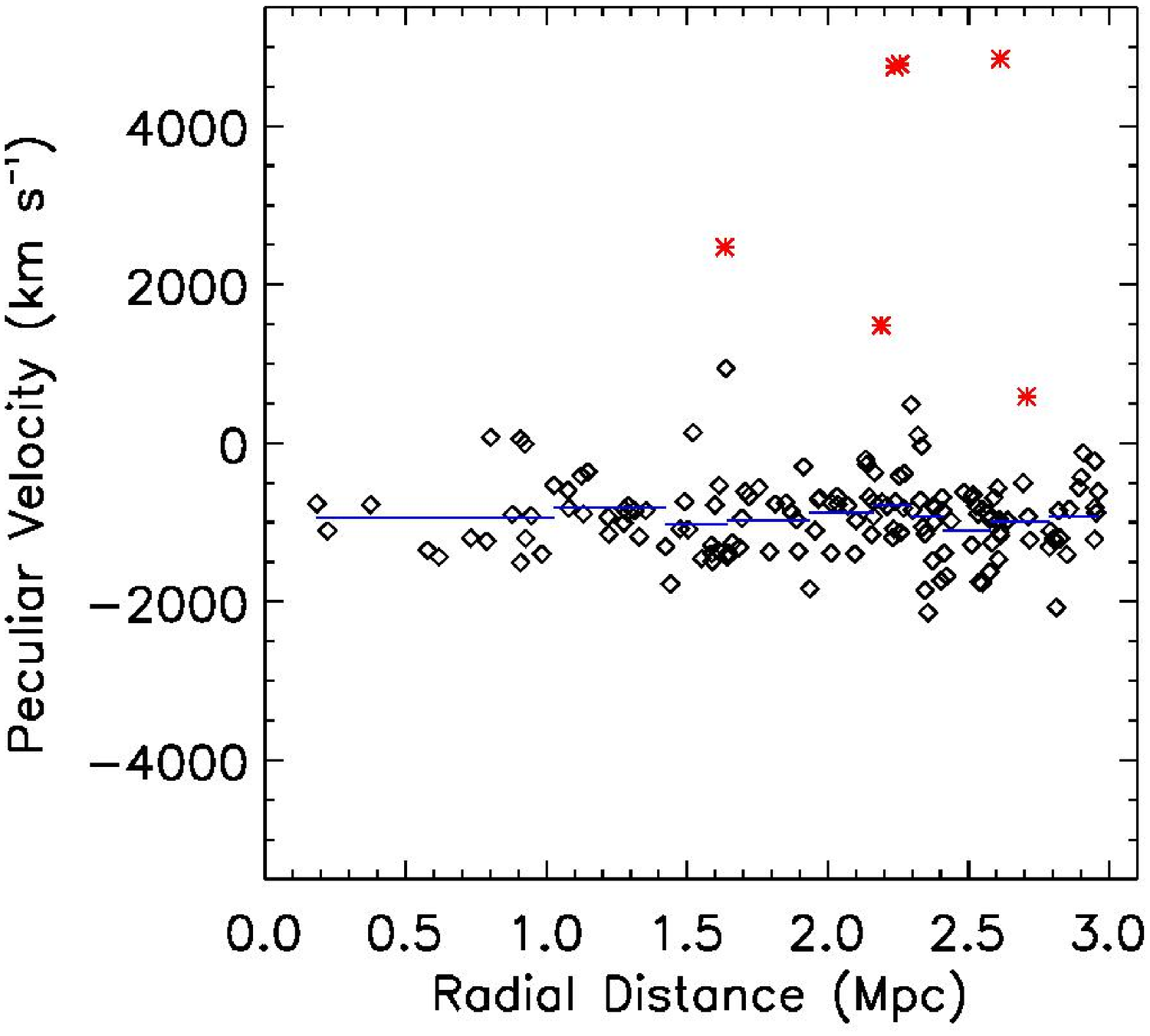}
\includegraphics[scale=0.4]{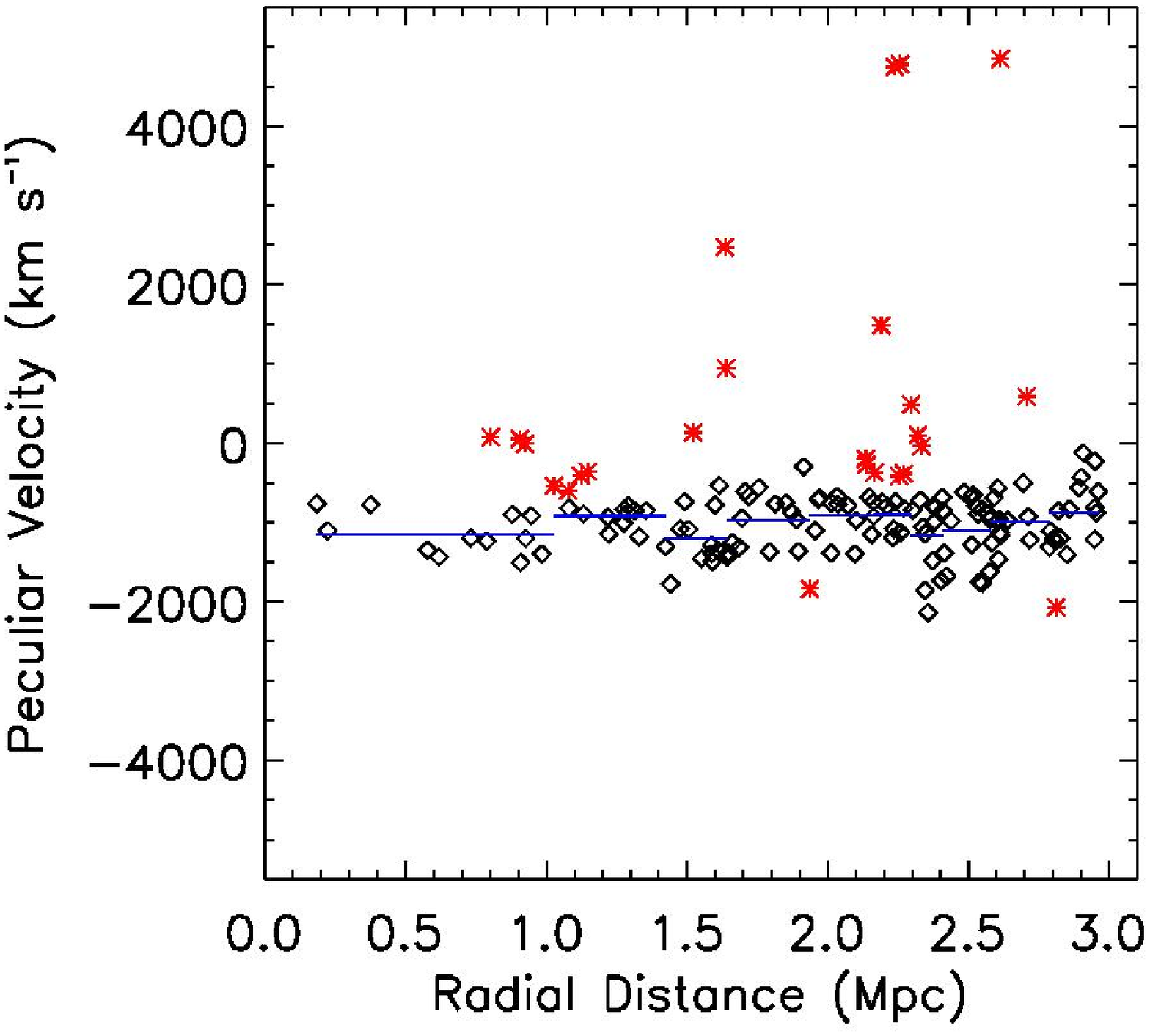}
\caption{A plot of the peculiar velocity of a given galaxy versus the radial distance from the center of the cluster.  The (red) asterisks represent sources rejected as interlopers.  The (blue) lines represent the bi-weight mean peculiar velocity for the galaxies within each of our radial distance bins.  Both of these plots are for the same cluster, the left-hand panel shows the results of interloper rejection as determined by the fixed gap method and the right-hand panel shows the results using the shifting gapper method.  (A color version of this figure is available online.)} \label{velspread}
\end{center}
\end{figure*}

\index{The Clusters!Interloper Rejection}
It is expected that some of these sources will be interlopers and not actual members of the cluster.  For this work, we use two different interloper rejection methods and compare the results.  The first method is referred to as the fixed gap method.  In this method, we binned all of the potential cluster members as a function of radial distance from the cluster center.  This is similar to the procedure used by \citet{fadda1996}.  We require that each bin have a minimum of $15$ galaxies, and that there are between $2$ and $10$ bins.  This requires a minimum of $30$ galaxies with spectroscopically measured redshifts within the angular area as described in \S~\ref{cluster_center}.  If a potential cluster has fewer than $30$ members, the cluster is rejected and not analyzed further.  Within each of these bins we sort the galaxies by their peculiar velocities.  For completeness, we define the peculiar velocity of a galaxy within a cluster as:
\begin{equation}
\Delta v_{i} = c \frac{\left(z_i - \bar{z}\right)}{\left(1 + \bar{z}\right)}, \label{equation_pec_velocity}
\end{equation}
where $\Delta v_i$ is the peculiar velocity of galaxy $i$, $z_i$ is the redshift of galaxy $i$, and $\bar{z}$ is the average redshift of the cluster.  Any galaxy within a given bin whose gap with an adjacent (in peculiar velocity) galaxy is greater than $1000$ km s$^{-1}$ is rejected as an interloper.  If, after the interloper rejection, a bin is left with fewer than 5 galaxies, the cluster is rejected and not analyzed further.  In these cases, a minimum of $10$ galaxies within that radial bin have been rejected as interlopers and it is likely that either a large fraction of the presumed cluster galaxies are actually interlopers or the distribution of the galaxies within the cluster is too far from Gaussian to be analyzed with normal substructure tests.  These rejected interlopers are seen as (red) asterisks in the left panel of \cref{velspread}.

We also employed a second method to reject cluster interlopers.  We refer to this as the shifting gapper method.  We follow the procedure outlined by \citet{owers2009,owers2011}.  Again, galaxies are sorted into bins as a function of radial distance from the center of the cluster, the same as with the fixed method.  Within each radial distance bin, galaxies are sorted by their peculiar velocity with respect to the velocity of the cluster.  In each bin, the ``f-pseudosigma" \citep{beers1990} is determined and used as the velocity gap to reject outliers.  The value of f-pseudosigma corresponds to the normalized difference between the upper and lower fourths of a data set and is good for quick calculations.  It can be calculated for any data set as follows:
\begin{equation}
f = F_u - F_l,
\end{equation}
and,
\begin{equation}
S_f = f / 1.349,
\end{equation}
where $F_u$ is the value of the upper fourth (or quartile) of the data set, $F_l$ is the value of the lower fourth of the data set, $f$ is known as the f-spread, and $S_f$ is the value of f-pseudosigma.  \citet{beers1990} note that for large data sets, the value of $f$ is $1.349$ for standard normal distributions.  This process is repeated for each bin until either the number of sources stabilizes, the value of f-pseudosigma drops below $250$ km s$^{-1}$, or the value of f-pseudosigma begins to increase.  If, at any point, the number of sources in a bin drops below 5, the entire cluster is not used in any further analysis.  The sources rejected as cluster interlopers using this shifting gapper method are shown as (red) asterisks in the right panel of \cref{velspread}.

We have chosen to use two different interloper rejection methods throughout this paper because the galaxies identified as interlopers are different for the two methods.  The fixed gap method uses the same peculiar velocity gap for every radial distance bin in every cluster.  This method consistently eliminates galaxies with large peculiar velocity gaps, but tends to identify fewer interlopers than the shifting gapper method.  On the other hand, the shifting gapper method uses a different peculiar velocity gap for each radio distance bin based on the dispersion of galaxies within that bin.  In many cases, it identifies interlopers with smaller peculiar velocity gaps than the fixed gap method uses.  In some cases, when there is a large peculiar velocity spread in a radial distance bin, the shifting gapper method is less likely to identify interlopers than the fixed gap method.  We see no physical reason to prefer one method over the other, and the galaxies identified as interlopers are different enough between the two methods that we go forward with analyzing our results using both interloper rejection methods.  Any cluster in which the radio-host galaxy is rejected as an interloper is removed from the samples and from further study.  Of course, it is possible that sources rejected as interlopers using the fixed gap method are not rejected using the shifting gapper method, and vice versa, so the samples of galaxy clusters hosting radio sources are different when comparing the fixed gap interloper rejection method with the shifting gapper interloper rejection method.

\begin{figure*}
\begin{center}
\capstart
\includegraphics[scale=0.4]{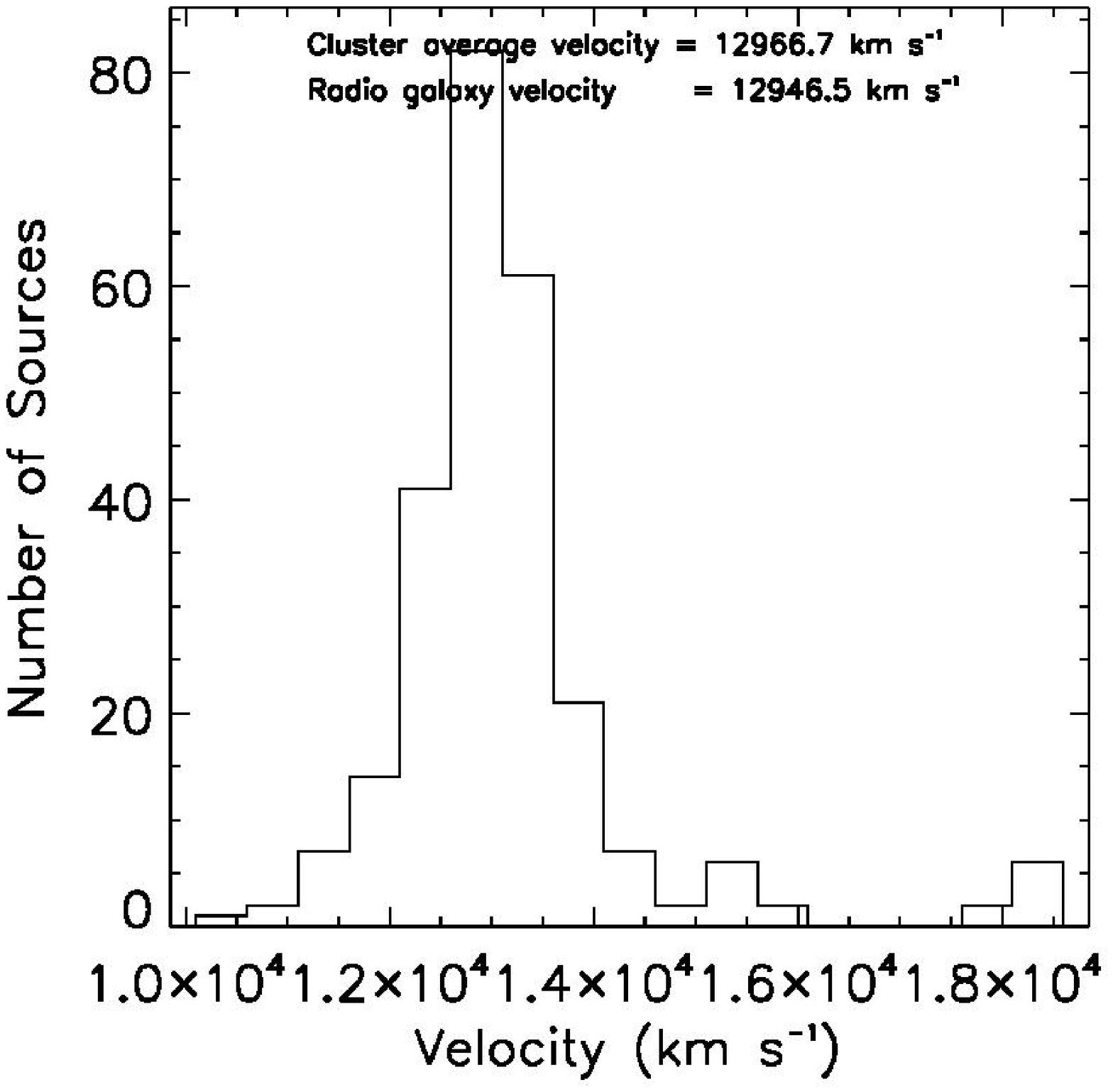}
\includegraphics[scale=0.4]{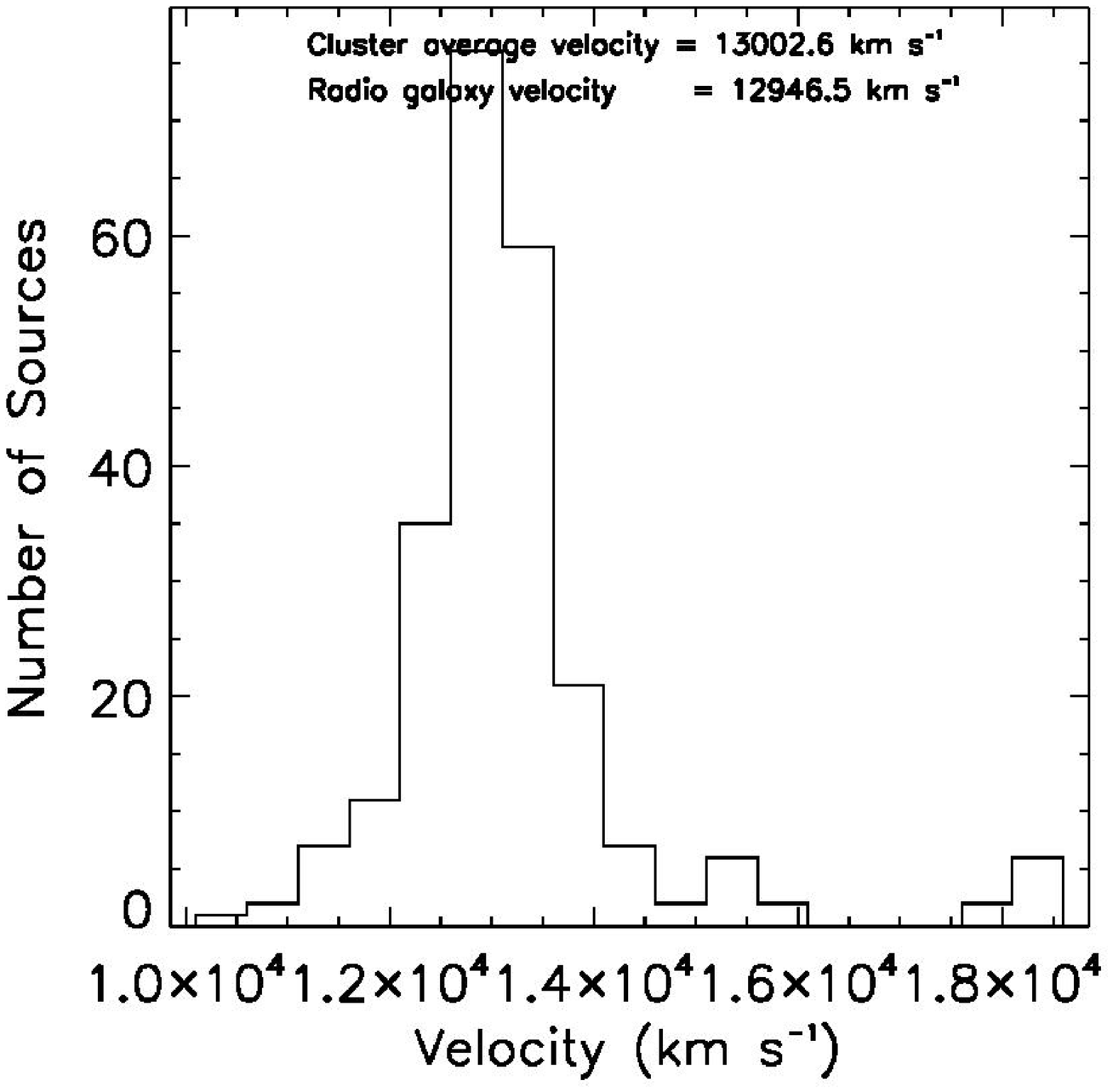}
\caption{Histograms of the galaxy velocities for one of the clusters in our samples.  The left-hand panel shows the histogram produced using the fixed gap interloper rejection method and the right-hand panel shows the histogram produced using the shifting gapper interloper rejection method.  These histogram plots are generated for each cluster.} \label{velhist}
\end{center}
\end{figure*}

After rejecting interlopers, we created histograms of the recessional velocites of each of our clusters.  We show an example of one cluster in \cref{velhist}.
These histograms allow for a quick examination of the velocity distribution of potential galaxy clusters.  After eliminating interlopers from within each cluster and removing any cluster with fewer than $30$ remaining spectroscopically measured member galaxies, we are left with 9, 7, 9, and 5 clusters from the visual-bent, auto-bent, straight, and single-component samples, respectively, using the fixed gap interloper rejection method.  Using the shifting gapper interloper rejection method, we are left with 11, 8, 13, and 7 clusters from the visual-bent, auto-bent, straight, and single-component samples, respectively.  We refer to the number of spectroscopically confirmed cluster members as N$_{3.0}^z$, the number of galaxies within $3$ Mpc and $\pm 5000$ km s$^{-1}$ of the center of the cluster.  This is a very large peculiar velocity range.  Most of the member galaxies have peculiar velocities much lower than this.  However, we wanted to be sure to detect any possible member galaxies or infalling sub-clusters and relied on our interloper rejection methods to eliminate those galaxies at large peculiar velocities not associated with the cluster.

\subsection{Cluster Richness Measurements} \label{cluster_richness_measurements} \index{The Clusters!Cluster Richness Measurements}
We are employing two different metrics to classify the richness of each cluster.  We use the N$_{1.0}^{-19}$ richness measurement that we used in \citet{wing2011} that counts the number of sources more luminous than $M_r=-19$ within $1.0$ Mpc of the radio source, corrected to account for background sources.  This is calculated at the redshift of the radio source.  This richness metric was developed to determine the nature of the surrounding environment of extended radio sources and making use of only photometric measurements.  The other richness metric we utilize within this paper, N$_{3.0}^z$ discussed above, uses the bi-weight mean center of the cluster in Right Ascension and Declination as well as redshift (as discussed in \S~\ref{cluster_center}) and counts the number of spectroscopically confirmed \SDSS\/ sources with $3.0$ Mpc and $\pm 5000$ km s$^{-1}$ of that center.  The radius and peculiar velocity spread is calculated using the redshift of the center of the cluster.

We go forward using both of these richness metrics because they measure different quantities and the N$_{1.0}^{-19}$ metric was used in \citet{wing2011} in the original definition of the clusters and their richnesses.  The N$_{1.0}^{-19}$ metric measures the local environment of the radio source.  We expect for clusters where the radio source is near the center of the cluster and the extent of the cluster is smaller than $1.0$ Mpc, the N$_{1.0}^{-19}$ richness metric and the N$_{3.0}^z$ richness metric should directly correlate.  However, in this paper, we want to examine clusters that are potentially host to sub-clusters and could have significant concentrations of galaxies beyond $1.0$ Mpc from the radio source.  We also do not wish to assume that the radio galaxy is located at the center of cluster.  For cases where the radio galaxy is located a significant distance from the center of the cluster, or there exist large sub-clusters at distances beyond $1.0$ Mpc from the radio galaxy, the N$_{1.0}^{-19}$ richness metric will underestimate the richness of the cluster.  Therefore, we utilize the N$_{3.0}^z$ richness metric in an attempt to more accurately determine the richness of the cluster.

\section{Substructure Analysis} \label{substructure_analysis} \index{Substructure Analysis}
We followed the prescriptions of \citet{pinkney1996} to determine substructure.  We have used a suite of tests, each of which have strengths and weaknesses for identifying substructure.  We have run one two-dimensional test (position in the sky) and three three-dimensional tests (positional as well as velocity information).  Based on these results, we seek to determine whether there is a significant amount of substructure present in a given cluster.  Throughout this section, when discussing averages, centers, and velocity dispersions, these have been found using a robust, outlier-resistant, bi-weight mean.

\subsection{Normalization of Tests} \label{normalization} \index{Substructure Analysis!Normalization of Tests}
Each of the tests was normalized by performing Monte Carlo simulations with randomized inputs.  We were investigating whether there was significant substructure present in these clusters, which requires a comparison to a null-hypothesis.  For our two-dimensional test, the null-hypothesis is a smooth distribution of galaxies within a plane with no preference to the azimuthal angle of the galaxy with respect to the center of the cluster \citep{west1988}.  Further, the surface density of galaxies within the cluster will decrease smoothly fitting a gaussian profile (although gaussianity does not have to be assumed) as the distance to the cluster center increases.  For our three-dimensional tests, the null-hypothesis is a lack of correlation between the position and velocity of member galaxies.  This means that there should be no difference between the velocity mean and dispersion of the galaxy cluster as a whole and any local area within the cluster \citep{pinkney1996}.

For our two-dimensional test (the $\beta$ test), galaxy positions were randomized by preserving the distance between the galaxy and the cluster center but randomly varying the azimuthal angle of the galaxy with respect to the cluster center.  For our three-dimensional tests (the $\Delta$, $\alpha$, and $\epsilon$ tests), we randomized the peculiar velocity with respect to the galaxy positions.  For each cluster, we perform $10,000$ Monte Carlo simulations.  The values for our various substructure tests with these randomized inputs allows us to determine the significance of the measurement implying substructure.  We can also use the average value of these simulations to give a normalization constant for each substructure test and each cluster.

Because we utilize $10,000$ Monte Carlo simulations, we are able to distinguish significance up to $\sim3.9\sigma$.  Some of the clusters had a significance of $100\%$.  For these clusters, we assigned a value of $3.9\sigma$ confidence.  However, it is important to keep in mind that the confidence values for these clusters is only a lower limit.  For most of our figures, we plot the significance of the presence of substructure in terms of $\sigma$ confidence level.

We have listed in \cref{table_sources_fixed1} and \cref{table_sources_fixed2} (using the fixed gap interloper rejection method) and \cref{table_sources_gapper1} and \cref{table_sources_gapper2} (using the shifting gapper interloper rejection method) the physical properties of the radio sources and host clusters (\cref{table_sources_fixed1,table_sources_gapper1}) and the substructure measurements for each cluster (\cref{table_sources_fixed2,table_sources_gapper2}).

Column 1 of \cref{table_sources_fixed1,table_sources_gapper1} lists the name of the radio source associated with the cluster, column 2 lists the sample that the corresponding radio source is in (V for the visual-bent sample, A for the auto-bent sample, S for the straight sample, and C for the single-component sample), columns 3 and 4 give the right ascension and declination of the radio source, columns 5 and 6 give the right ascension and declination of the center of the cluster, column 7 gives the FR classification \citep{fanaroff1974} of the radio source as determined using the optical magnitude and radio power criteria given in \citet{ledlow1996}, column 8 gives the FR classification of the radio source as determined visually, column 9 gives the redshift of the radio source, column 10 gives the redshift of the center of the cluster, column 11 gives the richness of the cluster surrounding the radio source as measured using the N$_{1.0}^{-19}$ metric in \citet{wing2011} (the number of galaxies brighter than $M_r=-19$ and within 1 Mpc of the radio source), column 12 gives the richness of the cluster using the N$_{3.0}^z$ values as described above, and column 13 gives the value of the velocity dispersion ($\sigma$) of the cluster.

Column 1 of \cref{table_sources_fixed2,table_sources_gapper2} again lists the name of the radio source associated with the cluster, column 2 lists the sample that the corresponding radio source is in (same as above), column 3 lists the normalized value of the $\beta$ substructure statistic and column 4 gives the $\sigma$ confidence value of that substructure measurement.  Columns 5 and 6 list the normalized value of the $\Delta$ substructure test and the $\sigma$ confidence value, columns 7 and 8 are the same for the $\alpha$ substructure test, and columns 9 and 10 are the same for the $\epsilon$ substructure test.  Column 11 lists nearby Abell clusters (within $3.0$ Mpc from the center of the cluster, using the redshift of the cluster).

We also analyzed the correlations between various cluster parameters by means of a Spearman correlation.  The values for Spearman correlations fall between $-1$ and $1$, with the bounds implying that both parameters are monotonically related, either decreasing or increasing.  A value of zero implies that there is no correlation between the two parameters.  The significance of this correlation is calculated and given as the number of standard deviations by which the correlation coefficient deviates from the null-hypothesis.  We have listed the values of the Spearman correlation coefficient and the significance for all of the different parameters we evaluated in \cref{table_spearman_fixed} (for the fixed gap method) and \cref{table_spearman_gapper} (for the shifting gapper method).  The first two columns of each of these tables gives the names of the parameters that we are examining for a potential correlation.  The third column lists the value of the Spearman correlation for those two parameters, and the fourth column gives the significance of the correlation, in units of standard deviation from the null-hypothesis.

The various parameters we examined include the redshift (listed as $z$ in the tables) of the radio source, the redshift (listed as $z_{clust}$ in the tables) of the center of the cluster, the average de-reddened $r-i$, $g-r$, and $g-i$ colors of the spectroscopically confirmed cluster members, the velocity dispersion of the cluster (listed as $\sigma$ in the tables), the opening angle of the radio source, the peculiar velocity of the radio source (listed as $\Delta v_{radio}$ in the tables), the richness of the cluster as measured by N$_{1.0}^{-19}$ (from \citet{wing2011}) and N$_{3.0}^z$ (as described above), the $r$-band absolute magnitude of the brightest cluster galaxy (BCG), the average recessional velocity of the cluster members (listed as $\bar{v}$ in the tables), and the difference in recessional velocity between the radio-host-galaxy and the average of the cluster members (listed as $\bar{v} - v$ in the tables).  We have also examined correlations between many of these parameters and the values of the different substructure tests, both normalized and as a function of their significance.  The different substructure tests are listed by name in the tables and, if the correlation is done using the significance of the substructure detection,  ``significance'' follows the name of the substructure test.  In addition, we also examined correlations involving the physical separation between the radio source and the BCG (as measured at the redshift of the cluster center), and the difference between our two cluster richness measurements, normalized by the N$_{3.0}^z$ richness measurement.

\subsection{The \texorpdfstring{$\beta$}{beta} Test} \label{beta} \index{Substructure Analysis!The $\beta$ Test}
The $\beta$ test is our only two-dimensional test.  The $\beta$ test checks for mirror symmetry and was proposed by \citet{west1988}.  The $\beta$ test is sensitive to deviations from mirror symmetry, but not circular symmetry.  Thus, the $\beta$ test will not report the presence of substructure in a cluster that is elongated but with a smooth distribution of galaxies.  The $\beta$ test will identify clusters with clumps of galaxies as having a high likelihood of substructure, as long as the galaxy clumps are not distributed symmetrically about the center of the cluster.  The $\beta$ test will fail to find substructure in clusters where the galaxies form clumps in peculiar velocity space.  \citet{pinkney1996} find that the $\beta$ test is more sensitive to the presence of substructure than any of the other two-dimensional tests they examined, with the exception of the Fourier Elongation test.

As described in \citet{pinkney1996}, we determined the mean distance between each galaxy, $i$, and its five nearest neighbors.  This value is compared to the same value of the point diametrically opposite (that is, the same position but on the other side of the cluster, through the cluster center) of the galaxy.  Thus, the value of asymmetry for the galaxy $i$ is:
\begin{equation}
\beta_i = {\rm log}_{10}\/\left(\frac{d_o}{d_i}\right), \label{equation_beta}
\end{equation}
where $d_i$ is the mean distance for galaxy $i$ and $d_o$ is the mean distance for the point opposite (through the center of the cluster) galaxy $i$.  The value of $\beta$ for the entire cluster is the average of the $\beta_i$ values.  For perfectly symmetric clusters, $\beta=0$.  We normalize our value of $\beta$ for each cluster by dividing it by the average $\beta$ values from our $10,000$ Monte Carlo simulations, as described in \S~\ref{normalization}.  We also use these Monte Carlo simulations to determine the $\sigma$ confidence level of the presence of substructure within the cluster.

\begin{figure*}
\begin{center}
\capstart
\includegraphics[scale=0.4]{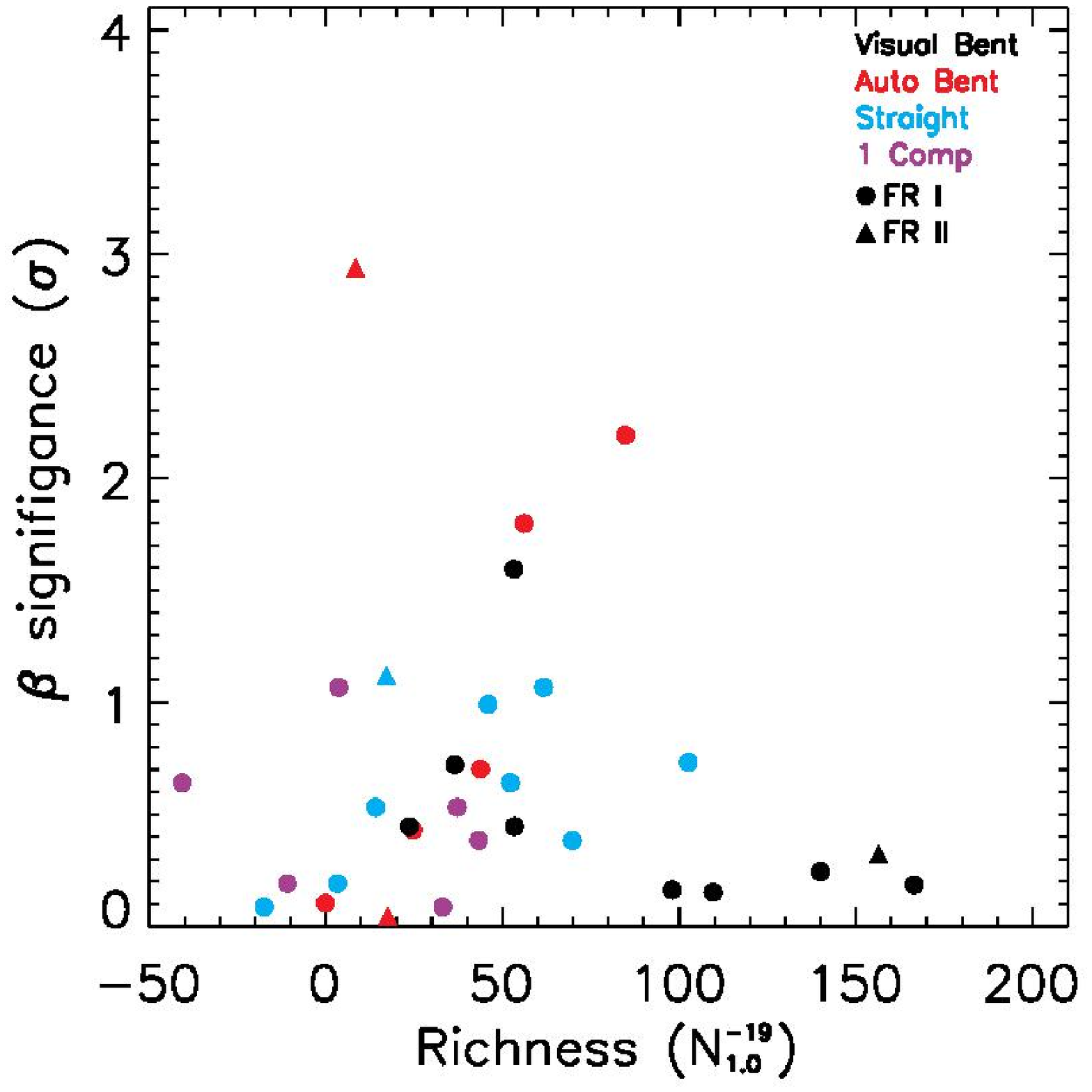}
\includegraphics[scale=0.4]{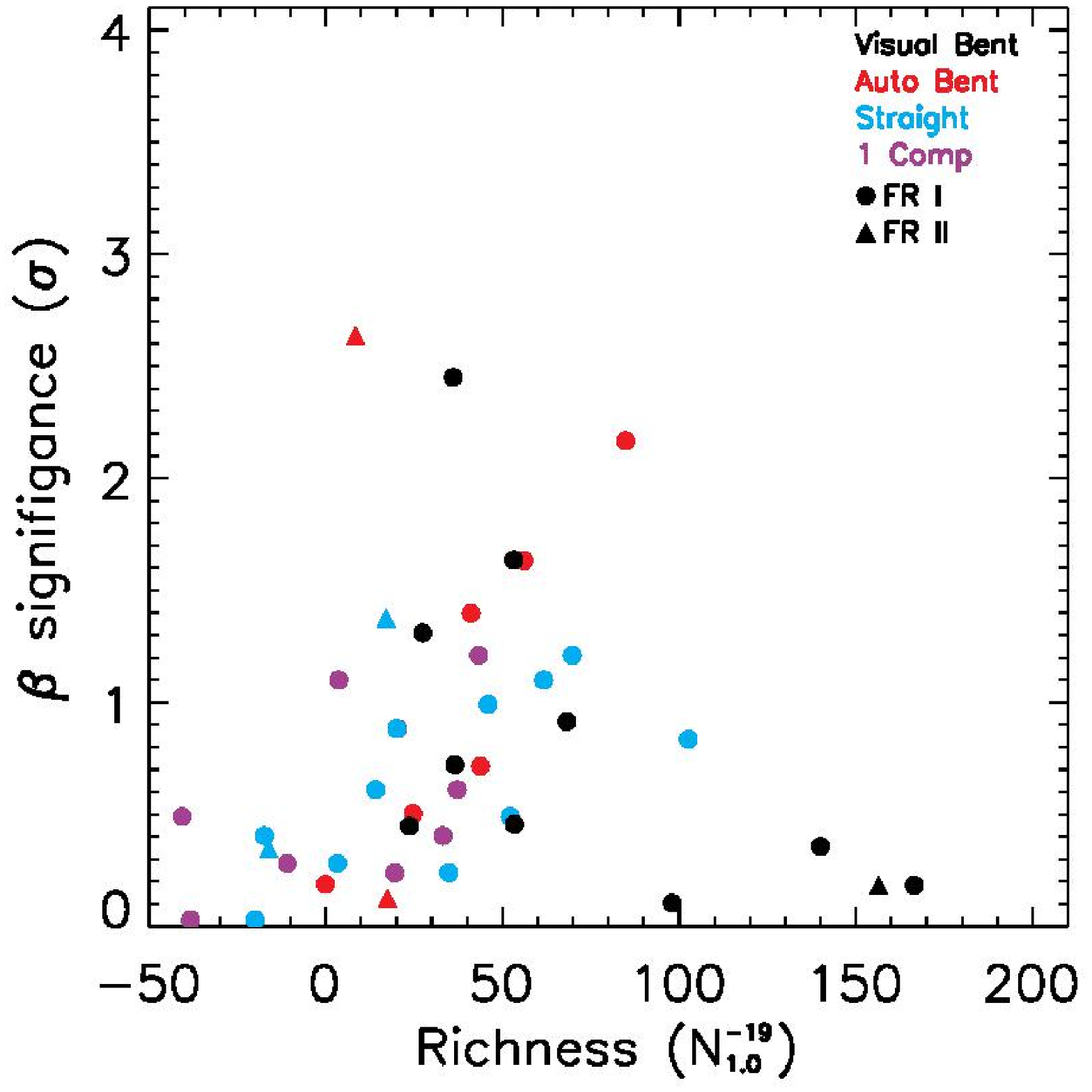}
\caption{Richness of the cluster, as measured by N$^{-19}_{1.0}$, as a function of the significance of the presence of substructure as measured by the $\beta$ test.  There appears to be a possible correlation between the richness of the cluster and the significance of the presence of substructure.  However, upon examination of Spearman correlation coefficients, we find that there is a slight negative correlation.  This result is most likely influenced heavily by the richest clusters in the visual-bent sample.  The left-hand panel shows the plot for the sources using the fixed gap interloper rejection method.  The right-hand panel shows the plot for the sources using the shifting gapper interloper rejection method.  Lightest-gray (purple) filled symbols represent the single-component sample, light-gray (blue) filled symbols represent the straight sample, dark-gray (red) filled symbols represent the auto-bent sample, and black filled symbols represent the visual-bent sample.  The FR I sources are represented by circles and the FR II sources are represented by triangles.  (A color version of this figure is available online.)} \label{clustersizevsbeta}
\end{center}
\end{figure*}

\cref{table_spearman_fixed,table_spearman_gapper} show that there is no clear correlation between the value of $\beta$ and any of the cluster properties that we measured, with the exception of the richness of the cluster \citep[as measured in ][]{wing2011}.  \cref{clustersizevsbeta} shows the richness of the cluster versus the significance of the presence of substructure as measured by the $\beta$ statistic.  An examination of \cref{table_spearman_fixed,table_spearman_gapper} shows that there is a slight negative, but not very significant, correlation.  A possible explanation for this is that the rich clusters tend to be regular and relaxed compared to the irregular poor clusters.

Substructure measurements were made for a total of $30$ clusters using the fixed gap interloper rejection method.  Using the $\beta$ test, $4$ of these clusters were identified as having substructure at greater than $2\sigma$ confidence.  For the $39$ clusters detected using the shifting gapper method, the $\beta$ test detected substructure with greater than $2\sigma$ confidence in $4$ of them.  At $13$\% and $10\%$, the $\beta$ test detects significant substructure in only a small fraction of the clusters in our samples, and the least out of all of the substructure tests we use in this paper.  In \S~\ref{results} we further break the results down into our bent and straight samples and look at only clusters with at least $50$ cluster members.  However, when we only look at these clusters, $4$ out of $18$, or $22$\%, of clusters identified using the fixed gap interloper rejection method, and $4$ out of $22$, or $18$\%, of clusters identified using the shifting gapper interloper rejection method have substructure at the $2\sigma$ confidence level according to the $\beta$ test.

\subsection{The \texorpdfstring{$\Delta$}{Delta} Test} \label{delta} \index{Substructure Analysis!The $\Delta$ Test}
The $\Delta$ test, first introduced by \citet{dressler1988}, is a three-dimensional test.  It is a measure of the deviation of local velocity dispersion and average velocity compared to the overall cluster values for both.  Areas with substructure will have local velocity dispersions and average velocities significantly different from those of the cluster as a whole.  The $\Delta$ test will be insensitive to the presence of substructure within the cluster in two cases.  The first case, identified by \citet{dressler1988}, involves clusters where the subclusters within the cluster are superimposed.  The second case, identified by \citet{pinkney1996}, occurs when the masses of subclusters are the same and the axis of merger is in the plane of the sky.  \citet{pinkney1996} note that the projection angles required for these cases to come to pass are rare and thus are unlikely to significantly influence the measurement of substructure within a cluster.  \citet{dressler1988} found that the $\Delta$ test was resistant to outliers and robust to parametric uncertainty.  Further, \citet{pinkney1996} find that of all the substructure tests they examined, the $\Delta$ test is most sensitive to substructure and that it responded to deviations from the null hypothesis in all cases with substructure.

In their original paper, \citet{dressler1988} took local to mean the ten nearest galaxies.  More recently, \citet{bird1995} and others \citep{pinkney1996,brainerd1998,solanes1999,aguerri2010} have used $\sqrt{N}$, where $N$ is the total number of galaxies in the cluster, to define local.  This deviation of local values from the global value can be defined as:
\begin{equation}
\delta_i^2=\left(\frac{N_{nn}+1}{\sigma^2}\right)\left[\left(\bar{v}_{local }- \bar{v}\right)^2 + \left(\sigma_{local} - \sigma\right)^2\right], \label{equation_delta}
\end{equation}
where $N_{nn} = \sqrt{N}$, $\bar{v}$ and $\bar{v}_{local}$ are the global and local average recessional velocity, and $\sigma$ and $\sigma_{local}$ are the global and local velocity dispersions.  \citet{dressler1988} define $\Delta$ as $\Sigma \delta_i$ and note that, for clusters with no substructure, $\Delta$ approaches $N$.  For comparison between clusters of different sizes, we will define $\Delta$ here as the averaged value of $\delta_i$ over the entire cluster.  Thus, for clusters with no substructure, $\Delta$ approaches $1$.  As with the $\beta$ test, we have normalized this value by means of $10,000$ Monte Carlo simulations.  We also use these Monte Carlo simulations to determine the $\sigma$ confidence level of the presence of substructure within the cluster.

\begin{figure*}
\begin{center}
\capstart
\includegraphics[scale=0.4]{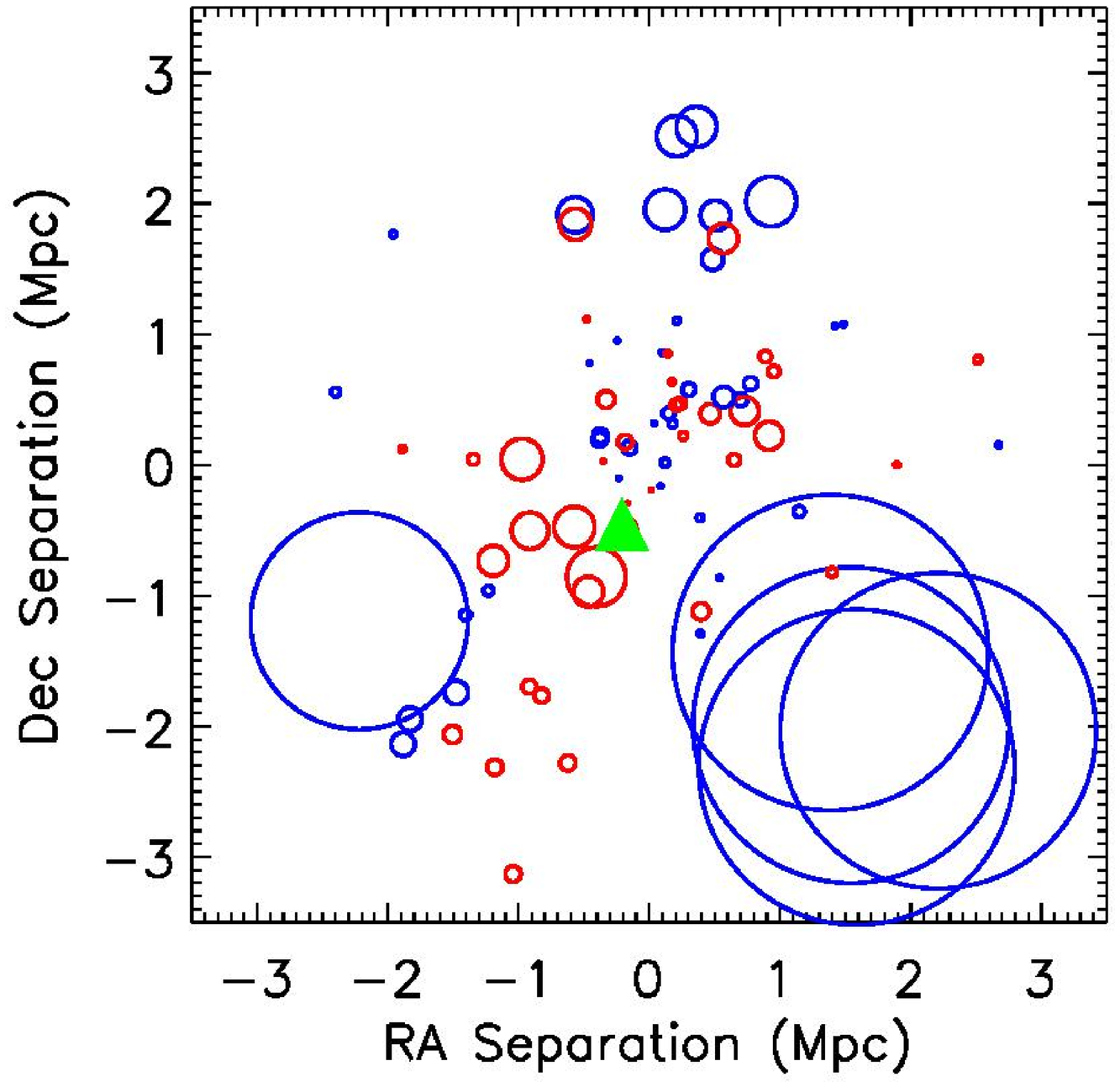}
\includegraphics[scale=0.4]{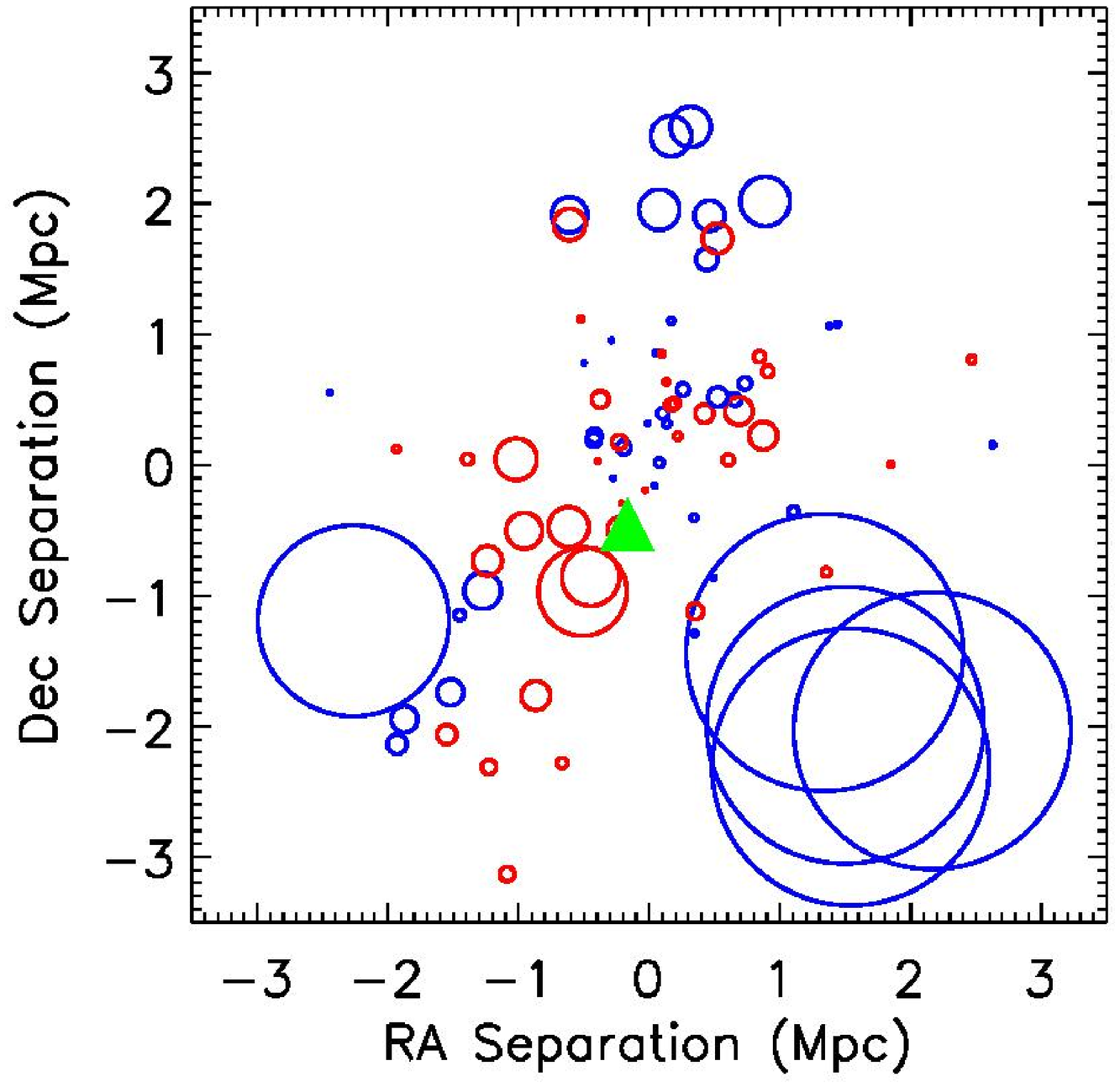}
\caption{A bubble plot visualization of the $\Delta$ statistic for one of the sources in our sample.  Each galaxy in the cluster is located at the center of a circle whose radius is proportional to $e^{\delta_i}$.  The larger the radius of the circle, the more likely it is to be located in a clump of galaxies within the galaxy cluster.  Light-gray (red) circles indicate that the galaxy has a redshift greater than the average of the cluster and black (blue) a smaller redshift.  The black (green) triangle represents the location of the radio galaxy.  The left-hand panel shows the plot for the sources using the fixed gap interloper rejection method.  The right-hand panel shows the plot for the sources using the shifting gapper interloper rejection method.  (A color version of this figure is available online.)} \label{bubble}
\end{center}
\end{figure*}

The $\Delta$ test allows us to visualize the substructure of the cluster by looking at both the position of each galaxy within the cluster (including its recessional velocity relative to the center of the cluster) and the deviation of that galaxy from the global values of velocity dispersion and average velocity.  \cref{bubble} shows an example of the visualization possible with bubble plots.
Each circle represents a galaxy in the cluster, with the radius of the circle proportional to $e^{\delta_i}$.  Sources with redshifts greater than the average cluster redshift are plotted in red and those with redshifts less than the average cluster redshift are plotted in blue.  The larger the radius of the circle, the more likely the galaxy is positioned in a clump of galaxies within the galaxy cluster that have significantly different local values of velocity dispersion and average velocity compared to the global values.

\begin{figure*}
\begin{center}
\capstart
\includegraphics[scale=0.4]{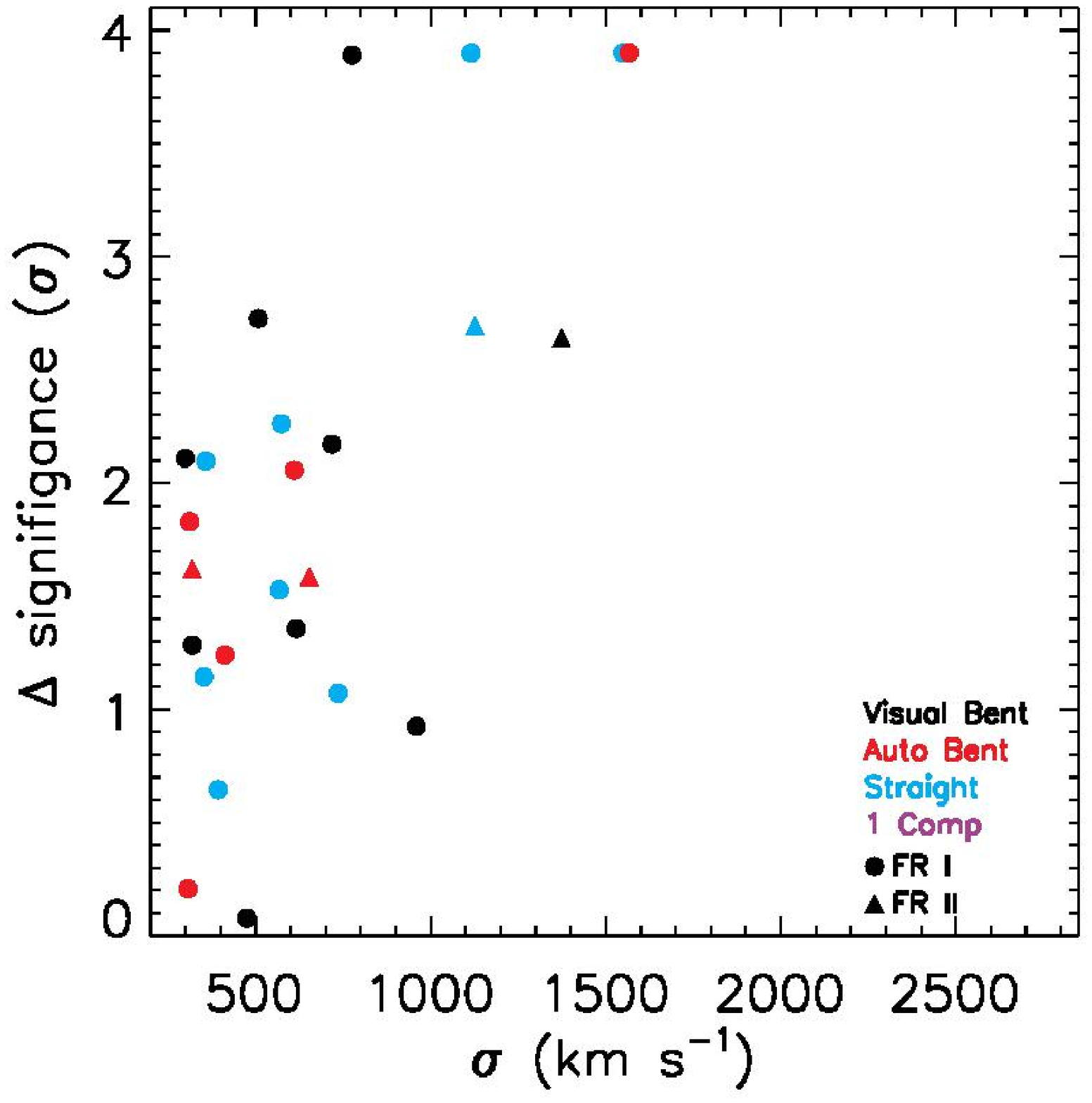}
\includegraphics[scale=0.4]{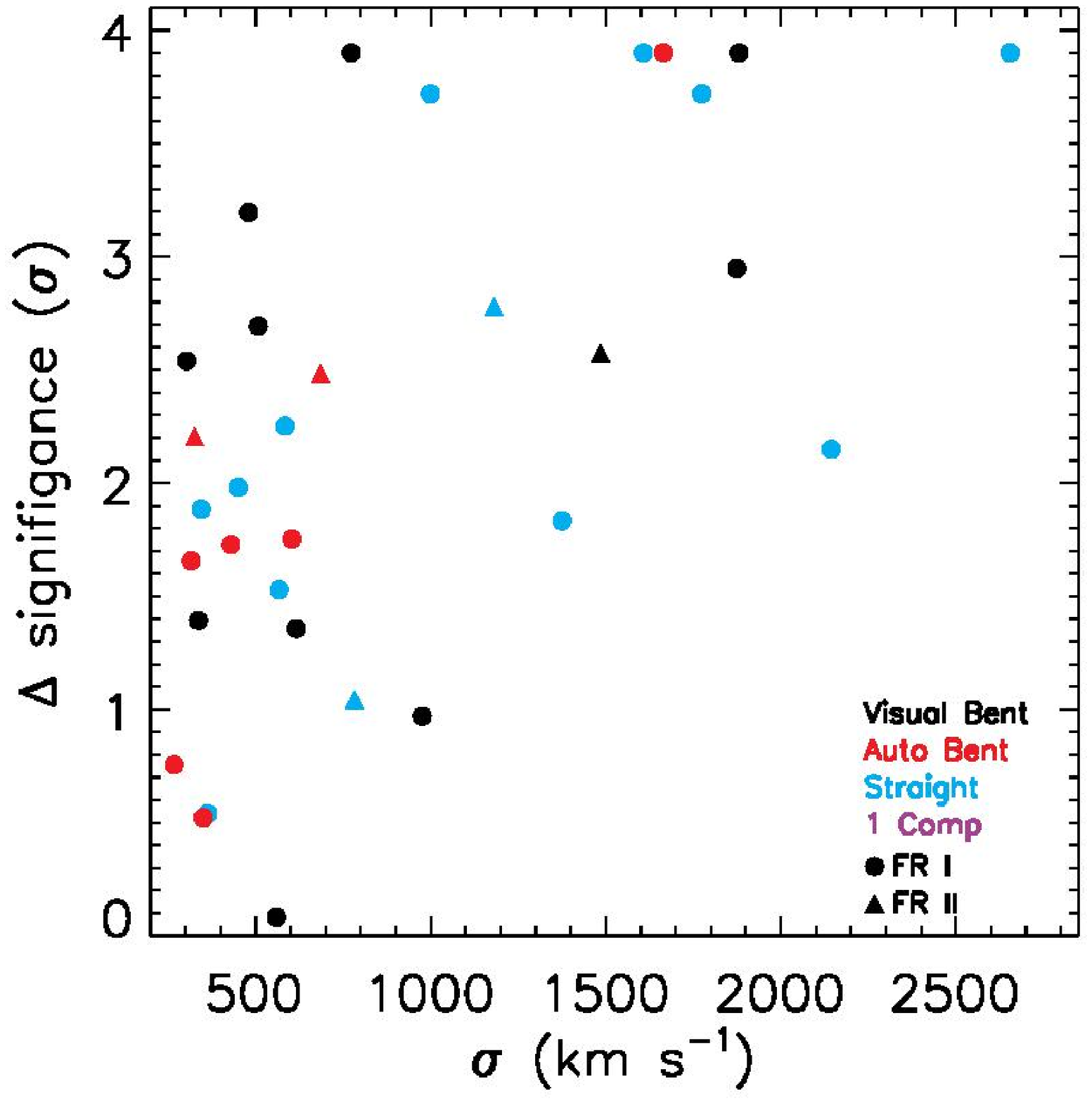}
\caption{Values of the significance of the presence of substructure as measured by $\Delta$ as a function of cluster velocity dispersion.  The symbols are the same as in \cref{clustersizevsbeta}.  We see a clear positive trend of $\Delta$ as a function of velocity dispersion.  The left-hand panel shows the plot for the sources using the fixed gap interloper rejection method.  The right-hand panel shows the plot for the sources using the shifting gapper interloper rejection method.  (A color version of this figure is available online.)} \label{veldisvsds}
\end{center}
\end{figure*}

\cref{veldisvsds} shows the significance of the presence of substructure, measured by the $\Delta$ statistic, as a function of the velocity dispersion of the cluster.
There appears to be a trend of increasing significance of the presence of substructure as the velocity dispersion of the cluster increases.  This is to be expected.  The greater the amount of substructure within the cluster, the more likely the presence of velocity outliers within the member galaxies.  This will increase the velocity dispersion of the cluster.  Further, the definition of the $\Delta$ statistic - see \cref{equation_delta} - relies heavily on the velocity dispersion of the cluster.  Thus, it is expected that the presence of significant substructure and the velocity dispersion of the cluster will be correlated.  We find a Spearman correlation coefficient of $0.58$ with a significance of $3.15\sigma$ for the fixed gap method and a correlation of $0.61$ with a significance of $3.73\sigma$ for the shifting gapper method.  These are strong correlations.

\begin{figure*}
\begin{center}
\capstart
\includegraphics[scale=0.4]{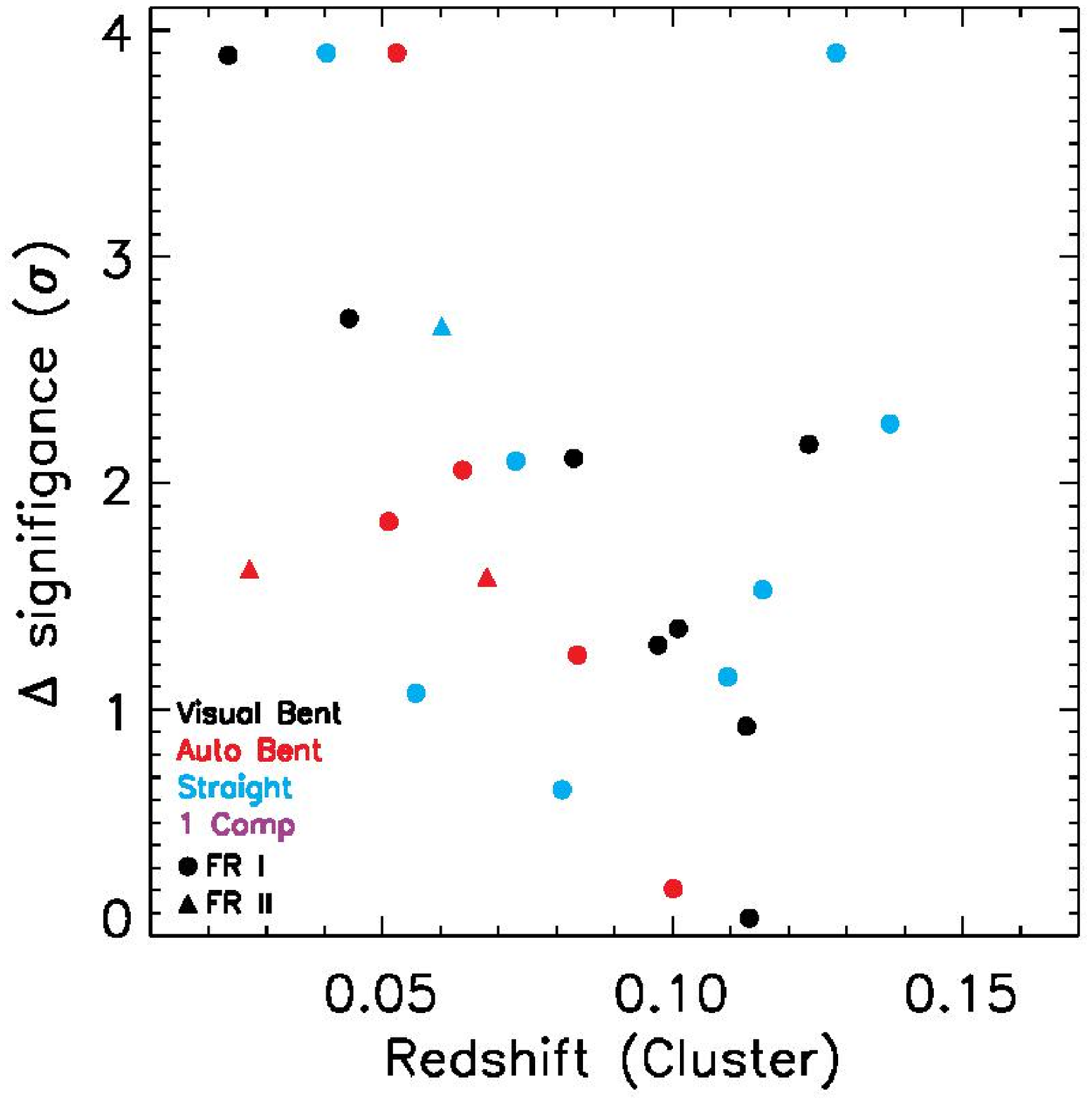}
\includegraphics[scale=0.4]{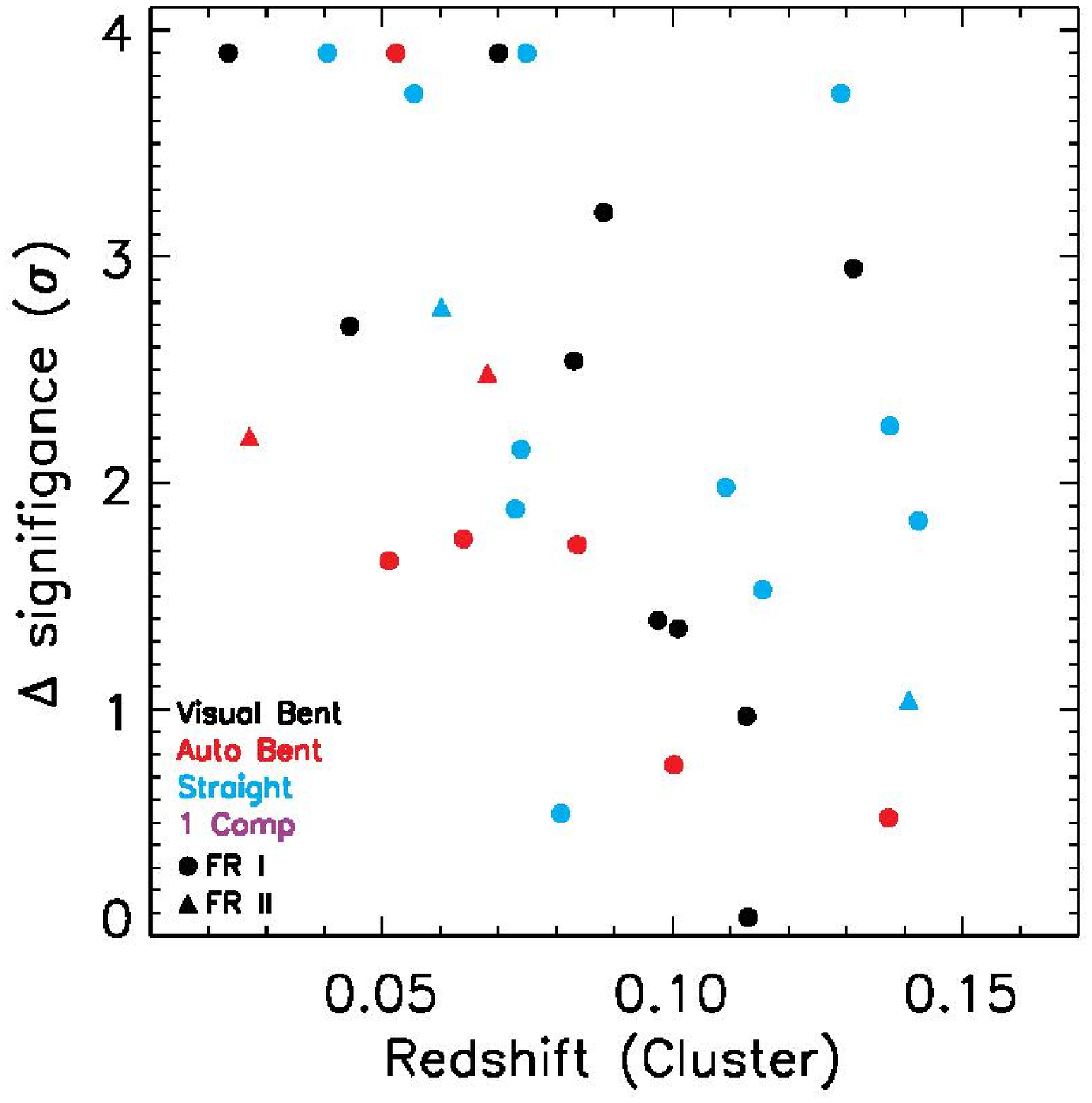}
\caption{Redshift versus $\Delta$.  The symbols are the same as in \cref{clustersizevsbeta}.  The left-hand panel shows the plot for the sources using the fixed gap interloper rejection method.  The right-hand panel shows the plot for the sources using the shifting gapper interloper rejection method.  We see a negative correlation between the significance of the presence of substructure as measured by the $\Delta$ test and the redshift.  Our observations do not go far enough in redshift space to begin to expect to see a correlation as a result of the period of increased cluster-cluster mergers.  Otherwise we would expect to see substructure within clusters increase as the redshift increases.  (A color version of this figure is available online.)} \label{zvsds}
\end{center}
\end{figure*}

\cref{zvsds} shows the presence of significant substructure shown as $\sigma$ confidence as measured by the $\Delta$ statistic as a function of redshift.
If our redshift range were larger, we might expect to see a trend towards increasing substructure with increasing redshift.  Instead, there is a negative trend with the clusters with greatest likelihood of having substructure occurring at the nearest redshifts.  The Spearman correlation coefficient is $-0.41$ with a significance of $2.18\sigma$ and $-0.50$ with a significance of $3.08\sigma$ for the fixed gap and shifting gapper method, respectively, using the redshift of the center of the cluster.  This is likely a result of the detection limits of the \SDSS.  Clusters at lower redshifts are more likely to have more cluster members with spectroscopically measured redshifts, meaning that we will be able to measure substructure more significantly for these objects.  This, in turn, leads us to detect a negative correlation between redshift and the presence of significant cluster substructure.

Of the $30$ clusters identified using the fixed gap interloper rejection method, the $\Delta$ test identifies substructure at greater than $2\sigma$ confidence in $15$ of them, including $5$ clusters where none of the $10,000$ Monte Carlo simulations produced a substructure measurement as high as the actual cluster.  Using the fixed gap method, $22$ of the $39$ clusters contained substructure at the $2\sigma$ or greater confidence level using the $\Delta$ test, $7$ of which were more significant than $10,000$ Monte Carlo simulations can show.  At $50$\% and $56$\%, the $\Delta$ test identifies substructure in a large fraction of the clusters we studied.  The $\Delta$ test identifies more clusters with significant substructure than any of the other substructure tests that we utilize.  We further examine the substructure fraction of clusters with more than $50$ members in \S~\ref{results}, but we find that $10$ of $18$, or $56$\%, of clusters identified using the fixed gap interloper rejection method, and $14$ of $22$, or $64$\%, of clusters identified using the shifting gapper interloper rejection method have substructure detected at the $2\sigma$ confidence level using the $\Delta$ test.

\subsection{The \texorpdfstring{$\alpha$}{alpha} Test} \label{alpha} \index{Substructure Analysis!The $\alpha$ Test}
The $\alpha$ test was introduced by \citet{west1990}.  This test measures the shift in the center of the cluster with unweighted and weighted positions.  The positions are weighted using the local velocity dispersion.  The $\alpha$ test can be affected by Poisson noise as well as the underlying velocity and spatial distributions of the member galaxies in cases where there is no substructure present.  Thus it is possible that the $\alpha$ test will report the presence of substructure in cases where there is none present.  However, even though it would seem that the $\alpha$ test is likely to suffer from effects due to superposition of cluster members, this is not necessarily the case.  A superposition of galaxies within a subcluster that has a peculiar velocity substantially different from the primary cluster galaxies will have large local standard deviations.  This has the effect of reducing the centroid shift for galaxies with similar peculiar velocities to the subcluster.  At the same time, it is likely that most of the galaxies with similar peculiar velocities to the subcluster will also be members of the subcluster.  These two effects counteract each other, and it is not necessarily clear which has a greater impact on the final value of $\alpha$.

The first step in determining $\alpha$ is to find the spatial center of the cluster:
\begin{equation}
x_c=\frac{1}{N} \sum_{i=1}^{N} x_i,\;\;\;y_c=\frac{1}{N} \sum_{i=1}^{N} y_i. \label{equation_alpha_1}
\end{equation}
The weight for each galaxy is defined as $w_i=1/\sigma_i$, where $\sigma_i$ is the line-of-sight velocity dispersion for the galaxy $i$ and its $N_{nn}$ nearest neighbors.  With these weights we can now calculate the weighted center for each group of $N_{nn}$ galaxies using:
\begin{equation}
x_c'=\frac{\sum_{i=1}^{N_{nn}+1} x_i w_i}{\sum_{i=1}^{N_{nn}+1} w_i},\;\;\;y_c'=\frac{\sum_{i=1}^{N_{nn}+1} y_i w_i}{\sum_{i=1}^{N_{nn}+1} w_i}. \label{equation_alpha_2}
\end{equation}
We can then use this difference between the center of these nearest neighbor velocity groups and the unweighted center of the cluster as a whole:
\begin{equation}
\gamma_i = \left[\left(x_c - x_c'\right)^2 + \left(y_c - y_c'\right)^2\right]. \label{equation_alpha_3}
\end{equation}
The value of $\alpha$ is the average of the $\gamma$ values.  Again, we have normalized $\alpha$ by dividing by the average of the $\alpha$ values in the $10,000$ Monte Carlo simulations.  Larger normalized values correspond to a greater amount of substructure.  We are also able to measure the significance of the presence of substructure within each cluster as a function of $\sigma$ confidence.

An examination of \cref{table_spearman_fixed,table_spearman_gapper} shows that the only cluster property that the $\alpha$ substructure statistic is correlated with is the richness of the cluster.  As with the $\beta$ and $\Delta$ substructure tests, we see a negative correlation between the presence of significant substructure as measured by the $\alpha$ test, and the richness of the cluster.

Using the fixed gap interloper rejection method, the $\alpha$ test identifies $9$ out of $30$ clusters as having substructure at greater than $2\sigma$ confidence levels.  Of the $39$ clusters identified using the shifting gapper interloper rejection method, $11$ are identified as having substructure at greater than $2\sigma$ confidence using the $\alpha$ test.  This includes $1$ cluster where the $10,000$ Monte Carlo simulations do not produce a cluster with the level of substructure as the actual cluster.  At $30$\% and $28$\%, the alpha test identifies significant substructure in nearly a third of the clusters in our samples.  This is second only to the $\Delta$ test in terms of the number of clusters identified as containing significant substructure.  When we examine only the clusters with $50$ or more cluster members (see \S~\ref{results}), we find that $6$ of $18$, or $33$\%, of clusters identified using the fixed gap method, and $7$ of $22$, or $32$\%, of clusters identified using the shifting gapper interloper rejection method have substructure detected at the $2\sigma$ confidence level with the $\alpha$ test.

\subsection{The \texorpdfstring{$\epsilon$}{epsilon} Test} \label{epsilon} \index{Substructure Analysis!The $\Epsilon$ Test}
The $\epsilon$ test, introduced by \citet{bird1993}, quantifies the correlation between the position of a galaxy within the cluster and the projected mass estimator.  Both \citet{bird1993} and \citet{pinkney1996} find that the $\epsilon$ test is sensitive to mergers where there is no redshift separation.  However, the $\epsilon$ test is not as sensitive as the $\Delta$ test.  \citet{pinkney1996} find that the $\epsilon$ test is less sensitive to the detection of substructure than both the $\Delta$ and $\alpha$ tests but that it reacted strongly to dispersion gradients in isothermal clusters.  The likelihood of the $\epsilon$ test detecting the presence of substructure is increased if the subcluster has a velocity separation with the main cluster.  \citet{pinkney1996} found that the $\epsilon$ test was more sensitive to the presence of substructure than the $\Delta$ test in clusters where a substantial amount of time ($2$-$4$ Gyr) has passed after the core crossing of the merger.

The projected mass estimator \citep[PME, see][]{heisler1985} is defined as:
\begin{equation}
M_{PME} = \xi\left(\frac{24}{\pi GN}\right) \sum_{j=1}^{N_{nn}} v_{zj}^2 r_j, \label{equation_epsilon_1}
\end{equation}
where $v_{zj}$ is the radial peculiar velocity with respect to the local average velocity, $r_j$ is the projected distance from the galaxy to the center of the $N_{nn}$ nearest galaxies, and $\xi$ is a constant equal to $4/3$ for isotropic orbits.  The center of the group of $N_{nn}$ galaxies is defined as the position of galaxy $i$.  The substructure statistic is defined as:
\begin{equation}
\epsilon=\frac{1}{N_{gal}} \sum_{i=1}^{N} M_{PME}. \label{equation_epsilon_2}
\end{equation}
The units of $\epsilon$ are in solar masses.  Again we have normalized our values of $\epsilon$ by averaging the values of $10,000$ Monte Carlo simulations and dividing our observed $\epsilon$ value by this normalization constant.  Clusters with substructure will have groups with smaller projected separations than clusters with no substructure, leading to smaller values of $\epsilon$.  Thus, for this test, normalized values less than $1$ correspond to the presence of substructure.  We are also able to measure the significance of the presence of substructure within each cluster as a function of $\sigma$ confidence.

An examination of \cref{table_spearman_fixed,table_spearman_gapper} shows that the only cluster property that the $\epsilon$ substructure statistic is correlated with is the richness of the cluster.  As with the $\beta$, $\Delta$, and $\alpha$ substructure tests, we see a negative correlation between the presence of significant substructure as measured by the $\epsilon$ test, and the richness of the cluster.

The $\epsilon$ test identifies $8$ of the $30$ clusters selected with the fixed gap interloper rejection method as having substructure at greater than $2\sigma$ confidence.  This includes $1$ cluster where the actual cluster has more significant substructure than any of the $10,000$ Monte Carlo simulations.  Of the $39$ clusters we examined identified with the shifting gapper method, $10$ were found to have substructure at the $2\sigma$ or higher confidence level with the $\epsilon$ test.  Thus $27$\% and $26$\% of the clusters are identified as having significant substructure with the $\epsilon$ test.  This is a higher fraction than the $\beta$ test, but less than the other two three-dimensional substructure tests.  Examining only the clusters with at least $50$ cluster members (we discuss this in greater detail in \S~\ref{results}), we find that $4$ out of $18$, or $22$\%, of clusters identified using the fixed gap interloper rejection method, and $7$ out of $22$, or $32$\%, of clusters identified using the shifting gapper interloper rejection method have substructure detected at greater than $2\sigma$ confidence using the $\epsilon$ test.

\subsection{Summary of Substructure Tests} \label{substructure_summary} \index{Substructure Analysis!The Substructure Tests}
We have used these four tests to determine if significant levels of substructure exist within our clusters.  Each of the substructure tests examines a different aspect of substructure within the cluster, so we do not necessarily expect to see an obvious correlation among the different tests.  However, using the tests in conjunction with one another, we are able to arrive at a determination for the substructure within a cluster.

\cref{table_spearman_fixed,table_spearman_gapper} show that for both the fixed gap and shifting gapper interloper rejection methods, there are a few parameters that have significant correlations.  We see that the confidence level for the presence of optical substructure as measured by each of the substructure tests is correlated negatively with the N$_{1.0}^{-19}$ richness metric, implying that as the significance of optical substructure increases, the richness of the cluster (within a 1 Mpc radius from the radio source) decreases.  In addition, the $\Delta$ substructure measurement is strongly correlated with the velocity dispersion of the cluster.  This is expected, as discussed in \S~\ref{delta}, because the calculation of $\Delta$ contains the velocity dispersion of the cluster.  We also see evidence of a negative correlation between the normalized value of the $\epsilon$ test statistic and the redshift of the cluster, as well as the opening angle of the radio source.  We see a positive correlation between the normalized value of $\epsilon$ and the $M_r$ magnitude of the radio-host-galaxy, as well as the confidence level for the presence of optical substructure measured by the $\epsilon$ test and the velocity dispersion of the cluster.  Lastly, we also see a strong positive correlation between the fractional difference between our two richness metrics and the separation between the radio source and the center of the cluster.

We also find a slight negative correlation between the redshift of the cluster and the significance of the presence of substructure within that cluster.  It appears as though each of the tests shows a slight trend towards the lowest redshift sources being located in the clusters with the most significant substructure.  It is probable that we are seeing a trend with the number of galaxies used to measure the substructure, and the measured presence of substructure.  Lower-redshift clusters have more cluster members with spectroscopically measured redshifts within the \SDSS\/, and are thus more likely to have detected substructure.

\begin{figure*}
\begin{center}
\capstart
\includegraphics[scale=0.4]{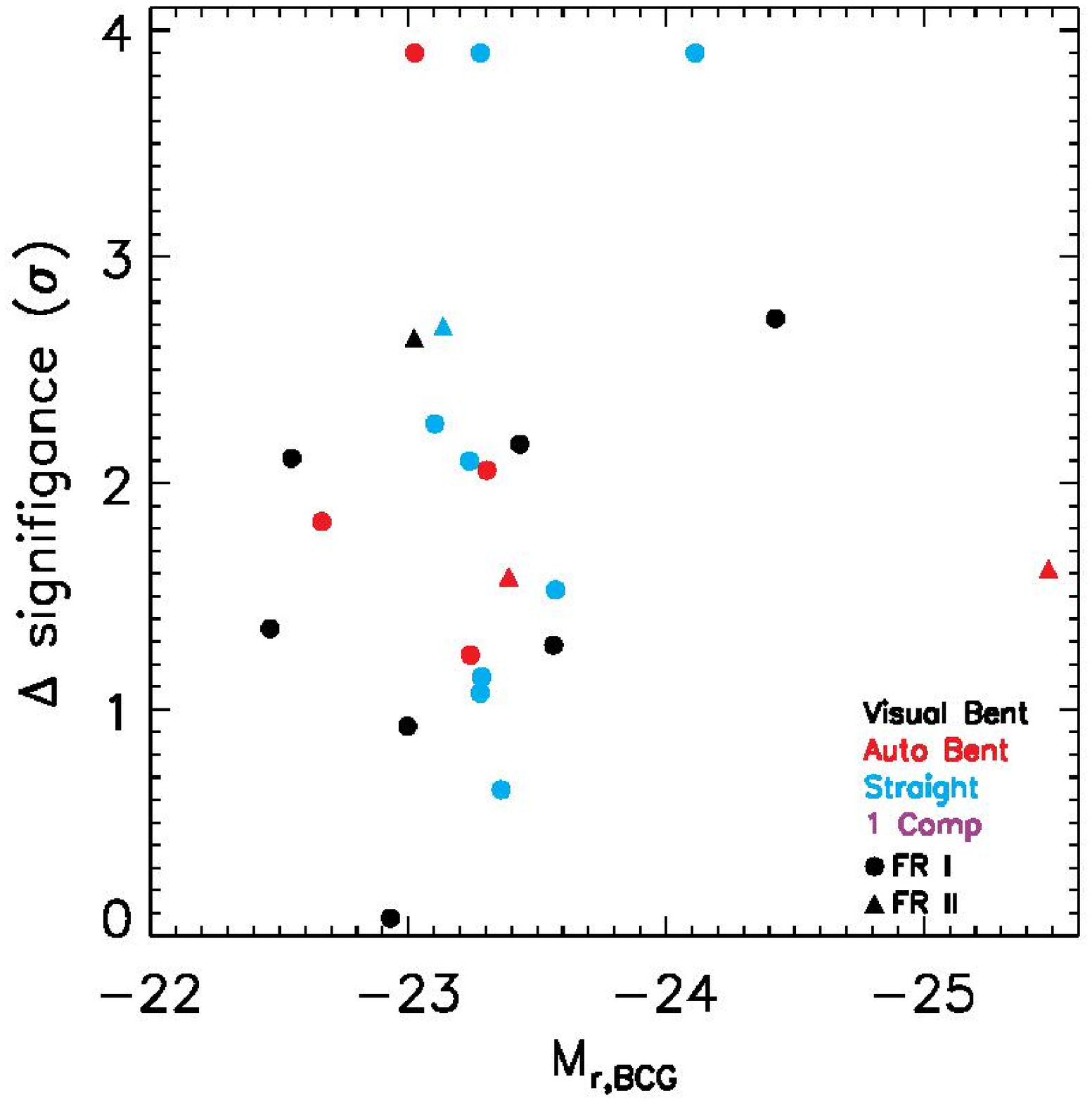}
\includegraphics[scale=0.4]{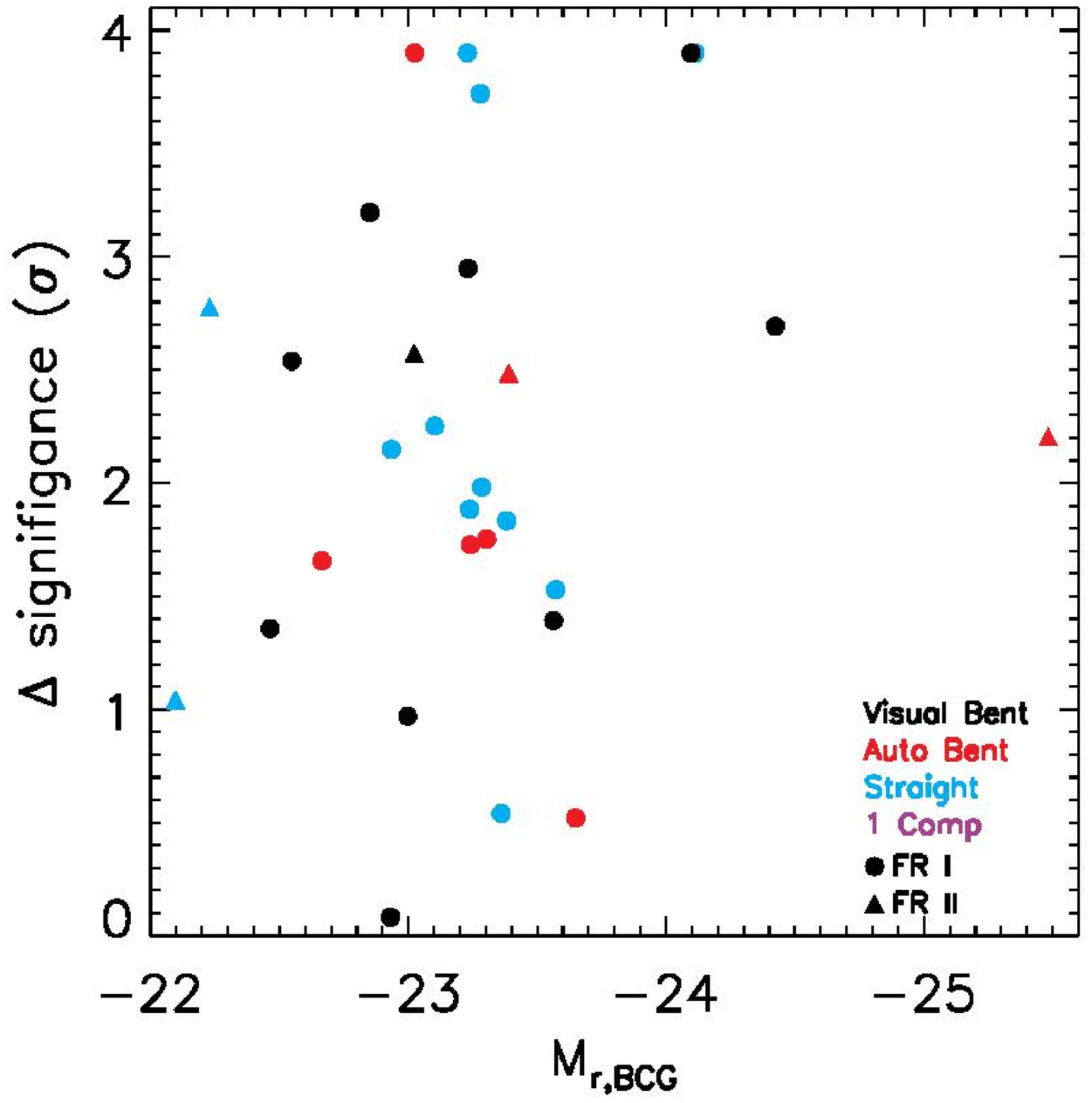}
\caption{Magnitude of the BCG versus $\Delta$.  The symbols are the same as in \cref{clustersizevsbeta}.  The left-hand panel shows the plot for the sources using the fixed gap interloper rejection method.  The right-hand panel shows the plot for the sources using the shifting gapper interloper rejection method.  There is a possible trend (for the $\Delta$ test, the other tests are less clear) for the clusters with brighter BCGs (and thus more massive) having less substructure.  (A color version of this figure is available online.)} \label{deltavsbcgmag}
\end{center}
\end{figure*}

We examined the relationship between the presence of substructure and the $r$-band absolute magnitude of the BCG.  \citet{ramella2007} found that the presence of substructure within the cluster is related to the luminosity of the BCG.  Specifically, they found that the clusters with the most luminous BCGs were preferentially located in clusters lacking substructure.  We do not observe this effect with the clusters in our samples.  \cref{deltavsbcgmag}
and \cref{table_spearman_fixed,table_spearman_gapper} show that, depending on the substructure test, the correlation between the presence of substructure and the $r$-band absolute magnitude of the BCG is at times positive \citep[implying that as substructure increases, the absolute magnitude of the BCG also increases, i.e. gets fainter, agreeing with ][]{ramella2007}, and other times negative.  In no case is the correlation confidence high enough to warrant a determination one way or the other.

There are a number of differences between our samples and the sample used by \citet{ramella2007}.  The sample of clusters from \citet{ramella2007} was selected from ROSAT All-Sky Survey X-ray data and consists of $77$ clusters within a redshift range of $0.04 < z < 0.07$.  These ranges are different than those for the sources in our samples.  Because of the lower redshift range and targeted spectroscopic observations, most of the clusters in the \citet{ramella2007} sample contain between $600$ and $1200$ members.  Only a few of the clusters in our samples have that many spectroscopically confirmed cluster members.  In addition, \citet{ramella2007} used a different substructure detection method than any of those that we used in this paper.  \citet{ramella2007} detected substructure within their sample of clusters using a program called DEDICA \citep{pisani1993,pisani1996}.  Essentially, the DEDICA program identifies substructure within a cluster by finding regions with high densities of cluster galaxies.  It is possible that the differences between our samples and the \citet{ramella2007} sample can explain the difference in the correlation between the absolute magnitude of the BCG and the presence of substructure within the cluster.

\section{Results} \label{results} \index{Results}
As discussed above, our results are shown in \cref{table_sources_fixed1,table_sources_fixed2,table_sources_gapper1,table_sources_gapper2}.  The results from using the fixed gap method of interloper rejection are shown in \cref{table_sources_fixed1,table_sources_fixed2}.  \cref{table_sources_fixed1} shows the physical properties for each of the clusters we examined, and \cref{table_sources_fixed2} lists the substructure properties of those same sources.  \cref{table_sources_gapper1,table_sources_gapper2} are the same, except that they show the results when using the shifting gapper interloper rejection method.  Because of the differences between the two methods, there are some clusters that meet the criteria  for the fixed gap method and not for the shifting gapper method, and vice versa.

One of the questions we are investigating is whether bent double-lobed radio sources are preferentially found in clusters with significant substructure.  Multiple methods can explain how double-lobed radio sources can be bent to the extents that they have been observed.  One method involves a large-scale cluster-cluster merger.  This large-scale merger will set the ICM in motion such that the ram pressure resulting from the relative velocity between the radio-host-galaxy and the ICM may cause the observed bending of the radio lobes.  Another method involves a smaller-scale merger, off-axis enough to cause ``sloshing'' within the ICM.  It is possible that this sloshing is sufficient to cause the observed bending of the radio lobes \citep{ascasibar2006,mendygral2012}.  In the first scenario, one would expect to see a high correlation between the presence of significant optical substructure (evidence of a recent large-scale cluster-cluster merger) and the presence of bent double-lobed radio sources compared to the same correlation with straight double-lobed radio sources.  In the second ``sloshing'' scenario, we do not expect optical substructure to be as significant.

\cref{table_sig_thresh_fixed,table_sig_thresh_gapper} show the fraction of sources in each sample with substructure detected at the $2.0\sigma$ or higher confidence level for each of the tests we used and a minimum of $50$ galaxies within $3.0$ Mpc and $\pm5000$ km s$^{-1}$ of the cluster center.  \cref{table_sig_thresh_fixed} lists the fractions when using the fixed gap method, and \cref{table_sig_thresh_gapper} lists the fractions when using the shifting gapper method.  We set a limit of $N^z_{3.0}>50$ to compare with other published results, specifically \citet{einasto2012}, who use only clusters with a minimum of $50$ sources in their analysis.

To aid in our analysis, and to increase the number of sources in each sample, we grouped all of our bent sources together and all of the straight and single-component sources together.  We also created a sample of the most likely ``true''  bent double-lobed radio sources (rather than possible unassociated, projected radio components), based on visual examination of radio contours from both bent double-lobed samples as a comparison.  We see no clear association between the presence of bent double-lobed radio sources and the detection of significant optical substructure as compared with non-bent sources.  This implies that large-scale cluster-cluster mergers are likely not the sole explanation for the bending of the radio lobes.  While clusters with bent radio sources frequently have significant substructure, clusters with straight or single-component radio sources are just as likely to show substructure.

\subsection{Correlations Between Cluster Properties} \label{more_correlations} 
\begin{figure*}
\begin{center}
\capstart
\includegraphics[scale=0.4]{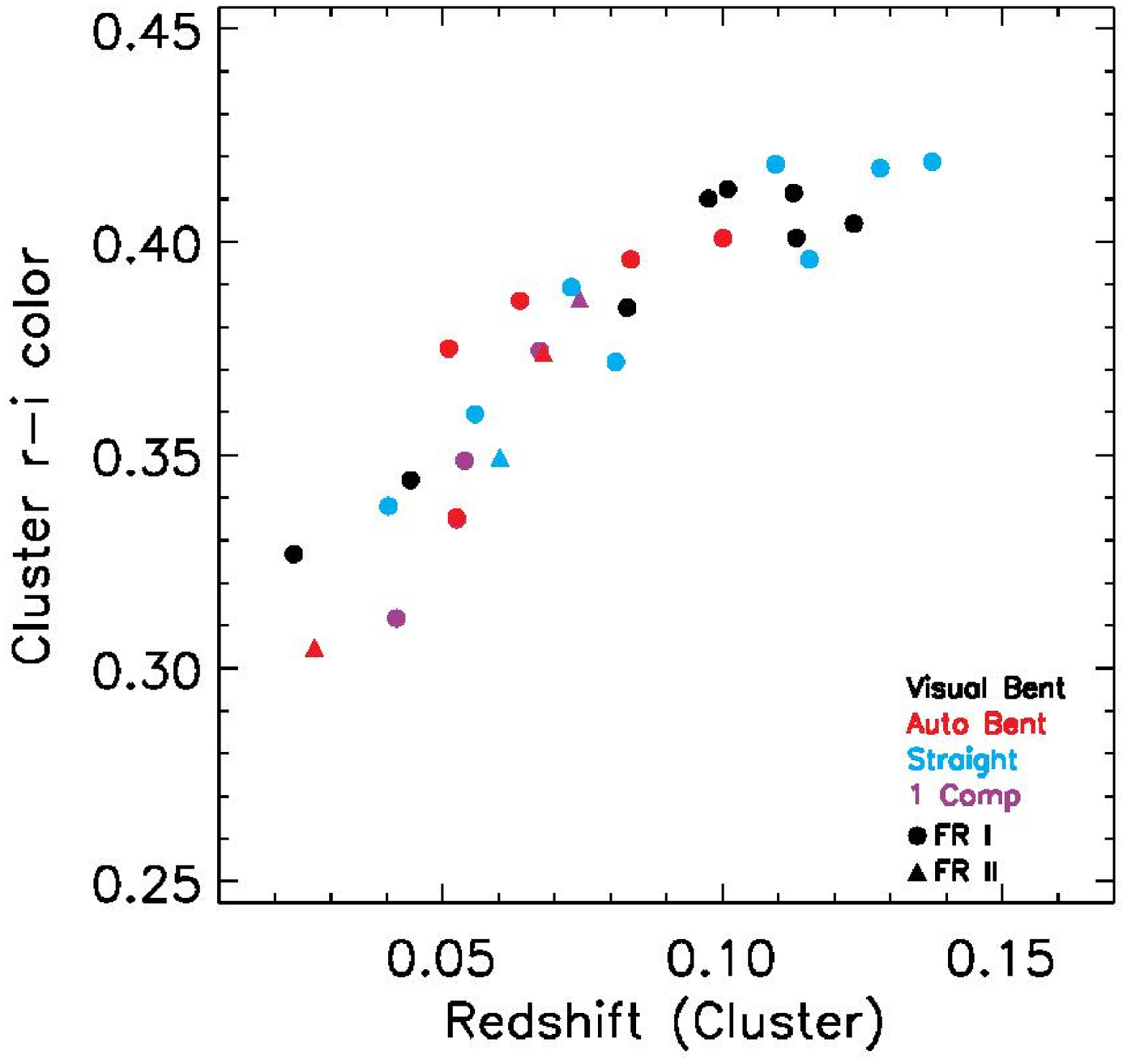}
\includegraphics[scale=0.4]{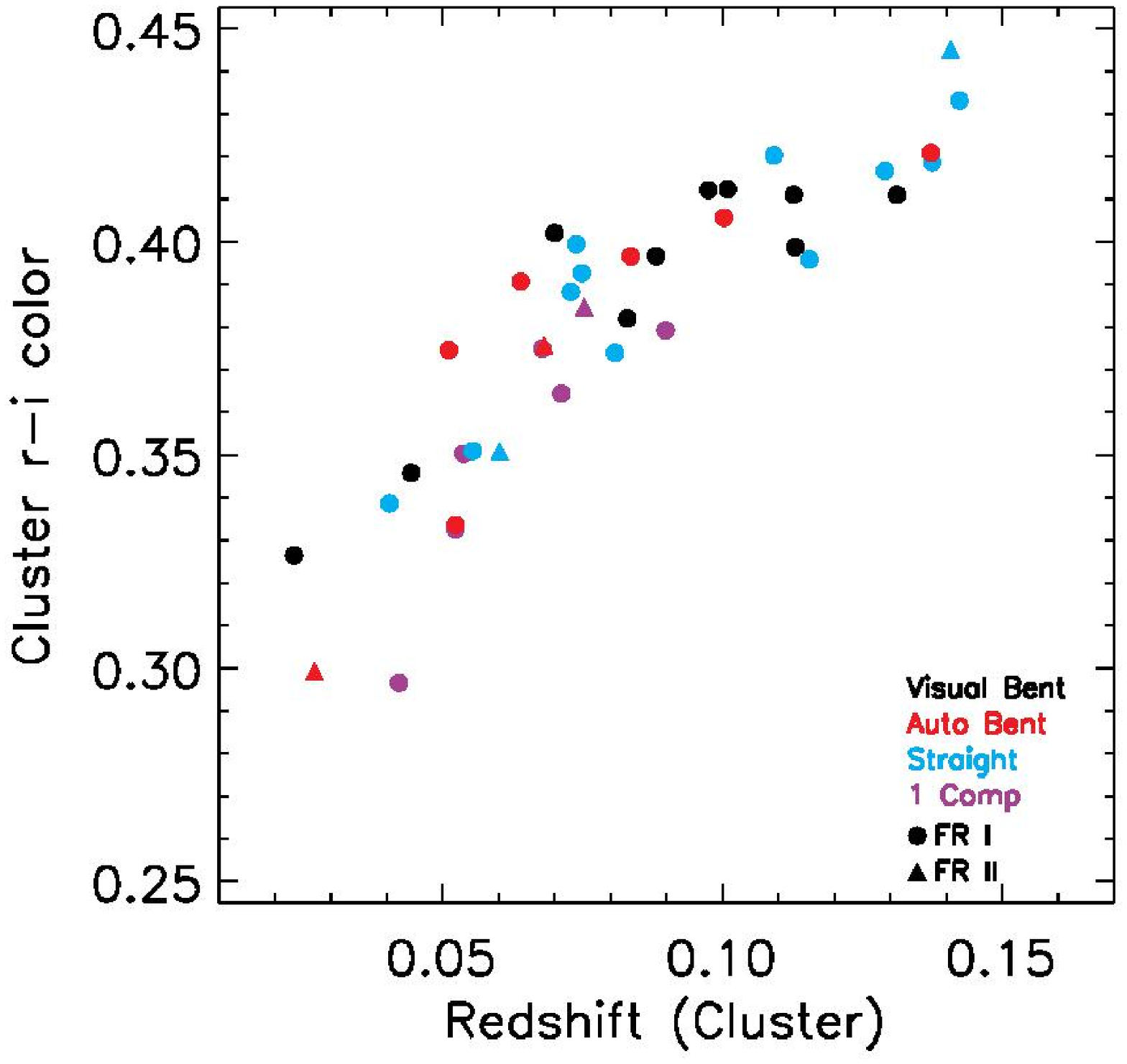}
\caption{Redshift versus $r-i$ average color for the clusters in our sample.  The symbols are the same as in \cref{clustersizevsbeta}.  The left-hand panel shows the results using the fixed gap interloper rejection method and the right panel shows the results using the shifting gapper interloper rejection method.  The redshift is the redshift of the center of the cluster.  There is a very clear correlation between the average de-reddened $r-i$ color of the galaxies within the cluster and the redshift of the cluster.  (A color version of this figure is available online.)} \label{redshiftvsricolor}
\end{center}
\end{figure*}

\index{Results!Additional Correlations Between Cluster Properties}
We examined the average colors of all of the galaxies within each cluster.  \cref{redshiftvsricolor} shows the relationship between the redshift of the cluster and the average $r-i$ color of the galaxies within the cluster.  There is a very strong correlation between the redshift of the cluster and the average $r-i$ color of the galaxies within the cluster.  This is, of course, expected as galaxies are reddened with distance.

\begin{figure*}
\begin{center}
\capstart
\includegraphics[scale=0.4]{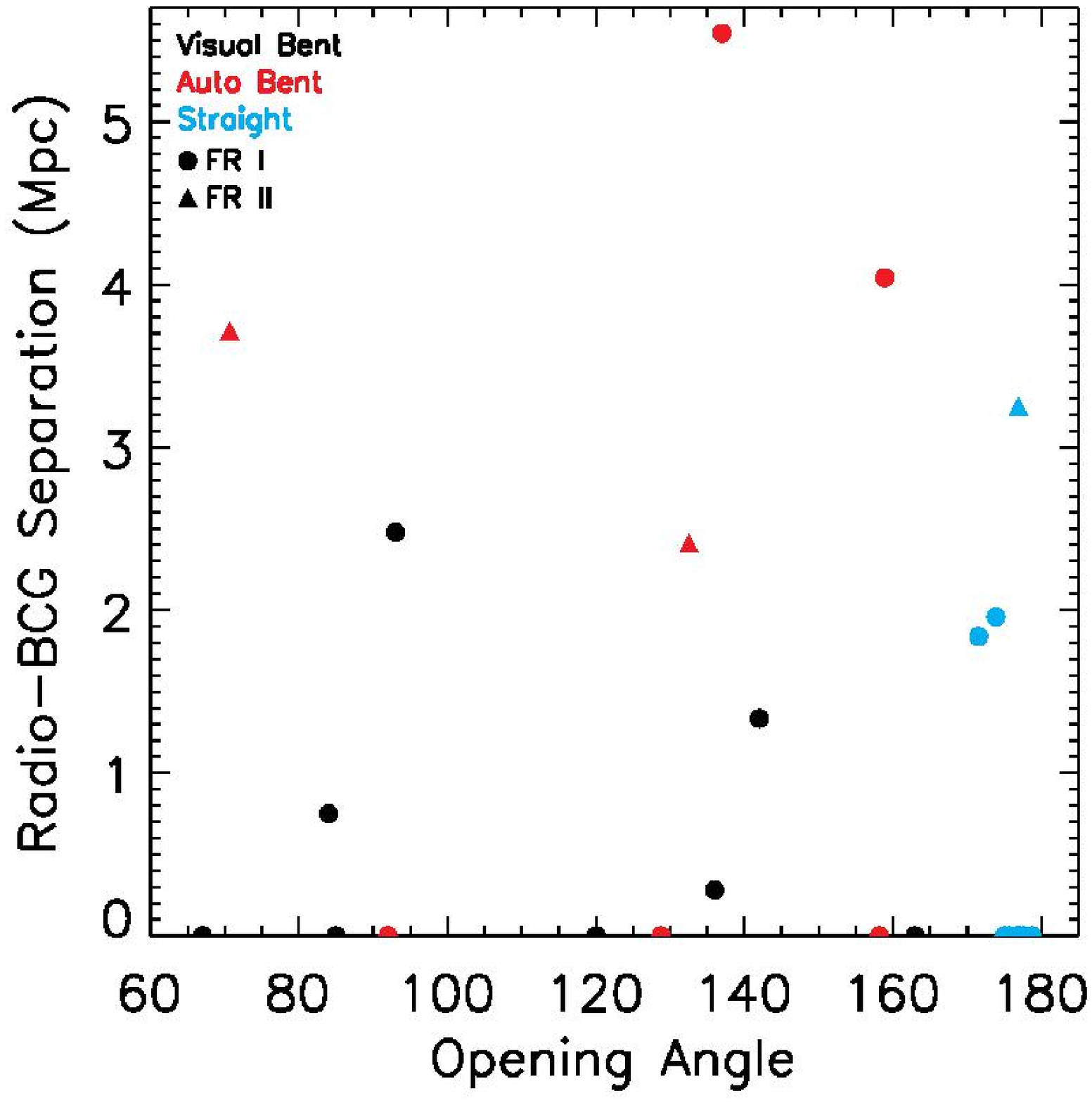}
\includegraphics[scale=0.4]{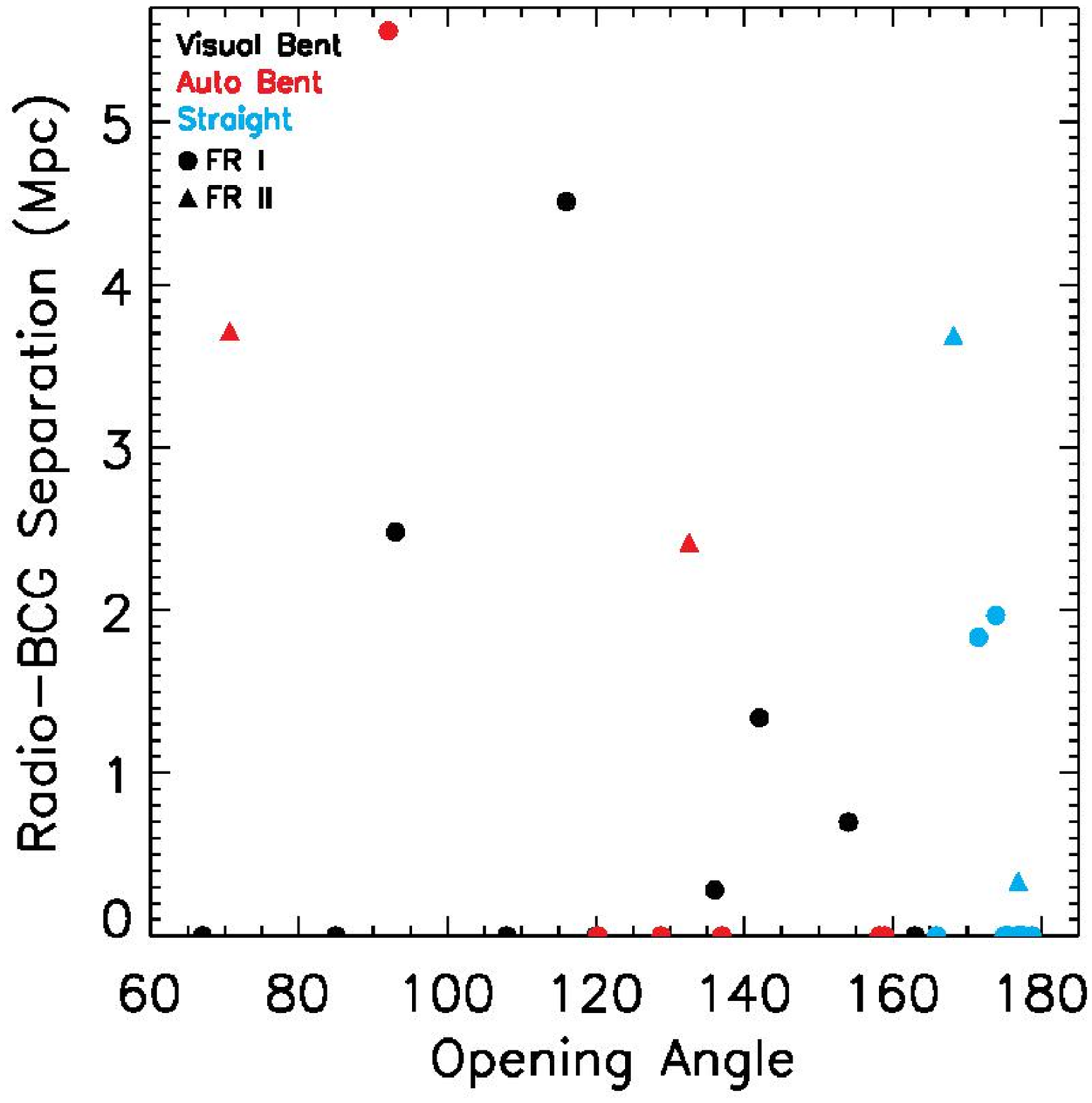}
\includegraphics[scale=0.4]{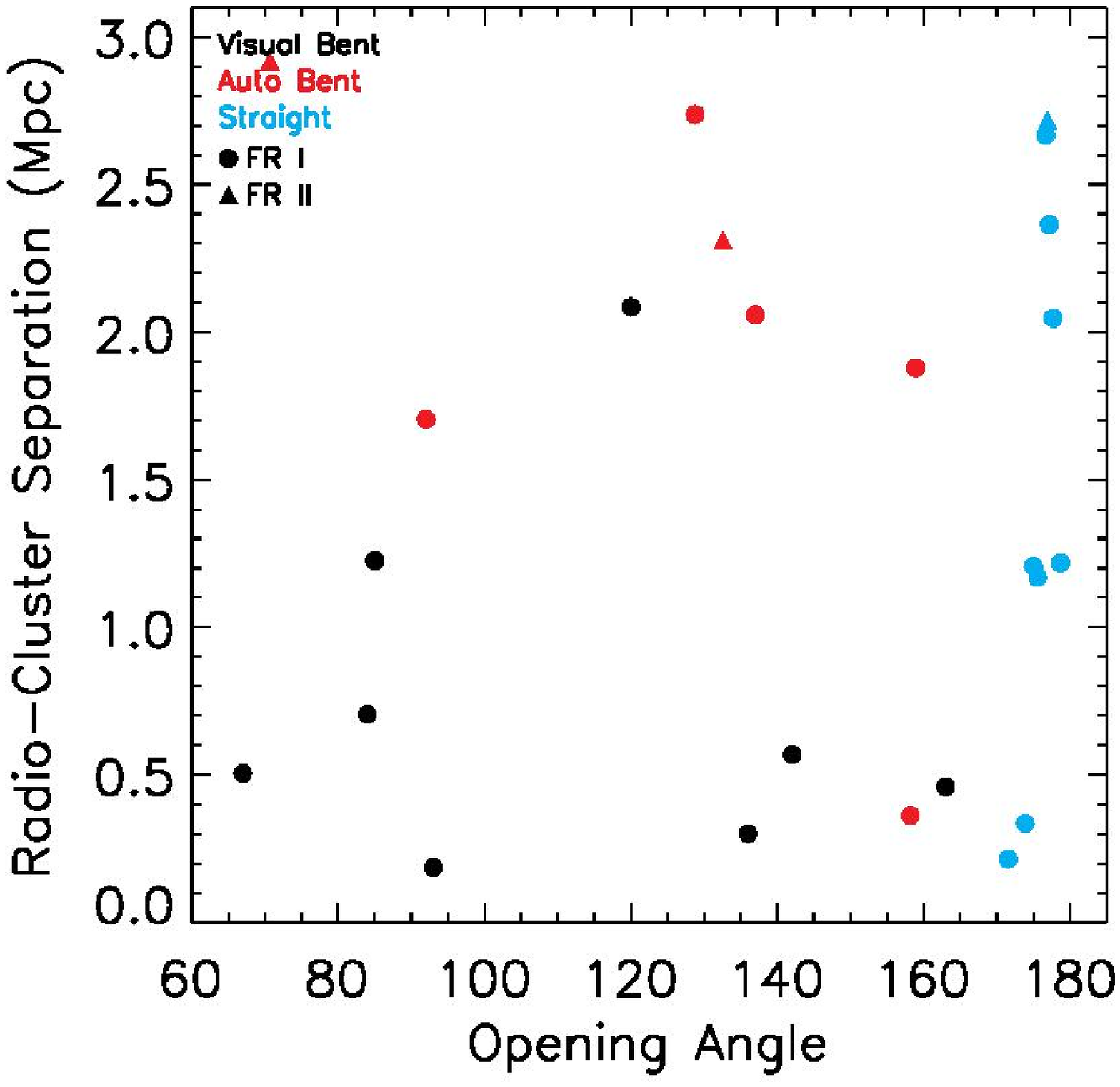}
\includegraphics[scale=0.4]{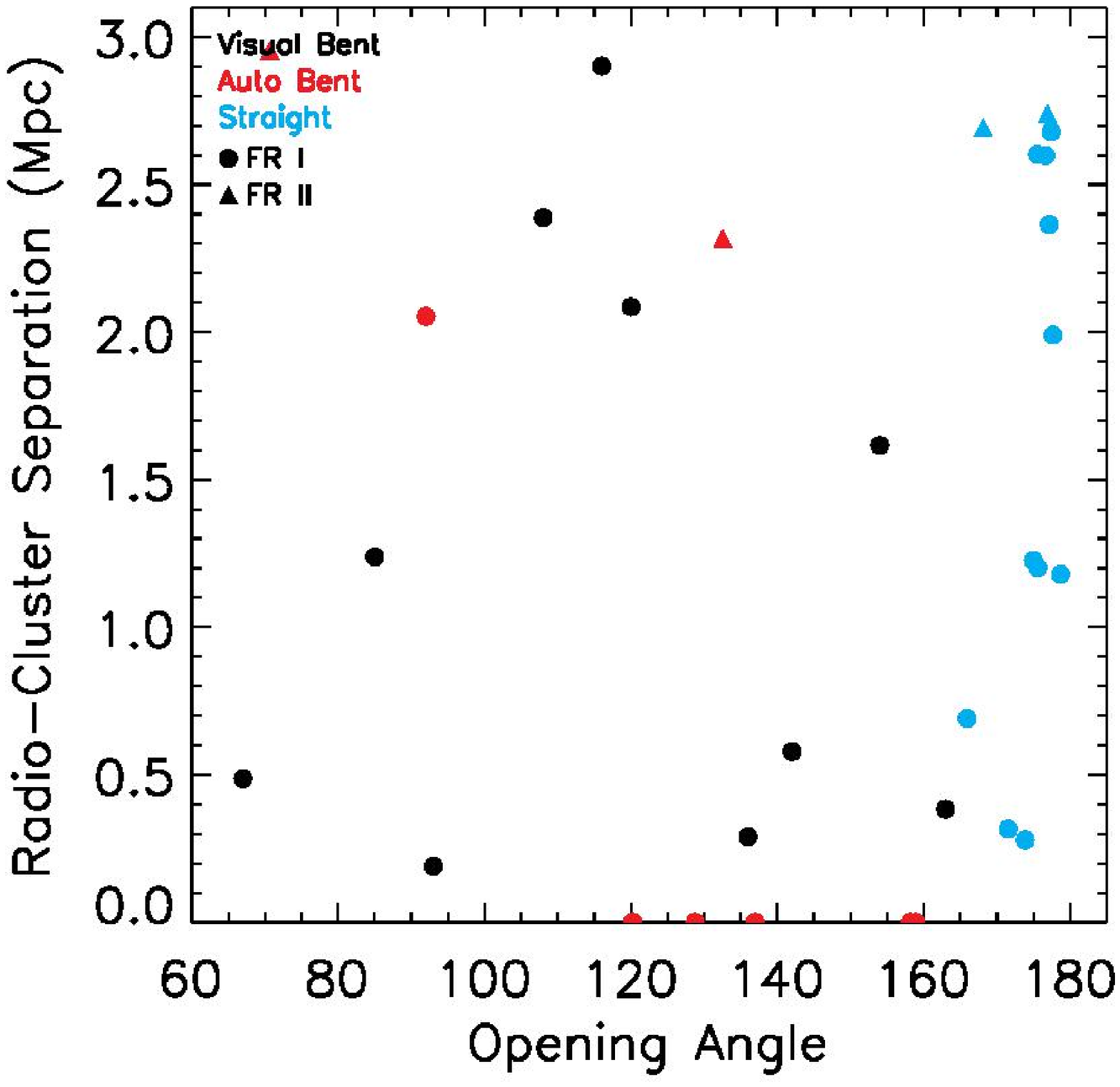}
\caption{Opening angle of the double-lobed radio source vs. the physical distance between the BCG and the radio-host-galaxy (top panels) as well as the physical distance between the bi-weight mean center of the cluster and the radio-host-galaxy (bottom panels).  Symbols are the same as in \cref{clustersizevsbeta}.  The left-hand panel shows the results from using the fixed gap interloper rejection method and the right-hand panel shows the results of the shifting gapper method.  There is no correlation between the opening angle of the radio source and the distance between the radio host galaxy and the BCG or the radio host galaxy and the bi-weight mean cluster center.  This implies that even very near the center of the cluster, radio lobes can be bent to very narrow angles.  (A color version of this figure is available online.)} \label{bendvsbcgsep}
\end{center}
\end{figure*}

We also examined the relationship between the opening angle of double-lobed radio sources and the distance between the radio source and the BCG.  A large separation between the radio host galaxy and the BCG implies that the radio galaxy is located near the outskirts of the cluster.  If it is easier to bend a double-lobed source located near the outskirts of a cluster, where galaxies have higher velocities, we might expect there to be a correlation between the opening angle of the radio lobes and the distance from the cluster center, with the more bent sources being preferentially located near the outskirts.  \cref{bendvsbcgsep} shows this relationship.
There is no correlation between the opening angle of the radio lobes and the projected physical separation between the BCG and the radio host galaxy.  Thus, even for radio sources located near the center of the cluster where the peculiar velocity is typically small, the lobes can be bent to small opening angles.  It is possible that this is due to projection effects.

\begin{figure*}
\begin{center}
\capstart
\includegraphics[scale=0.4]{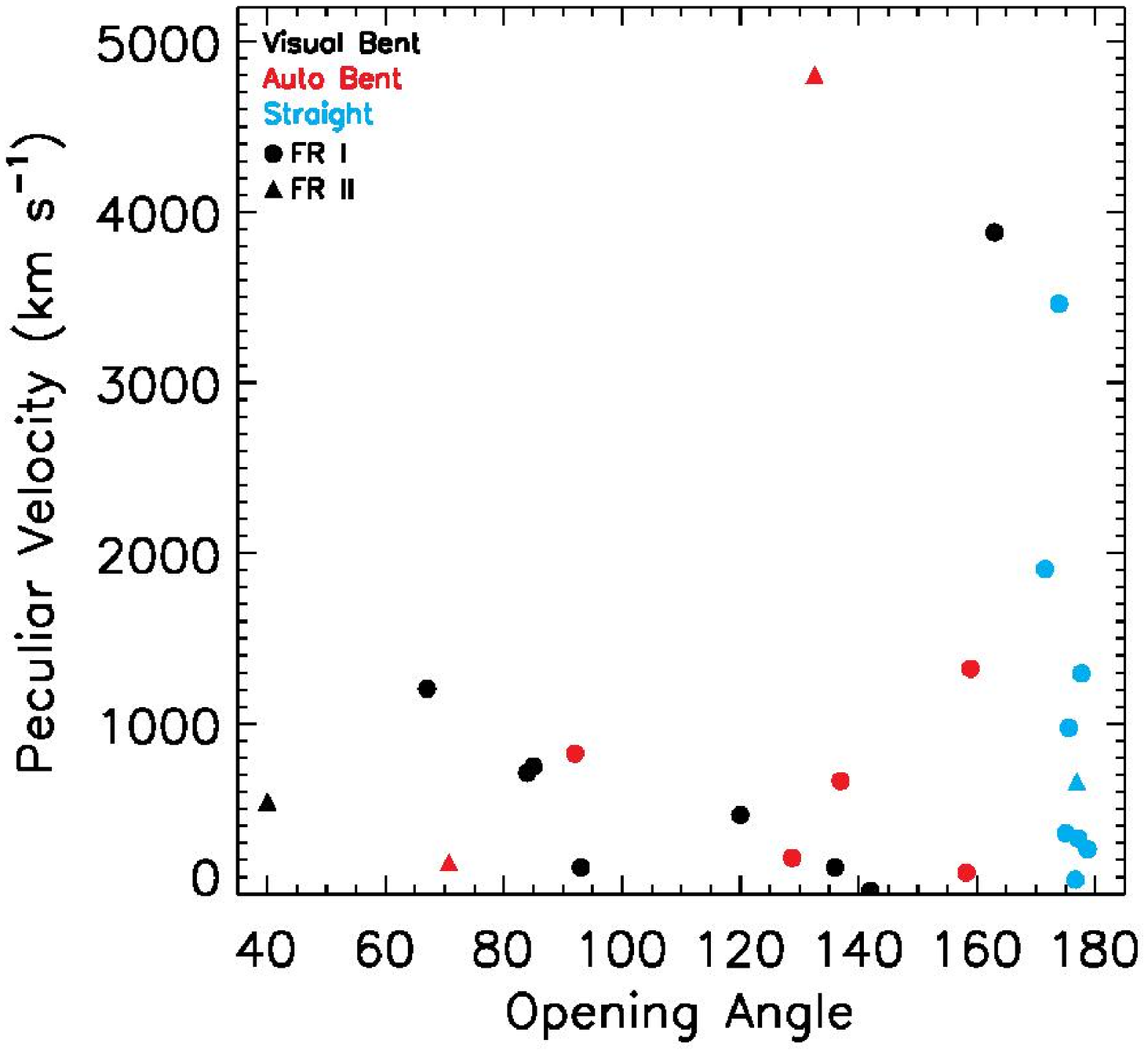}
\includegraphics[scale=0.4]{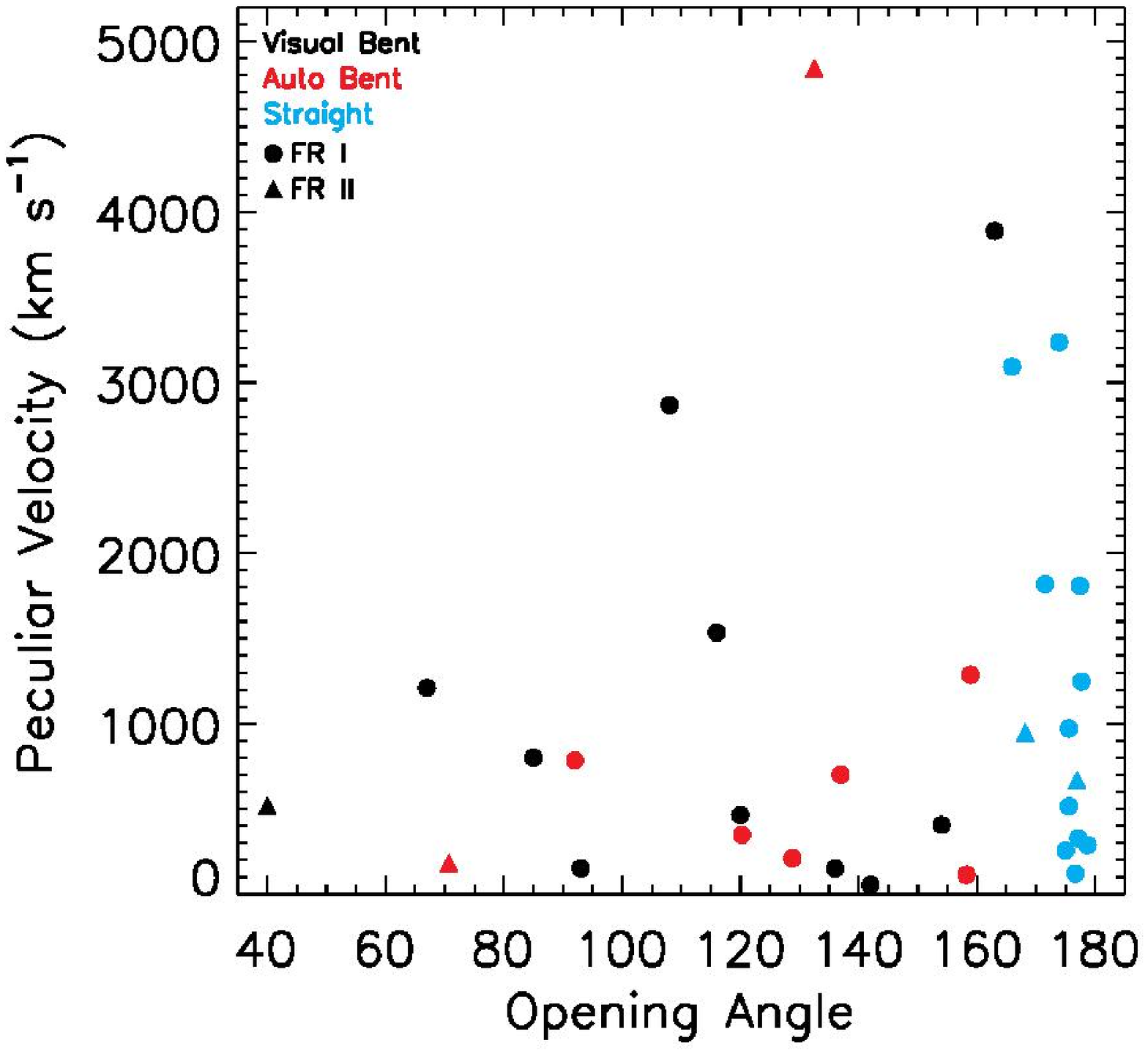}
\includegraphics[scale=0.4]{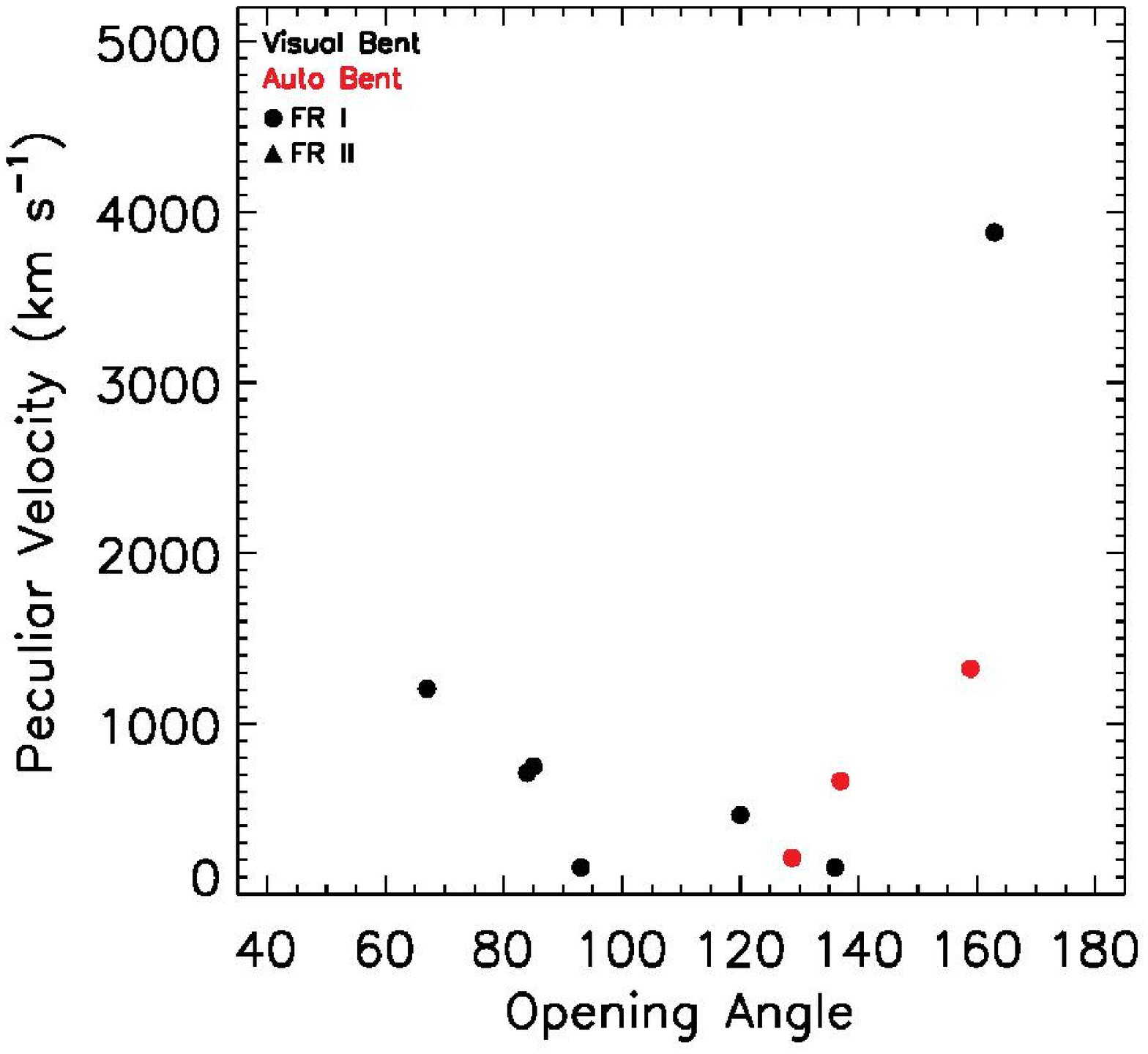}
\includegraphics[scale=0.4]{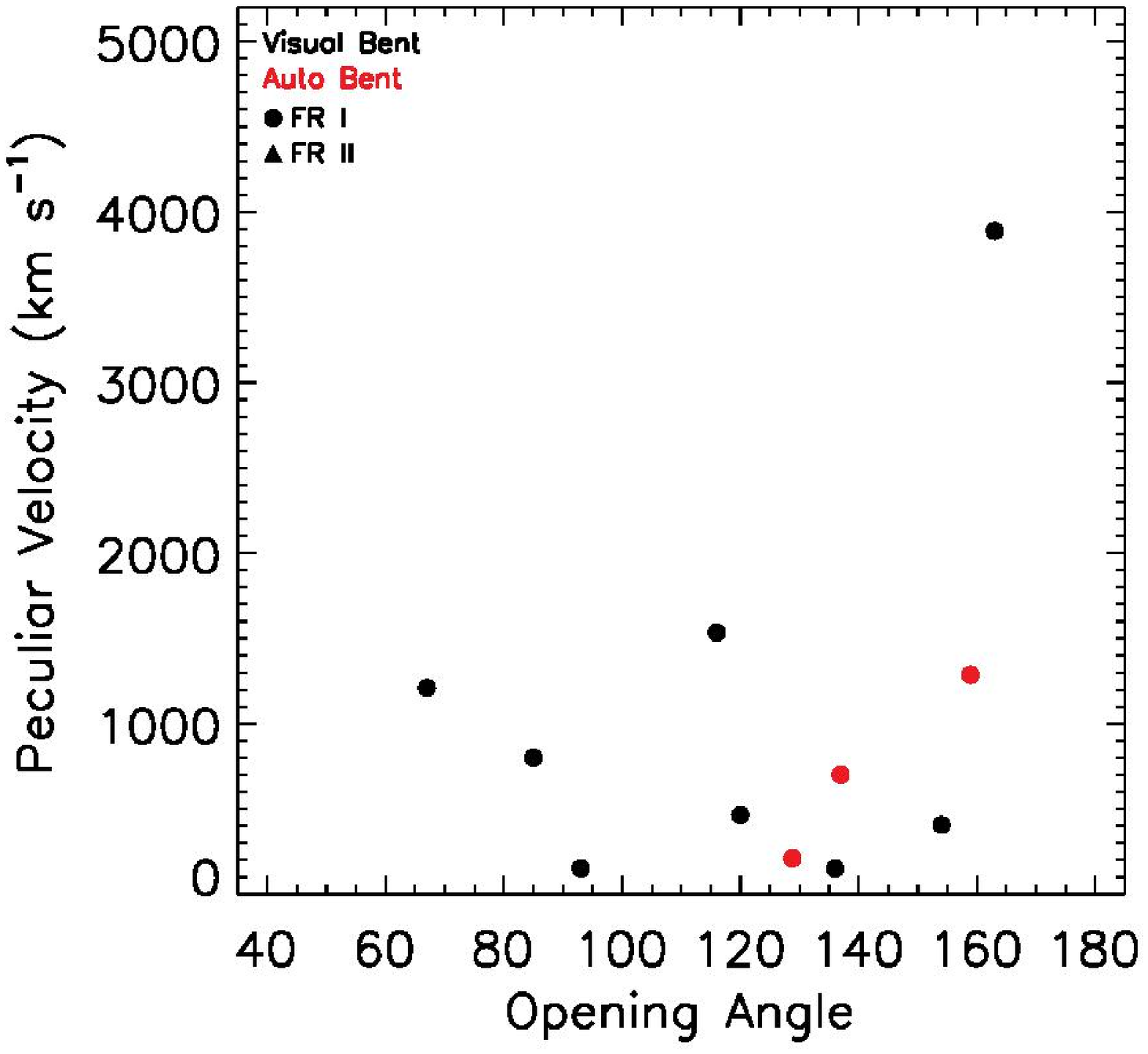}
\caption{Opening angle of the double-lobed radio source vs. the absolute value of the peculiar velocity of the radio host galaxy.  Symbols are the same as in \cref{clustersizevsbeta}.  The left-hand panels show the results with the fixed gap interloper rejection method and the right-hand panels show the results of the shifting gapper method.  The bottom panels show the correlation when only using those sources in the visual-bent and auto-bent samples that we identified to be the most likely true double-lobed radio sources.  There is no correlation between the opening angle of the radio source and the peculiar velocity of the radio host galaxy.(A color version of this figure is available online.)} \label{anglevspecvel}
\end{center}
\end{figure*}

We have also examined the relationship between the peculiar velocity of the radio source and the opening angle of the radio source.  \cref{anglevspecvel} shows this relationship.
There does not appear to be any correlation between these properties.  One possible reason for the lack of correlation is the angle at which we are viewing the double-lobed radio source.  Depending on the angle of inclination, the true opening angle of the radio source could be significantly different than what we have measured.  In general, a straight source will appear straight no matter the angle of inclination.  However, if the line-of-sight lies near the plane of the bend, it is possible for a bent source to appear straight.  If the axis is nearly parallel to the line-of-sight, a slightly bent source can appear bent at an exaggerated angle.  A closer examination of \cref{anglevspecvel} shows that most of our double-lobed radio sources with opening angles less than $100^{\circ}$ have peculiar velocities of less than $\sim1000$ km s$^{-1}$.  The peculiar velocities are line-of-sight components, so the motion in the plane of the sky that would lead to significant bending such as this is not measured.

\begin{figure*}
\begin{center}
\capstart
\includegraphics[scale=0.4]{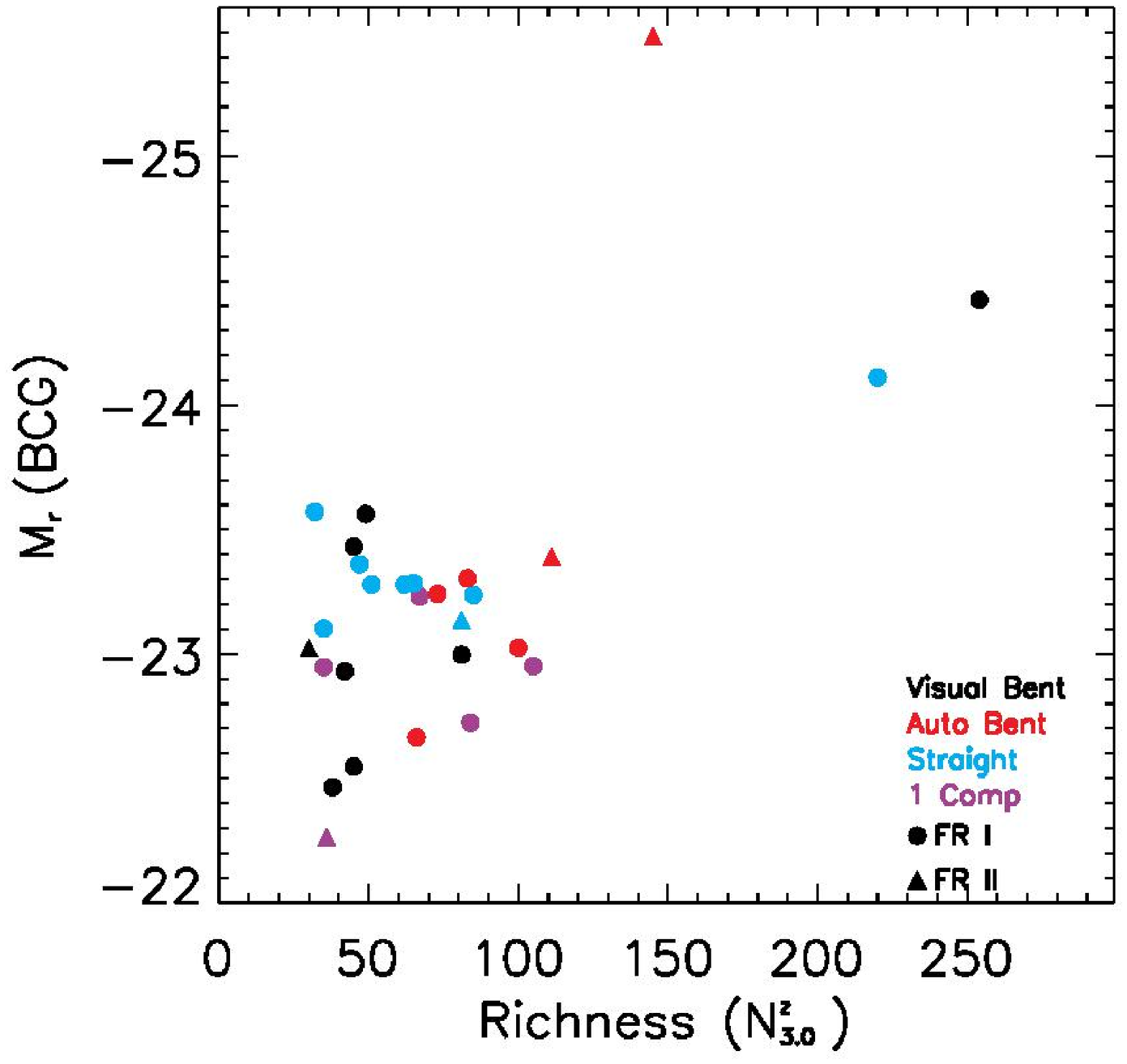}
\includegraphics[scale=0.4]{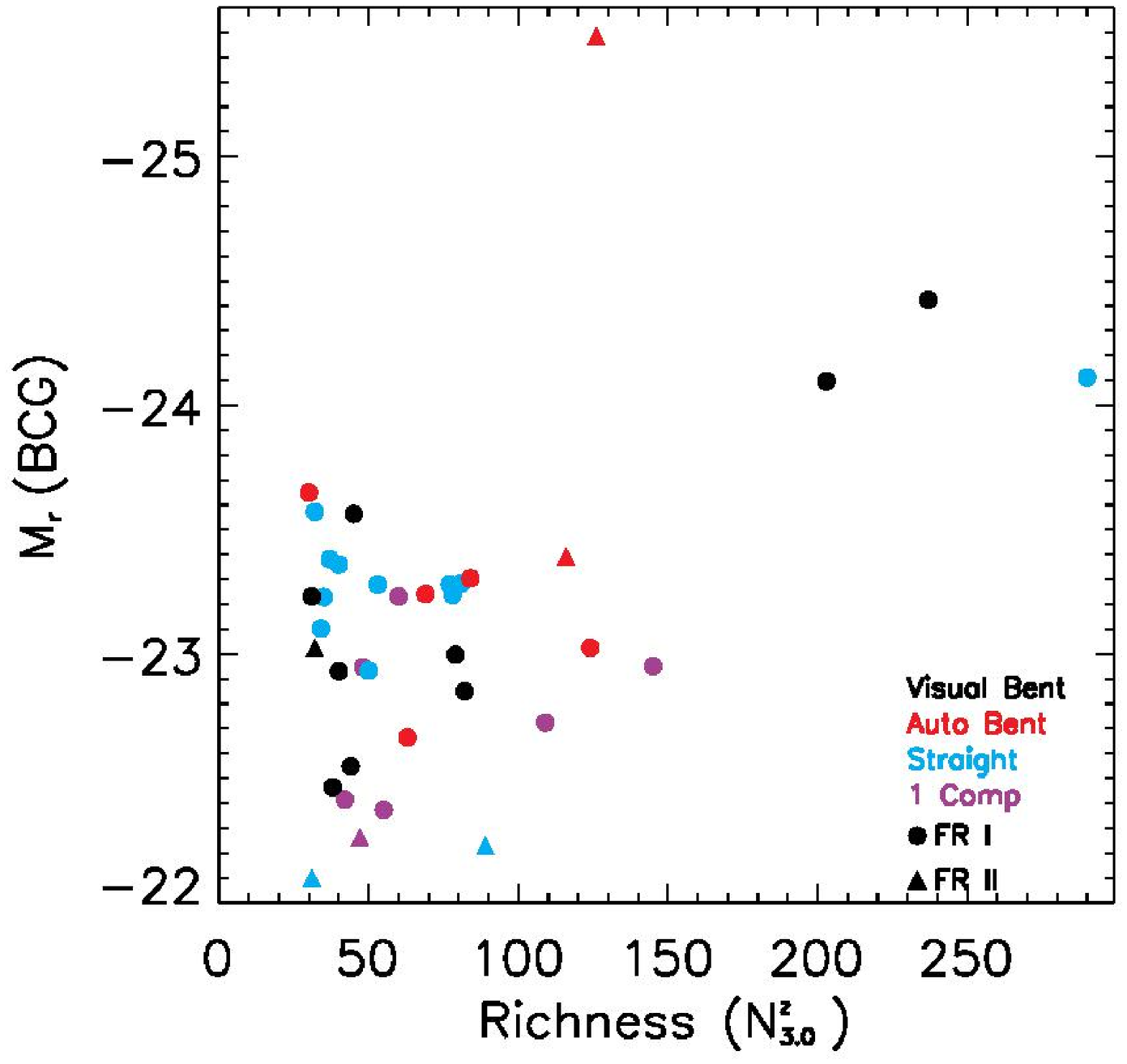}
\includegraphics[scale=0.4]{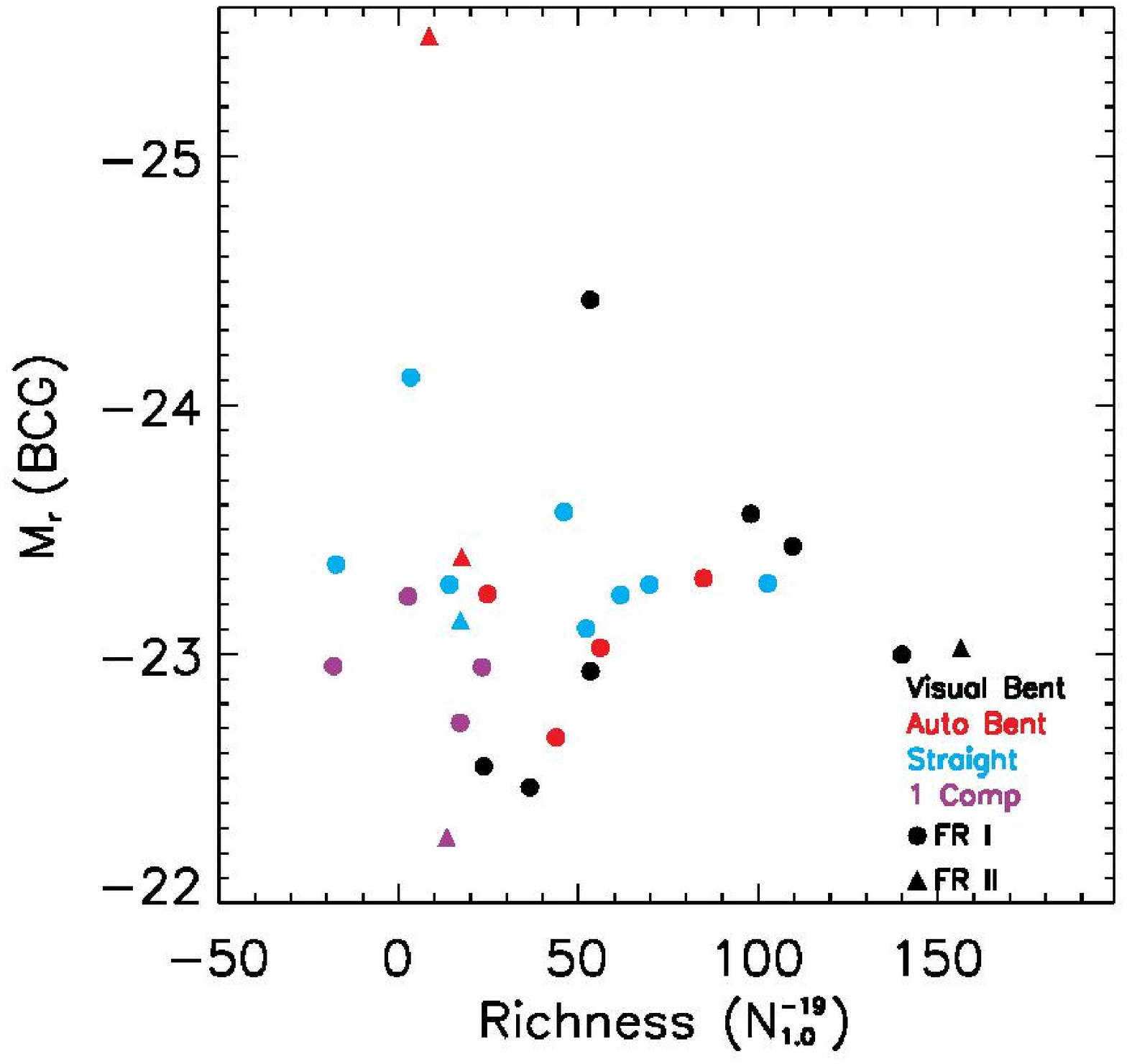}
\includegraphics[scale=0.4]{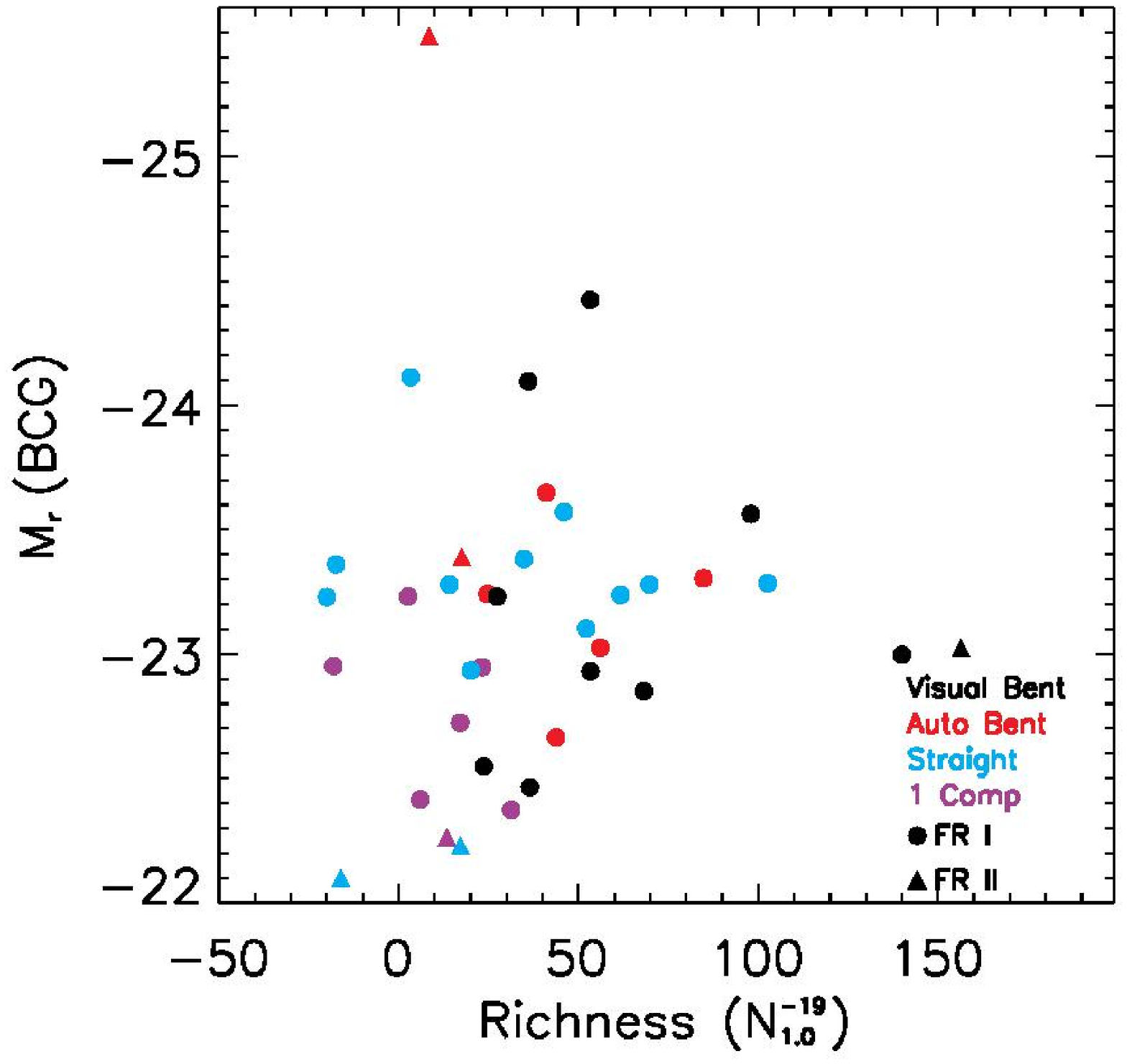}
\caption{Cluster richness vs $M_{r,BCG}$.  Symbols are the same as in \cref{clustersizevsbeta}.  The left-hand panels shows the results with the fixed gap interloper rejection method and the right-hand panels shows the results of the shifting gapper method.  The top panels use richness as defined by $N^z_{3.0}$ and the bottom panels use richness as defined by $N^{-19}_{1.0}$.  There is a relatively strong correlation between the richness of the cluster and the $r$-band absolute magnitude of the BCG.  The richest clusters also have the brightest BCGs.  (A color version of this figure is available online.)} \label{redclustersizevsbcgmag}
\end{center}
\end{figure*}

We also examined the correlation between the richness of the cluster and the $r$-band absolute magnitude of the BCG.  We expect that the richest clusters are more likely to have a more luminous BCG than poor clusters.  \cref{redclustersizevsbcgmag} shows the relationship between these parameters.
Each of these shows a slight negative correlation, implying that the richest clusters are also host to the most luminous BCGs.  Using the fixed gap method of interloper rejection, the Spearman correlation coefficient is $-0.37$, with a significance of $1.92\sigma$, when richness is measured by $N^z_{3.0}$.  For the shifting gapper method, the Spearman correlation coefficient is  $-0.23$, with a significance of $1.39\sigma$.

\begin{figure*}
\begin{center}
\capstart
\includegraphics[scale=0.4]{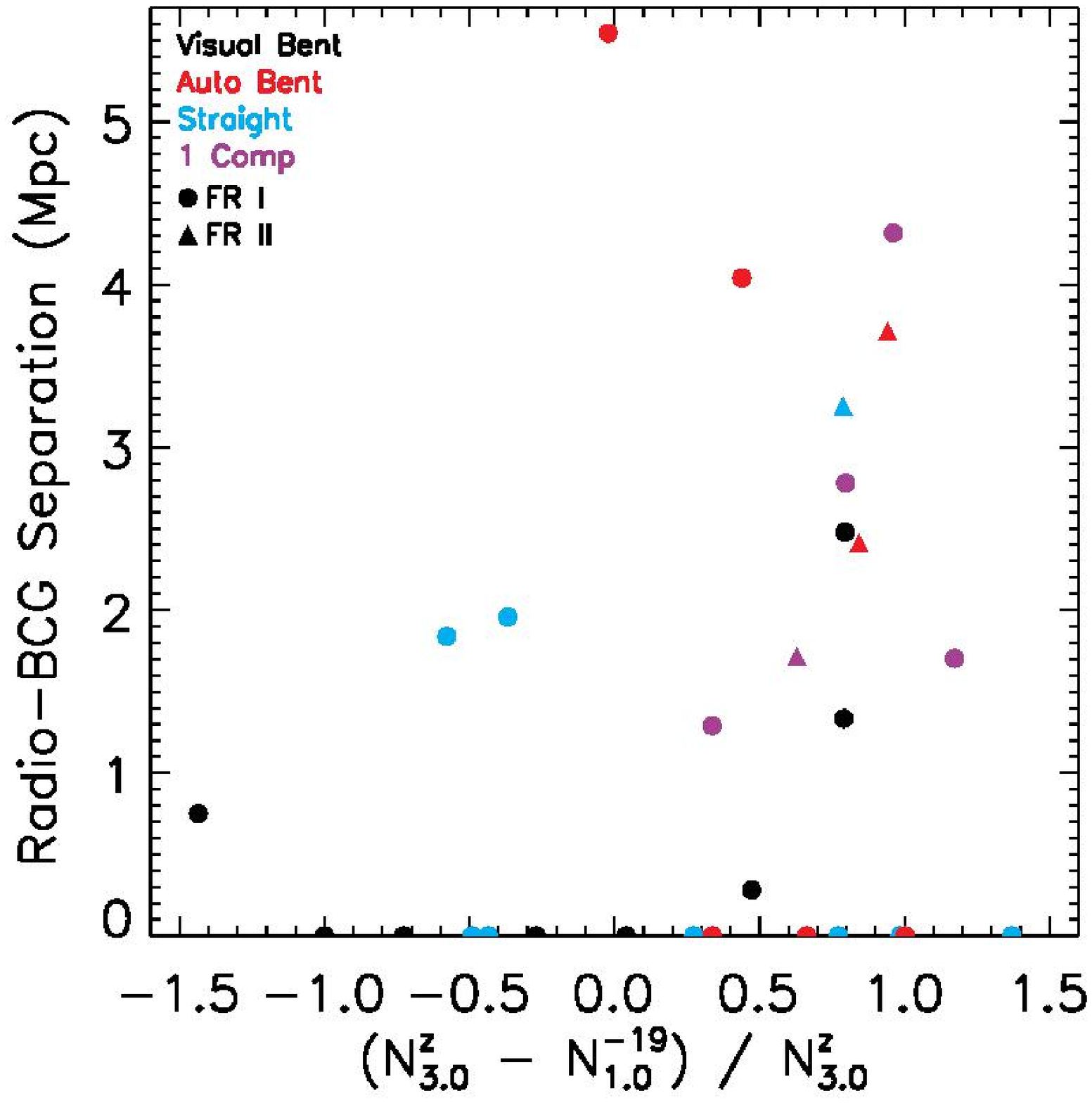}
\includegraphics[scale=0.4]{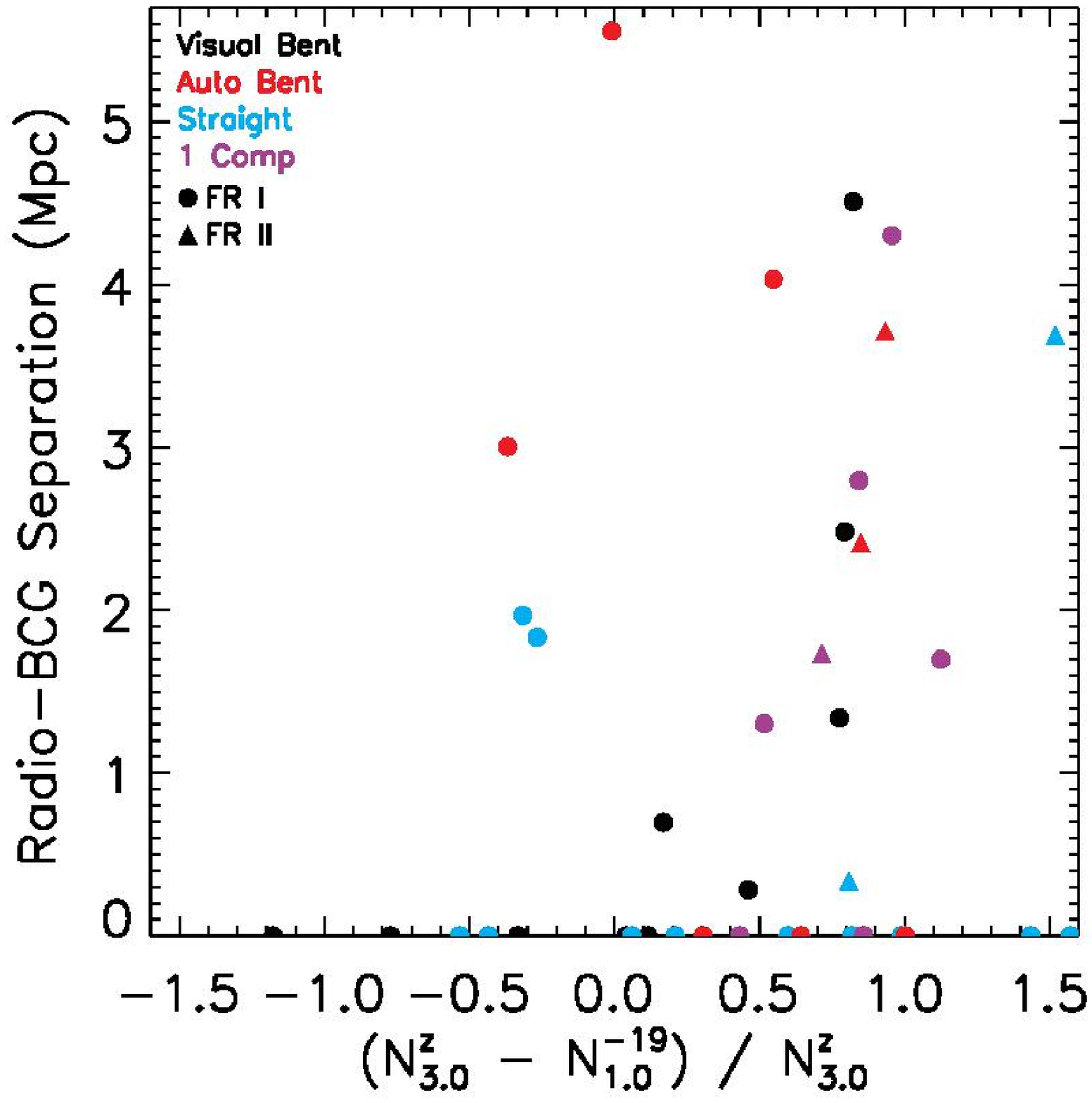}
\includegraphics[scale=0.4]{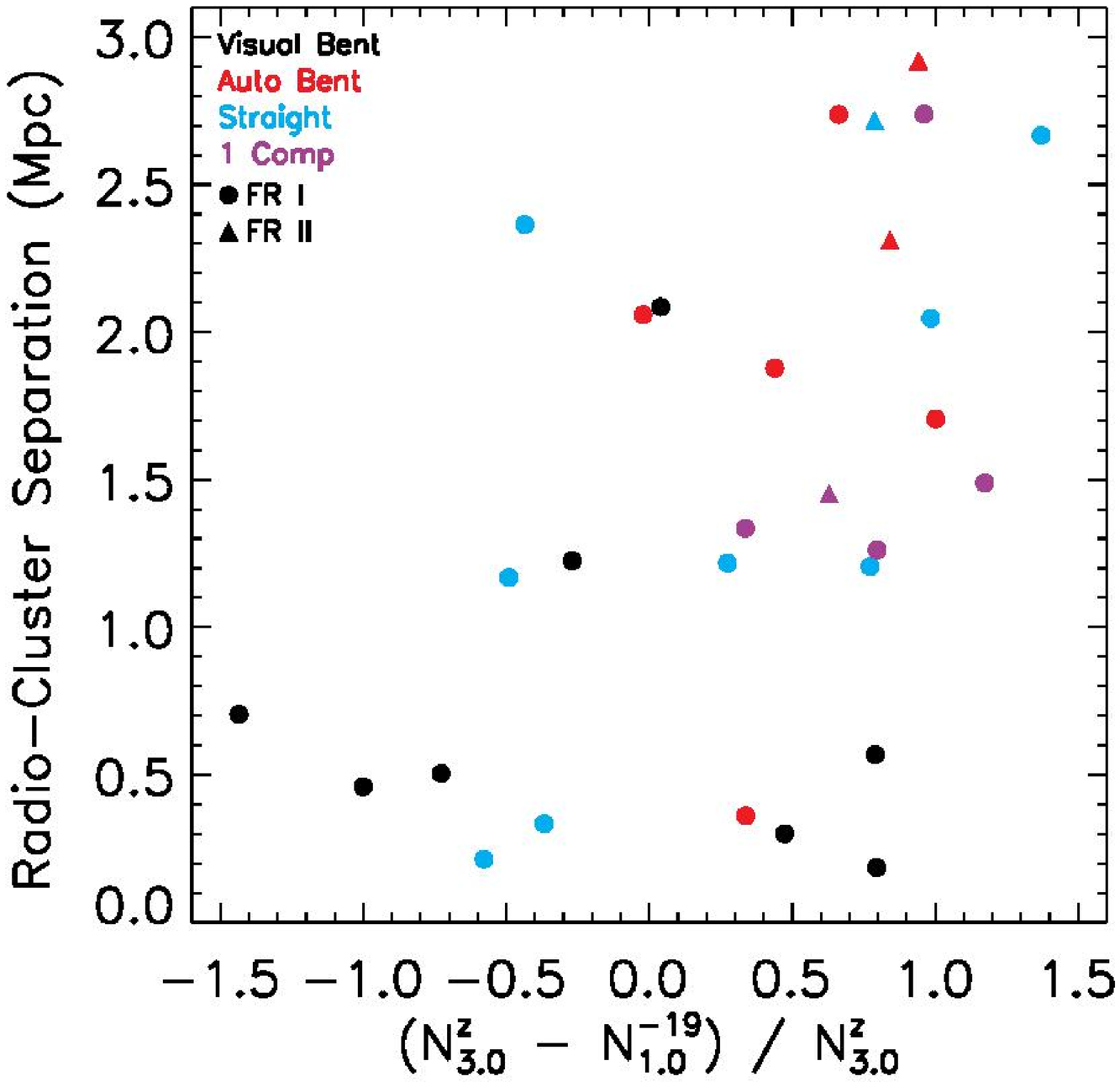}
\includegraphics[scale=0.4]{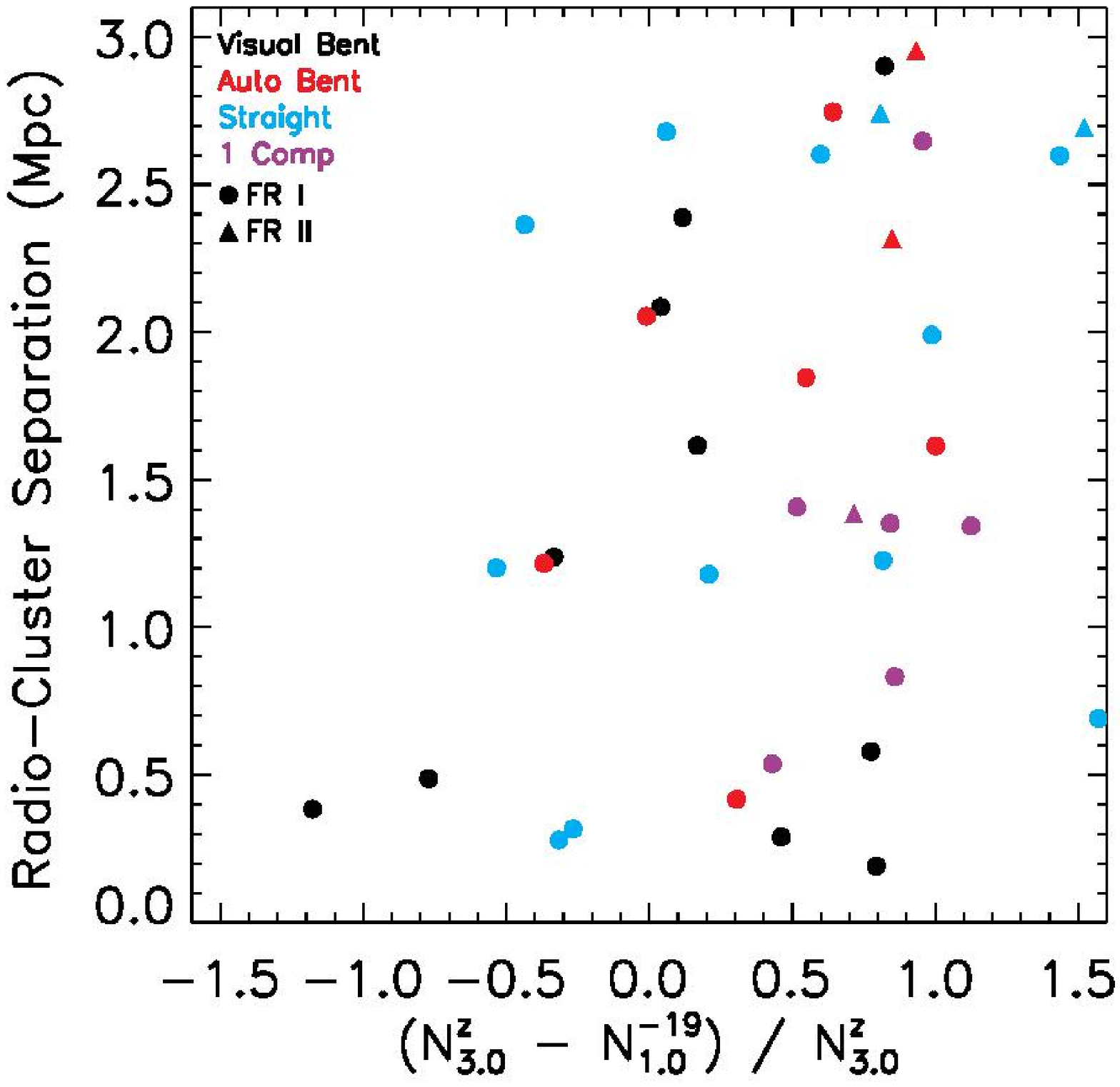}
\caption{Relationship between $N^z_{3.0} - N^{-19}_{1.0}/N^z_{3.0}$ and the projected physical separation between the BCG and the radio-host-galaxy (top panels), as well as the projected physical separation between the bi-weight mean center of the cluster and the radio-host-galaxy (bottom panels).  Symbols are the same as in \cref{clustersizevsbeta}.  The left-hand panel shows the results with the fixed gap interloper rejection method and the right-hand panel shows the results of the shifting gapper method.  In all of these plots, there is a positive correlation, implying that the farther the radio source is from the center of the cluster, the more the N$^{-19}_{1.0}$ richness metric underestimates the richness of the cluster.  (A color version of this figure is available online.)} \label{clusterfracvsbcgsep}
\end{center}
\end{figure*}

As discussed in \S~\ref{cluster_richness_measurements}, we use two different richness metrics.  The first metric, N$_{1.0}^{-19}$, counts the number of sources brighter than $M_r=-19$ within $1.0$ Mpc of the radio source, at the redshift of the radio source.  The second metric, N$_{3.0}^z$, is the number of spectroscopically confirmed \SDSS\/ sources within $\pm 5000$ km s$^{-1}$ of the center of the cluster and within a radius of $3.0$ Mpc of the center of the cluster, at the redshift of the cluster.  \cref{clusterfracvsbcgsep} shows the comparison between both the projected physical separation between the BCG and the radio-host galaxy and the normalized (by the N$_{3.0}^z$ richness metric) difference between the two different measured cluster richnesses, as well as the projected physical separation between the bi-weight mean center of the cluster and the radio-host galaxy.
There is a positive correlation between these values.  The implication of this positive correlation is that the difference between the N$_{1.0}^{-19}$ richness metric and the N$_{3.0}^z$ richness metric increases as the separation between the radio-host-galaxy and the BCG (or bi-weight mean center of the cluster) increases.  Thus, the farther the radio-host-galaxy is located from the center of the cluster, the more divergent the richness metrics become.  The N$^{-19}_{1.0}$ metric is clearly measuring a less rich cluster when the radio-host-galaxy is located near the outskirts of the cluster as compared to the N$_{3.0}^z$ richness metric.  This is not unexpected, as the N$^{-19}_{1.0}$ richness metric is centered on the radio-host-galaxy and the density of cluster members on the outskirts of clusters is lower than near the center.

\subsection{Sources with Abnormally High Velocity Dispersions}
\begin{figure*}
\begin{center}
\capstart
\includegraphics[scale=0.4]{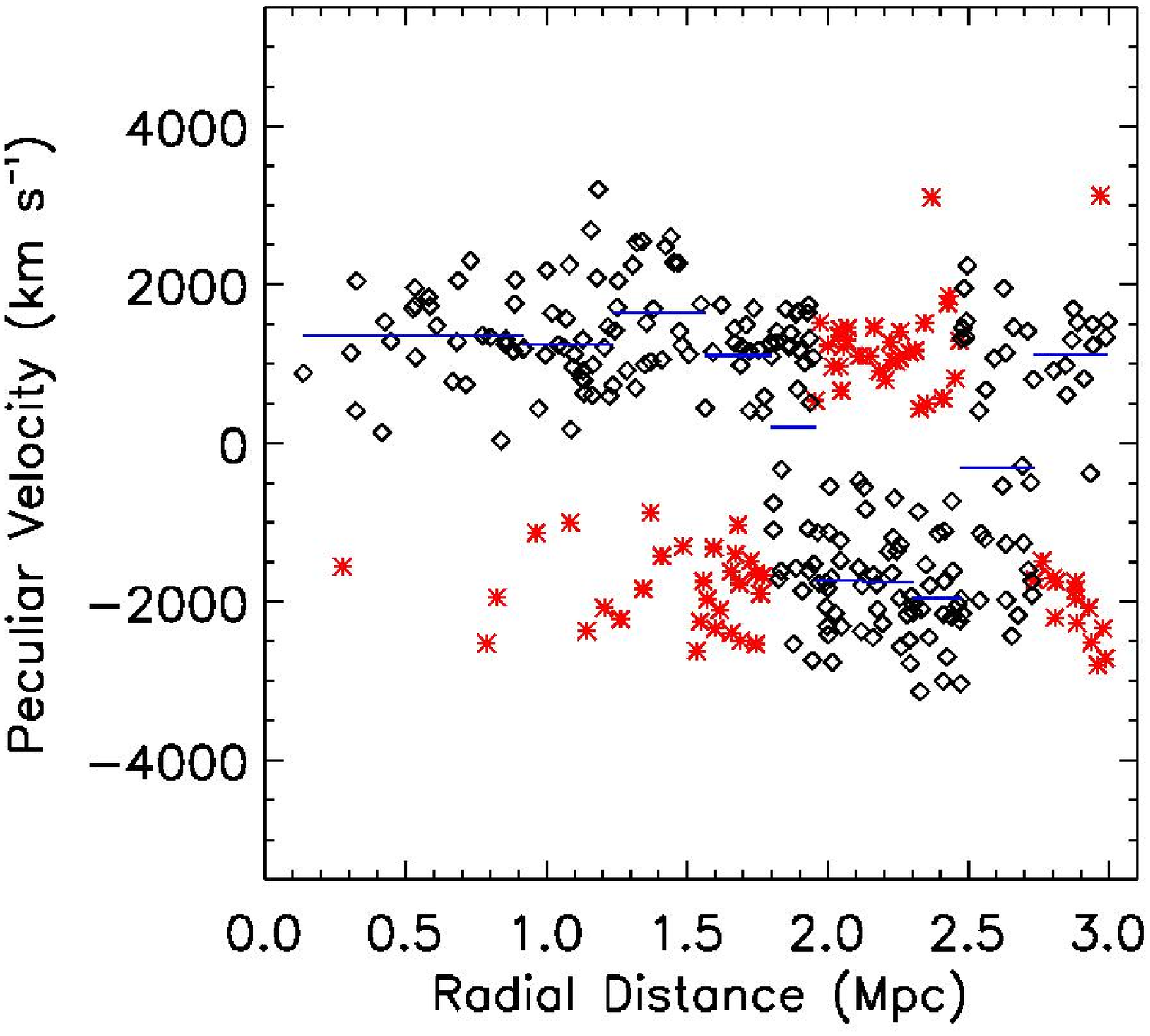}
\includegraphics[scale=0.4]{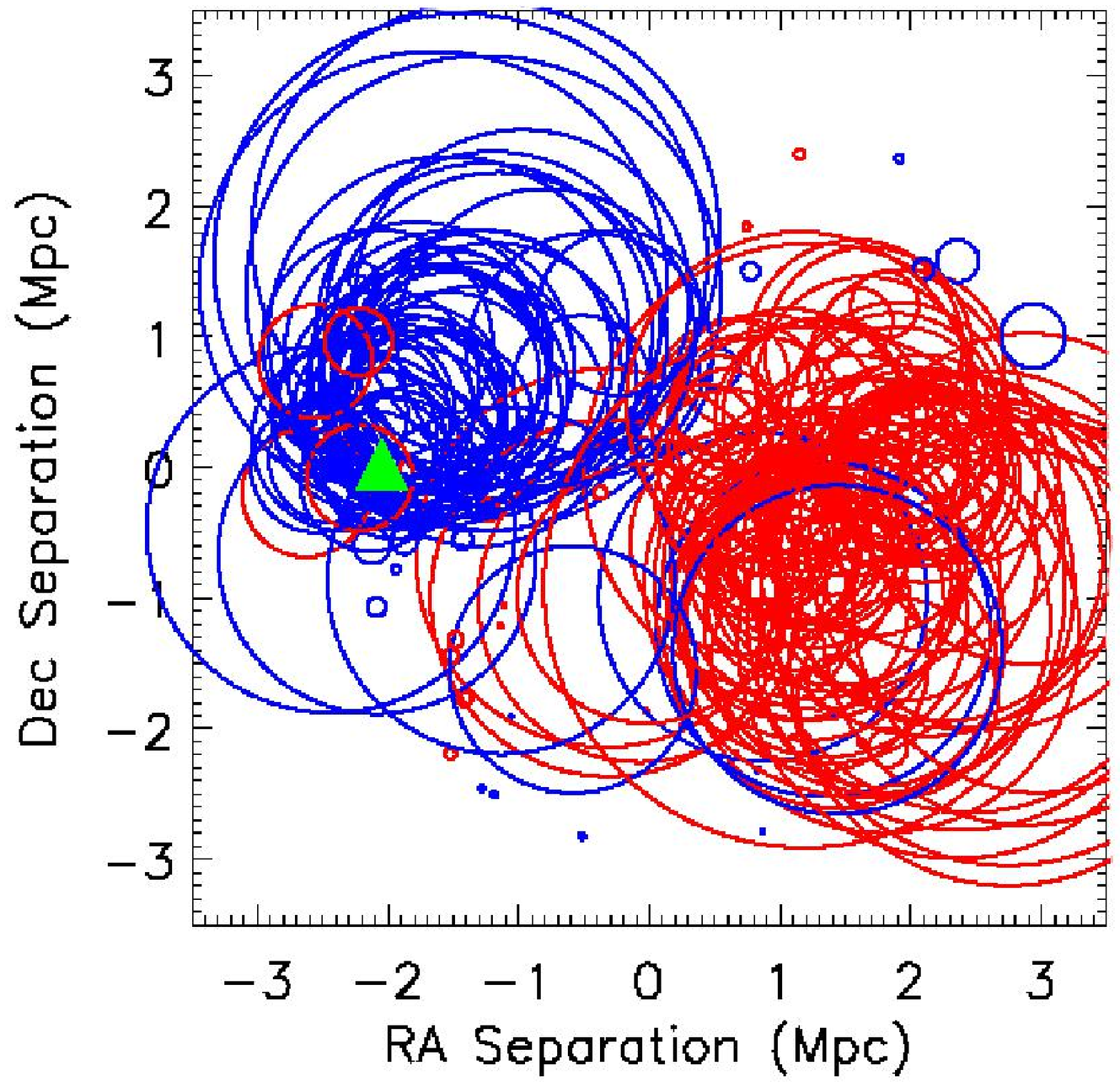}
\includegraphics[scale=0.4]{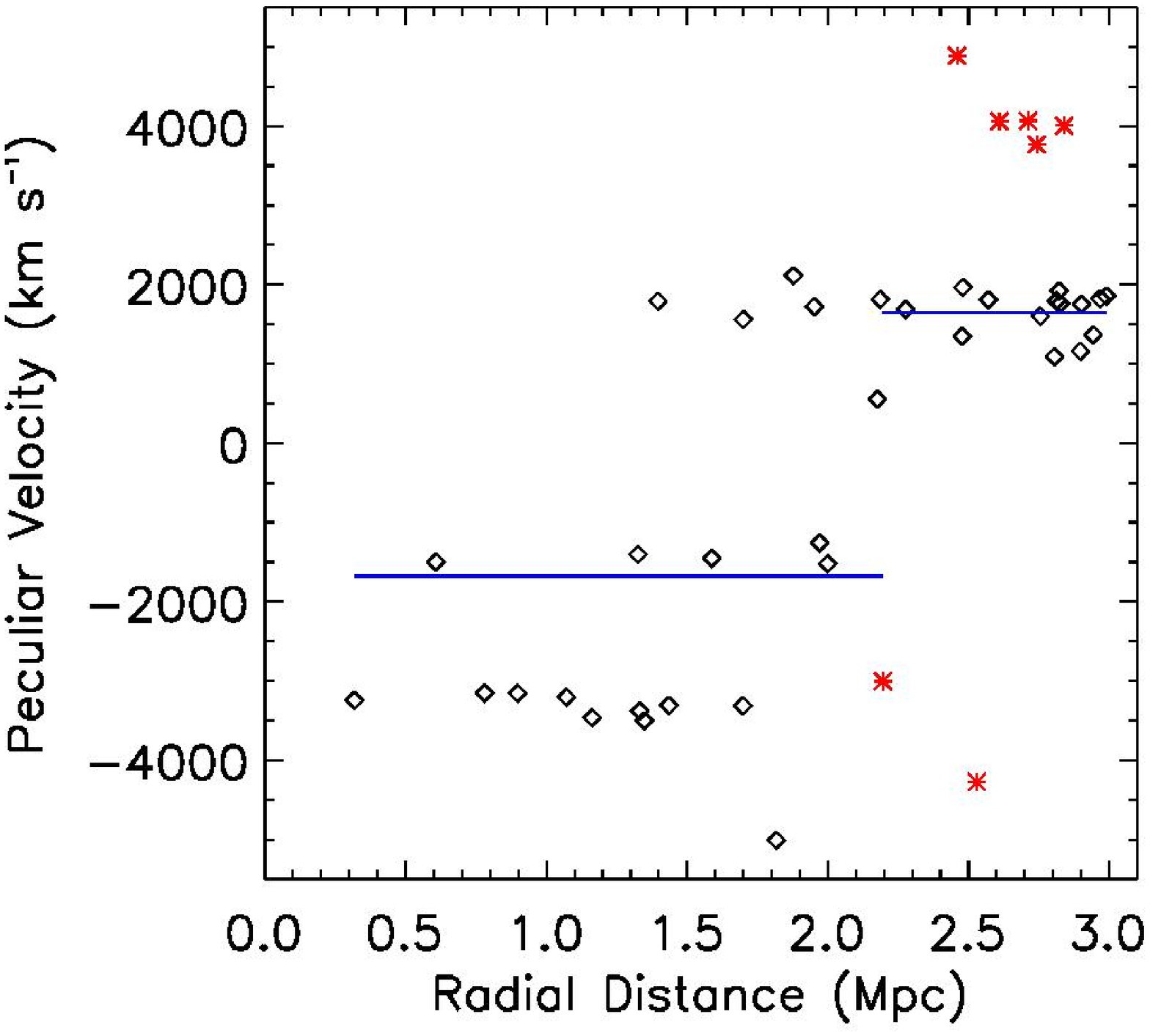}
\includegraphics[scale=0.4]{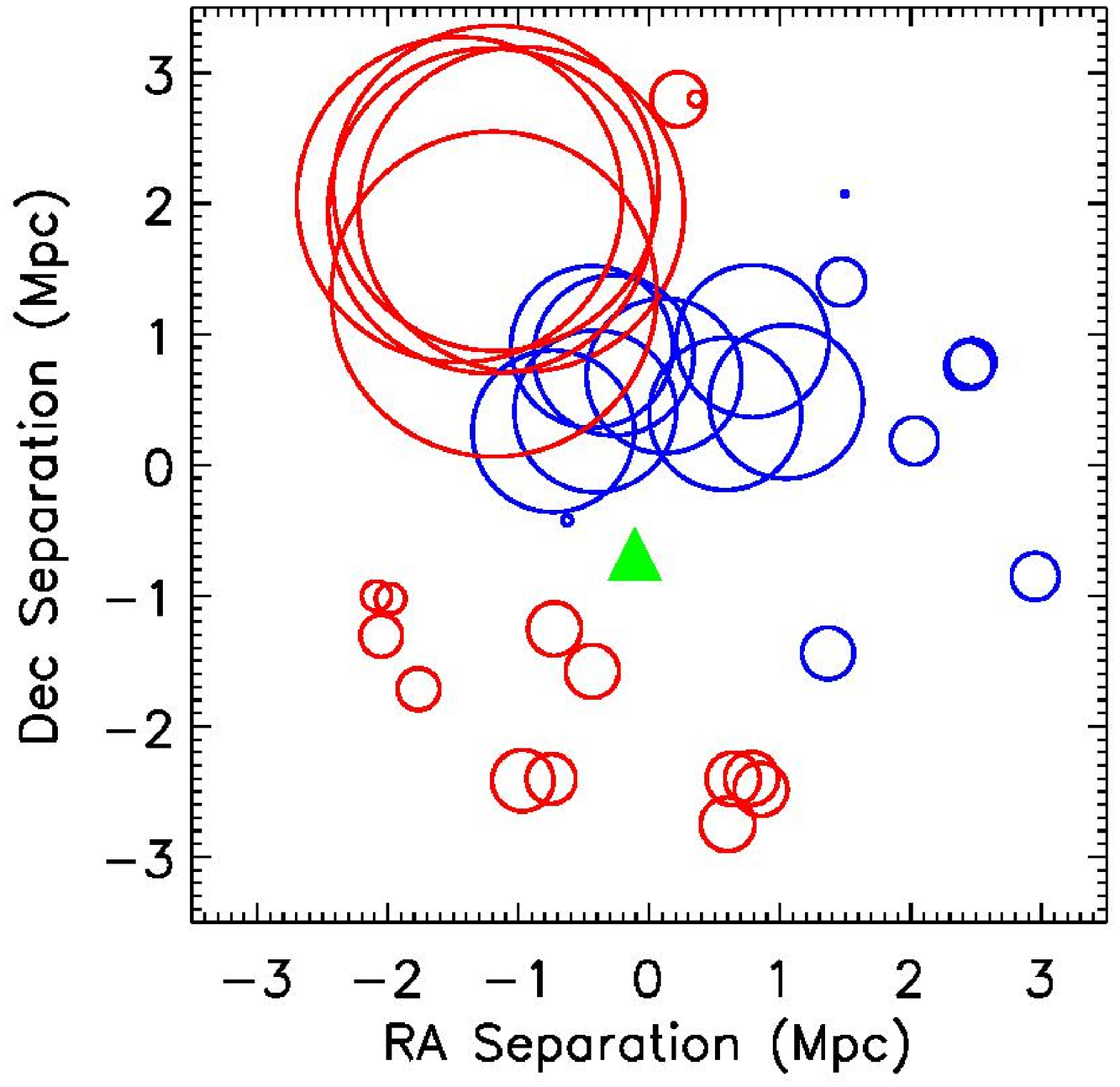}
\caption{The top left-hand panel shows the peculiar velocity of galaxies within the cluster associated with the radio source J151131.3+071506 as a function of distance from the center of the cluster.  This is the same as \cref{velspread}, except for a different cluster.  We see what appear to be two sub-clusters, located about $3000$ km s$^{-1}$ apart in peculiar velocity.  The top right-hand panel is the same as \cref{bubble}, except for this cluster.  Clearly there are two sub-clusters that are separated in both peculiar velocity and angular position that appear to be in the process of merging.  The bottom two panels are the same as the top panels, except for the source J121121.1+141439, with the shifting gapper interloper rejection method.  (A color version of this figure is available online.)} \label{mergingcluster}
\end{center}
\end{figure*}

Examination of \cref{table_sources_fixed1,table_sources_gapper1} reveals several clusters with velocity dispersions well above $1000$ km s$^{-1}$.  Rather than being necessarily massive clusters, closer examination shows them to be good candidates for merging cluster systems.  If we examine two of these systems in greater detail (source J151131.3+071506, which has a velocity dispersion of $1549$ km s$^{-1}$, with the fixed gap method, and J121121.1+141439, which has a velocity dispersion of $2655$ km s$^{-1}$, with the shifting gapper method), we see what appear to be merging clusters.  \cref{mergingcluster} shows the distribution of the galaxies within these clusters in peculiar velocity and distance from the center of the clusters, as well as a bubble plot diagram for the same clusters.  We see what is clearly (for both cases) two sub-clusters separated in both peculiar velocity and separation on the sky.

\section{Summary and Conclusions} \label{conclusions} \index{Summary and Conclusions}
\citet{aguerri2010} examined $88$ nearby clusters in \SDSS\/ and found that, using the $\Delta$ statistic, $\sim55\%$ of their clusters exhibited substructure in the outer cluster regions.  The region that they defined as the outer cluster is similar to the region that we searched for substructure.  Further, \citet{aguerri2010} found no correlation between cluster properties, such as the fraction of blue galaxies, cluster velocity dispersion, and the luminosity difference between the two brightest cluster galaxies, and the presence of substructure within the cluster.  We do find a very strong correlation between the normalized substructure measurement made using the $\Delta$ test and the velocity dispersion of the cluster.  This is not unexpected, as a quick check of \cref{equation_delta} shows us:  the velocity dispersion of the cluster is one of the main components that makes up the $\Delta$ measurement.

We find that, using the $\Delta$ substructure statistic, nearly $80\%$ of our clusters (in each different sample) have evidence of optical substructure at the $\ge 2\sigma$ level.  This is consistent with recent results from \citet{ramella2007} who find that $73\%$ of their clusters selected from the Wide-field Nearby Cluster Survey have substructure, and \citet{einasto2012} who, similar to us, searched \SDSS\/ DR8 and found that, using the $\Delta$ test, $\sim70\%$ of the clusters had significant substructure.  Looking at 25 clusters within the 2dF Galaxy Redshift Survey, \citet{burgett2004} found that 21 clusters ($84\%$) were detected to have substructure at the 99\% confidence level or higher in at least one of their substructure tests.

In general, our two-dimensional test ($\beta$) was the least sensitive to detecting optical substructure, identifying it less than $\sim30\%$ of the time for non-bent radio sources, but $\sim45\%$ of the time for bent double-lobed radio sources.  Our other three-dimensional tests ($\alpha$ and $\epsilon$) were similar in their detection of optical substructure, with generally $\sim40\%$ (but as low as $14\%$ and as high as $63\%$) of the clusters identified as having significant optical substructure, regardless of the morphology of the radio source contained within the cluster.

We have examined the optical substructure environments in a sample of clusters containing radio sources, either double-lobed radio sources or single-component radio sources.  The double-lobed radio sources that we examined can be either bent or straight.  One possible method to explain the bending of a double-lobed radio source is a recent cluster-cluster merger which disrupts the ICM.  This disruption will create velocities necessary to explain the bending of the radio lobes via ram pressure.  Mergers on this scale will also leave evidence in the form of increased substructure among the galaxies of the cluster.  We examined our samples for correlations between the presence of bent-lobed radio sources and substructure, as a test to determine if the merger scenario alone is enough to explain the bending of radio lobes.  In general, we found no preference for bent radio sources to be located in clusters with significant substructure as opposed to other types of radio sources.  Large-scale cluster-cluster mergers are likely not the sole mechanism responsible for creating conditions necessary to bend radio lobes. Another possible mechanism is ``sloshing,'' which is created by a previous small-scale off-axis merger \citep{mendygral2012}.

\acknowledgements
JDW and ELB were partially supported by NASA through the Astrophysics Data Analysis Program, grant number NNX10AC98G, and through NASA award RSA No. 1440385 issued by JPL/Caltech.

The FIRST project has been supported by grants from the National Geographic Society, the National Science Foundation, NASA, NATO, the Institute of Geophysics and Planetary Physics, Columbia University, and Sun Microsystems.

Funding for SDSS-III (DR8) has been provided by the Alfred P. Sloan Foundation, the Participating Institutions, the National Science Foundation, and the U.S. Department of Energy Office of Science. The SDSS-III web site is \url{http://www.sdss3.org/}.

SDSS-III is managed by the Astrophysical Research Consortium for the Participating Institutions of the SDSS-III Collaboration including the University of Arizona, the Brazilian Participation Group, Brookhaven National Laboratory, University of Cambridge, Carnegie Mellon University, University of Florida, the French Participation Group, the German Participation Group, Harvard University, the Instituto de Astrofisica de Canarias, the Michigan State/Notre Dame/JINA Participation Group, Johns Hopkins University, Lawrence Berkeley National Laboratory, Max Planck Institute for Astrophysics, Max Planck Institute for Extraterrestrial Physics, New Mexico State University, New York University, Ohio State University, Pennsylvania State University, University of Portsmouth, Princeton University, the Spanish Participation Group, University of Tokyo, University of Utah, Vanderbilt University, University of Virginia, University of Washington, and Yale University.

This research has made use of the NASA/IPAC Extragalactic Database (NED) which is operated by the Jet Propulsion Laboratory, California Institute of Technology, under contract with the National Aeronautics and Space Administration.

\bibliography{bibdeskbiblio}

\begin{thebibliography}{57}
\expandafter\ifx\csname natexlab\endcsname\relax\def\natexlab#1{#1}\fi

\bibitem[{{Aguerri} \& {S{\'a}nchez-Janssen}(2010)}]{aguerri2010}
{Aguerri}, J.~A.~L., \& {S{\'a}nchez-Janssen}, R. 2010, \aap, 521, A28+,
  1007.3497

\bibitem[{{Aihara} {et~al.}(2011){Aihara}, {Allende Prieto}, {An}, {Anderson},
  {Aubourg}, {Balbinot}, {Beers}, {Berlind}, {Bickerton}, {Bizyaev}, {Blanton},
  {Bochanski}, {Bolton}, {Bovy}, {Brandt}, {Brinkmann}, {Brown}, {Brownstein},
  {Busca}, {Campbell}, {Carr}, {Chen}, {Chiappini}, {Comparat}, {Connolly},
  {Cortes}, {Croft}, {Cuesta}, {da Costa}, {Davenport}, {Dawson}, {Dhital},
  {Ealet}, {Ebelke}, {Edmondson}, {Eisenstein}, {Escoffier}, {Esposito},
  {Evans}, {Fan}, {Femen{\'{\i}}a Castell{\'a}}, {Font-Ribera}, {Frinchaboy},
  {Ge}, {Gillespie}, {Gilmore}, {Gonz{\'a}lez Hern{\'a}ndez}, {Gott}, {Gould},
  {Grebel}, {Gunn}, {Hamilton}, {Harding}, {Harris}, {Hawley}, {Hearty}, {Ho},
  {Hogg}, {Holtzman}, {Honscheid}, {Inada}, {Ivans}, {Jiang}, {Johnson},
  {Jordan}, {Jordan}, {Kazin}, {Kirkby}, {Klaene}, {Knapp}, {Kneib},
  {Kochanek}, {Koesterke}, {Kollmeier}, {Kron}, {Lampeitl}, {Lang}, {Le Goff},
  {Lee}, {Lin}, {Long}, {Loomis}, {Lucatello}, {Lundgren}, {Lupton}, {Ma},
  {MacDonald}, {Mahadevan}, {Maia}, {Makler}, {Malanushenko}, {Malanushenko},
  {Mandelbaum}, {Maraston}, {Margala}, {Masters}, {McBride}, {McGehee},
  {McGreer}, {M{\'e}nard}, {Miralda-Escud{\'e}}, {Morrison}, {Mullally},
  {Muna}, {Munn}, {Murayama}, {Myers}, {Naugle}, {Neto}, {Nguyen}, {Nichol},
  {O'Connell}, {Ogando}, {Olmstead}, {Oravetz}, {Padmanabhan},
  {Palanque-Delabrouille}, {Pan}, {Pandey}, {P{\^a}ris}, {Percival},
  {Petitjean}, {Pfaffenberger}, {Pforr}, {Phleps}, {Pichon}, {Pieri}, {Prada},
  {Price-Whelan}, {Raddick}, {Ramos}, {Reyl{\'e}}, {Rich}, {Richards}, {Rix},
  {Robin}, {Rocha-Pinto}, {Rockosi}, {Roe}, {Rollinde}, {Ross}, {Ross},
  {Rossetto}, {S{\'a}nchez}, {Sayres}, {Schlegel}, {Schlesinger}, {Schmidt},
  {Schneider}, {Sheldon}, {Shu}, {Simmerer}, {Simmons}, {Sivarani}, {Snedden},
  {Sobeck}, {Steinmetz}, {Strauss}, {Szalay}, {Tanaka}, {Thakar}, {Thomas},
  {Tinker}, {Tofflemire}, {Tojeiro}, {Tremonti}, {Vandenberg}, {Vargas
  Maga{\~n}a}, {Verde}, {Vogt}, {Wake}, {Wang}, {Weaver}, {Weinberg}, {White},
  {White}, {Yanny}, {Yasuda}, {Yeche}, \& {Zehavi}}]{aihara2011}
{Aihara}, H. {et~al.} 2011, \apjs, 193, 29, 1101.1559

\bibitem[{{Allington-Smith} {et~al.}(1993){Allington-Smith}, {Ellis}, {Zirbel},
  \& {Oemler}}]{allington-smith1993}
{Allington-Smith}, J.~R., {Ellis}, R., {Zirbel}, E.~L., \& {Oemler}, A.~J.
  1993, \apj, 404, 521

\bibitem[{{Ascasibar} \& {Markevitch}(2006)}]{ascasibar2006}
{Ascasibar}, Y., \& {Markevitch}, M. 2006, \apj, 650, 102,
  arXiv:astro-ph/0603246

\bibitem[{{Baier}(1984)}]{baier1984}
{Baier}, F.~W. 1984, Astronomische Nachrichten, 305, 175

\bibitem[{{Becker} {et~al.}(1995){Becker}, {White}, \& {Helfand}}]{becker1995}
{Becker}, R.~H., {White}, R.~L., \& {Helfand}, D.~J. 1995, \apj, 450, 559

\bibitem[{{Beers} {et~al.}(1990){Beers}, {Flynn}, \& {Gebhardt}}]{beers1990}
{Beers}, T.~C., {Flynn}, K., \& {Gebhardt}, K. 1990, \aj, 100, 32

\bibitem[{{Beers} {et~al.}(1982){Beers}, {Geller}, \& {Huchra}}]{beers1982}
{Beers}, T.~C., {Geller}, M.~J., \& {Huchra}, J.~P. 1982, \apj, 257, 23

\bibitem[{{Bird}(1995)}]{bird1995}
{Bird}, C.~M. 1995, \apjl, 445, L81, arXiv:astro-ph/9503038

\bibitem[{{Bird} \& {Beers}(1993)}]{bird1993}
{Bird}, C.~M., \& {Beers}, T.~C. 1993, \aj, 105, 1596

\bibitem[{{Blanton}(2000)}]{blanton2000a}
{Blanton}, E.~L. 2000, PhD thesis, AA(COLUMBIA UNIVERSITY)

\bibitem[{{Blanton} {et~al.}(2001){Blanton}, {Gregg}, {Helfand}, {Becker}, \&
  {Leighly}}]{blanton2001}
{Blanton}, E.~L., {Gregg}, M.~D., {Helfand}, D.~J., {Becker}, R.~H., \&
  {Leighly}, K.~M. 2001, \aj, 121, 2915, arXiv:astro-ph/0102499

\bibitem[{{Blanton} {et~al.}(2000){Blanton}, {Gregg}, {Helfand}, {Becker}, \&
  {White}}]{blanton2000}
{Blanton}, E.~L., {Gregg}, M.~D., {Helfand}, D.~J., {Becker}, R.~H., \&
  {White}, R.~L. 2000, \apj, 531, 118, arXiv:astro-ph/9910099

\bibitem[{{Blanton} {et~al.}(2003){Blanton}, {Gregg}, {Helfand}, {Becker}, \&
  {White}}]{blanton2003}
------. 2003, \aj, 125, 1635, arXiv:astro-ph/0212407

\bibitem[{{Brainerd} {et~al.}(1998){Brainerd}, {Goldberg}, \&
  {Villumsen}}]{brainerd1998}
{Brainerd}, T.~G., {Goldberg}, D.~M., \& {Villumsen}, J.~V. 1998, \apj, 502,
  505, arXiv:astro-ph/9706165

\bibitem[{{Burgett} {et~al.}(2004){Burgett}, {Vick}, {Davis}, {Colless}, {De
  Propris}, {Baldry}, {Baugh}, {Bland-Hawthorn}, {Bridges}, {Cannon}, {Cole},
  {Collins}, {Couch}, {Cross}, {Dalton}, {Driver}, {Efstathiou}, {Ellis},
  {Frenk}, {Glazebrook}, {Hawkins}, {Jackson}, {Lahav}, {Lewis}, {Lumsden},
  {Maddox}, {Madgwick}, {Norberg}, {Peacock}, {Percival}, {Peterson},
  {Sutherland}, \& {Taylor}}]{burgett2004}
{Burgett}, W.~S. {et~al.} 2004, \mnras, 352, 605, arXiv:astro-ph/0405021

\bibitem[{{Burns}(1990)}]{burns1990}
{Burns}, J.~O. 1990, \aj, 99, 14

\bibitem[{{Carter} \& {Metcalfe}(1980)}]{carter1980}
{Carter}, D., \& {Metcalfe}, N. 1980, \mnras, 191, 325

\bibitem[{{Clarke} {et~al.}(2004){Clarke}, {Blanton}, \&
  {Sarazin}}]{clarke2004}
{Clarke}, T.~E., {Blanton}, E.~L., \& {Sarazin}, C.~L. 2004, \apj, 616, 178,
  arXiv:astro-ph/0408068

\bibitem[{{Colless} \& {Dunn}(1996)}]{colless1996}
{Colless}, M., \& {Dunn}, A.~M. 1996, \apj, 458, 435, arXiv:astro-ph/9508070

\bibitem[{{De Propris} {et~al.}(2002){De Propris}, {Couch}, {Colless},
  {Dalton}, {Collins}, {Baugh}, {Bland-Hawthorn}, {Bridges}, {Cannon}, {Cole},
  {Cross}, {Deeley}, {Driver}, {Efstathiou}, {Ellis}, {Frenk}, {Glazebrook},
  {Jackson}, {Lahav}, {Lewis}, {Lumsden}, {Maddox}, {Madgwick}, {Moody},
  {Norberg}, {Peacock}, {Percival}, {Peterson}, {Sutherland}, \&
  {Taylor}}]{de-propris2002}
{De Propris}, R. {et~al.} 2002, \mnras, 329, 87, arXiv:astro-ph/0109167

\bibitem[{{Dressler} \& {Shectman}(1988)}]{dressler1988}
{Dressler}, A., \& {Shectman}, S.~A. 1988, \aj, 95, 985

\bibitem[{{Eilek} {et~al.}(1984){Eilek}, {Burns}, {O'Dea}, \&
  {Owen}}]{eilek1984}
{Eilek}, J.~A., {Burns}, J.~O., {O'Dea}, C.~P., \& {Owen}, F.~N. 1984, \apj,
  278, 37

\bibitem[{{Einasto} {et~al.}(2012){Einasto}, {Vennik}, {Nurmi}, {Tempel},
  {Ahvensalmi}, {Tago}, {Liivam{\"a}gi}, {Saar}, {Hein{\"a}m{\"a}ki},
  {Einasto}, \& {Mart{\'{\i}}nez}}]{einasto2012}
{Einasto}, M. {et~al.} 2012, \aap, 540, A123, 1202.4927

\bibitem[{{Fadda} {et~al.}(1996){Fadda}, {Girardi}, {Giuricin}, {Mardirossian},
  \& {Mezzetti}}]{fadda1996}
{Fadda}, D., {Girardi}, M., {Giuricin}, G., {Mardirossian}, F., \& {Mezzetti},
  M. 1996, \apj, 473, 670, arXiv:astro-ph/9606098

\bibitem[{{Fanaroff} \& {Riley}(1974)}]{fanaroff1974}
{Fanaroff}, B.~L., \& {Riley}, J.~M. 1974, \mnras, 167, 31P

\bibitem[{{Fitchett} \& {Webster}(1987)}]{fitchett1987}
{Fitchett}, M., \& {Webster}, R. 1987, \apj, 317, 653

\bibitem[{{Flin} \& {Krywult}(2006)}]{flin2006}
{Flin}, P., \& {Krywult}, J. 2006, \aap, 450, 9

\bibitem[{{Geller} \& {Beers}(1982)}]{geller1982}
{Geller}, M.~J., \& {Beers}, T.~C. 1982, \pasp, 94, 421

\bibitem[{{Hardcastle} {et~al.}(2005){Hardcastle}, {Sakelliou}, \&
  {Worrall}}]{hardcastle2005}
{Hardcastle}, M.~J., {Sakelliou}, I., \& {Worrall}, D.~M. 2005, \mnras, 359,
  1007, arXiv:astro-ph/0502575

\bibitem[{{Heisler} {et~al.}(1985){Heisler}, {Tremaine}, \&
  {Bahcall}}]{heisler1985}
{Heisler}, J., {Tremaine}, S., \& {Bahcall}, J.~N. 1985, \apj, 298, 8

\bibitem[{{Hill} \& {Lilly}(1991)}]{hill1991}
{Hill}, G.~J., \& {Lilly}, S.~J. 1991, \apj, 367, 1

\bibitem[{{Hou} {et~al.}(2012){Hou}, {Parker}, {Wilman}, {McGee}, {Harris},
  {Connelly}, {Balogh}, {Mulchaey}, \& {Bower}}]{hou2012}
{Hou}, A. {et~al.} 2012, \mnras, 421, 3594, 1201.3676

\bibitem[{{Johnston-Hollitt} {et~al.}(2008){Johnston-Hollitt}, {Sato}, {Gill},
  {Fleenor}, \& {Brick}}]{johnston-hollitt2008}
{Johnston-Hollitt}, M., {Sato}, M., {Gill}, J.~A., {Fleenor}, M.~C., \&
  {Brick}, A.-M. 2008, \mnras, 390, 289, 0807.4579

\bibitem[{{Kriessler} \& {Beers}(1997)}]{kriessler1997}
{Kriessler}, J.~R., \& {Beers}, T.~C. 1997, \aj, 113, 80

\bibitem[{{Ledlow} \& {Owen}(1996)}]{ledlow1996}
{Ledlow}, M.~J., \& {Owen}, F.~N. 1996, \aj, 112, 9, arXiv:astro-ph/9607014

\bibitem[{{Mao} {et~al.}(2010){Mao}, {Sharp}, {Saikia}, {Norris},
  {Johnston-Hollitt}, {Middelberg}, \& {Lovell}}]{mao2010}
{Mao}, M.~Y., {Sharp}, R., {Saikia}, D.~J., {Norris}, R.~P.,
  {Johnston-Hollitt}, M., {Middelberg}, E., \& {Lovell}, J.~E.~J. 2010, \mnras,
  406, 2578, 1005.3649

\bibitem[{{Mao} {et~al.}(2011){Mao}, {Sharp}, {Saikia}, {Norris},
  {Johnston-Hollitt}, {Middelberg}, \& {Lovell}}]{mao2011}
------. 2011, Journal of Astrophysics and Astronomy, 32, 585

\bibitem[{{Markevitch} {et~al.}(1998){Markevitch}, {Forman}, {Sarazin}, \&
  {Vikhlinin}}]{markevitch1998}
{Markevitch}, M., {Forman}, W.~R., {Sarazin}, C.~L., \& {Vikhlinin}, A. 1998,
  \apj, 503, 77, arXiv:astro-ph/9711289

\bibitem[{{Mendygral} {et~al.}(2012){Mendygral}, {Jones}, \&
  {Dolag}}]{mendygral2012}
{Mendygral}, P.~J., {Jones}, T.~W., \& {Dolag}, K. 2012, \apj, 750, 166,
  1203.2312

\bibitem[{{Owers} {et~al.}(2009){Owers}, {Nulsen}, {Couch}, {Markevitch}, \&
  {Poole}}]{owers2009}
{Owers}, M.~S., {Nulsen}, P.~E.~J., {Couch}, W.~J., {Markevitch}, M., \&
  {Poole}, G.~B. 2009, \apj, 692, 702, 0810.4650

\bibitem[{{Owers} {et~al.}(2011){Owers}, {Randall}, {Nulsen}, {Couch}, {David},
  \& {Kempner}}]{owers2011}
{Owers}, M.~S., {Randall}, S.~W., {Nulsen}, P.~E.~J., {Couch}, W.~J., {David},
  L.~P., \& {Kempner}, J.~C. 2011, \apj, 728, 27, 1012.1315

\bibitem[{{Pinkney} {et~al.}(1996){Pinkney}, {Roettiger}, {Burns}, \&
  {Bird}}]{pinkney1996}
{Pinkney}, J., {Roettiger}, K., {Burns}, J.~O., \& {Bird}, C.~M. 1996, \apjs,
  104, 1

\bibitem[{{Pisani}(1993)}]{pisani1993}
{Pisani}, A. 1993, \mnras, 265, 706

\bibitem[{{Pisani}(1996)}]{pisani1996}
------. 1996, \mnras, 278, 697, arXiv:astro-ph/9508150

\bibitem[{{Ramella} {et~al.}(2007){Ramella}, {Biviano}, {Pisani}, {Varela},
  {Bettoni}, {Couch}, {D'Onofrio}, {Dressler}, {Fasano}, {Kj{\o}rgaard},
  {Moles}, {Pignatelli}, \& {Poggianti}}]{ramella2007}
{Ramella}, M. {et~al.} 2007, \aap, 470, 39, 0704.0579

\bibitem[{{Rhee} \& {Katgert}(1987)}]{rhee1987}
{Rhee}, G.~F.~R.~N., \& {Katgert}, P. 1987, \aap, 183, 217

\bibitem[{{Rhee} {et~al.}(1991){Rhee}, {van Haarlem}, \& {Katgert}}]{rhee1991}
{Rhee}, G.~F.~R.~N., {van Haarlem}, M.~P., \& {Katgert}, P. 1991, \aap, 246,
  301

\bibitem[{{Solanes} {et~al.}(1999){Solanes}, {Salvador-Sol{\'e}}, \&
  {Gonz{\'a}lez-Casado}}]{solanes1999}
{Solanes}, J.~M., {Salvador-Sol{\'e}}, E., \& {Gonz{\'a}lez-Casado}, G. 1999,
  \aap, 343, 733, arXiv:astro-ph/9812103

\bibitem[{{Springel} {et~al.}(2006){Springel}, {Frenk}, \&
  {White}}]{springel2006}
{Springel}, V., {Frenk}, C.~S., \& {White}, S.~D.~M. 2006, \nat, 440, 1137,
  arXiv:astro-ph/0604561

\bibitem[{{Venturi} {et~al.}(2000){Venturi}, {Bardelli}, {Morganti}, \&
  {Hunstead}}]{venturi2000}
{Venturi}, T., {Bardelli}, S., {Morganti}, R., \& {Hunstead}, R.~W. 2000,
  \mnras, 314, 594, arXiv:astro-ph/0001256

\bibitem[{{Venturi} {et~al.}(2001){Venturi}, {Bardelli}, {Zambelli},
  {Morganti}, \& {Hunstead}}]{venturi2001}
{Venturi}, T., {Bardelli}, S., {Zambelli}, G., {Morganti}, R., \& {Hunstead},
  R.~W. 2001, \mnras, 324, 1131, arXiv:astro-ph/0102248

\bibitem[{{West} \& {Bothun}(1990)}]{west1990}
{West}, M.~J., \& {Bothun}, G.~D. 1990, \apj, 350, 36

\bibitem[{{West} {et~al.}(1988){West}, {Oemler}, \& {Dekel}}]{west1988}
{West}, M.~J., {Oemler}, Jr., A., \& {Dekel}, A. 1988, \apj, 327, 1

\bibitem[{{Wing} \& {Blanton}(2011)}]{wing2011}
{Wing}, J.~D., \& {Blanton}, E.~L. 2011, \aj, 141, 88, 1008.1099

\bibitem[{{York} {et~al.}(2000){York}, {Adelman}, {Anderson}, {Anderson},
  {Annis}, {Bahcall}, {Bakken}, {Barkhouser}, {Bastian}, {Berman}, {Boroski},
  {Bracker}, {Briegel}, {Briggs}, {Brinkmann}, {Brunner}, {Burles}, {Carey},
  {Carr}, {Castander}, {Chen}, {Colestock}, {Connolly}, {Crocker}, {Csabai},
  {Czarapata}, {Davis}, {Doi}, {Dombeck}, {Eisenstein}, {Ellman}, {Elms},
  {Evans}, {Fan}, {Federwitz}, {Fiscelli}, {Friedman}, {Frieman}, {Fukugita},
  {Gillespie}, {Gunn}, {Gurbani}, {de Haas}, {Haldeman}, {Harris}, {Hayes},
  {Heckman}, {Hennessy}, {Hindsley}, {Holm}, {Holmgren}, {Huang}, {Hull},
  {Husby}, {Ichikawa}, {Ichikawa}, {Ivezi{\'c}}, {Kent}, {Kim}, {Kinney},
  {Klaene}, {Kleinman}, {Kleinman}, {Knapp}, {Korienek}, {Kron}, {Kunszt},
  {Lamb}, {Lee}, {Leger}, {Limmongkol}, {Lindenmeyer}, {Long}, {Loomis},
  {Loveday}, {Lucinio}, {Lupton}, {MacKinnon}, {Mannery}, {Mantsch}, {Margon},
  {McGehee}, {McKay}, {Meiksin}, {Merelli}, {Monet}, {Munn}, {Narayanan},
  {Nash}, {Neilsen}, {Neswold}, {Newberg}, {Nichol}, {Nicinski}, {Nonino},
  {Okada}, {Okamura}, {Ostriker}, {Owen}, {Pauls}, {Peoples}, {Peterson},
  {Petravick}, {Pier}, {Pope}, {Pordes}, {Prosapio}, {Rechenmacher}, {Quinn},
  {Richards}, {Richmond}, {Rivetta}, {Rockosi}, {Ruthmansdorfer}, {Sandford},
  {Schlegel}, {Schneider}, {Sekiguchi}, {Sergey}, {Shimasaku}, {Siegmund},
  {Smee}, {Smith}, {Snedden}, {Stone}, {Stoughton}, {Strauss}, {Stubbs},
  {SubbaRao}, {Szalay}, {Szapudi}, {Szokoly}, {Thakar}, {Tremonti}, {Tucker},
  {Uomoto}, {Vanden Berk}, {Vogeley}, {Waddell}, {Wang}, {Watanabe},
  {Weinberg}, {Yanny}, \& {Yasuda}}]{york2000}
{York}, D.~G. {et~al.} 2000, \aj, 120, 1579, arXiv:astro-ph/0006396

\bibitem[{{Zhao} {et~al.}(1989){Zhao}, {Burns}, \& {Owen}}]{zhao1989}
{Zhao}, J., {Burns}, J.~O., \& {Owen}, F.~N. 1989, \aj, 98, 64

\end{thebibliography}

\clearpage
\newpage

\LongTables

\begin{deluxetable}{lcccccrrrrrrr}
\tabletypesize{\tiny}
\tablewidth{0pt}
\tablecolumns{13}
\tablecaption{All Sources With Substructure Measurements (Using the Fixed Gap Method)}
\tablehead{\colhead{Source Name}&
		  \colhead{Smpl}&
		  \colhead{$\alpha$}&
		  \colhead{$\delta$}&
		  \colhead{$\alpha_{clus}$}&
		  \colhead{$\delta_{clus}$}&
		  \colhead{FR}&
		  \colhead{FR}&
		  \colhead{$z$}&
		  \colhead{$z_{clus}$}&
		  \colhead{N$^{-19}_{1.0}$}&
		  \colhead{N$^z_{3.0}$}&
		  \colhead{$\sigma$}
		  \cr
		   & & & & & & &
		   & & & & &
		  \colhead{(km s$^{-1}$)}
		  \cr
		  \colhead{(1)}&
		  \colhead{(2)}&
		  \colhead{(3)}&
		  \colhead{(4)}&
		  \colhead{(5)}&
		  \colhead{(6)}&
		  \colhead{(7)}&
		  \colhead{(8)}&
		  \colhead{(9)}&
		  \colhead{(10)}&
		  \colhead{(11)}&
		  \colhead{(12)}&
		  \colhead{(13)}}
\startdata
J005702.1$-$005231 & V & 00:57:02.10 & $-00$:52:31.0 & 00:56:42.47 & $-00$:42:48.9 & I & I & 0.044\tablenotemark{a} & 0.044 & $53$ & 254 & 508\\
J075917.1$+$270916 & V & 07:59:17.10 & $+27$:09:16.0 & 07:58:58.05 & $+27$:09:29.5 & I & I & 0.112\tablenotemark{b} & 0.098 & $98$ & 49 & 320\\
J080757.0$+$163716 & A & 08:07:57.05 & $+16$:37:16.2 & 08:08:00.54 & $+16$:21:53.9 & I & I? & 0.103\tablenotemark{b} & 0.100 & $0$ & 34 & 307\\
J091344.5$+$474216 & A & 09:13:44.54 & $+47$:42:16.4 & 09:13:11.57 & $+47$:39:54.2 & I & I? & 0.051\tablenotemark{a} & 0.051 & $44$ & 66 & 312\\
J102236.7$+$082238 & C & 10:22:36.72 & $+08$:22:37.7 & 10:24:25.25 & $+08$:19:33.8 & I & II & 0.044\tablenotemark{a} & 0.042 & $23$ & 35 & 1067\\
J102757.8$+$103345 & S & 10:27:57.86 & $+10$:33:45.9 & 10:27:50.78 & $+10$:33:21.6 & I & II & 0.102\tablenotemark{b} & 0.109 & $103$ & 65 & 353\\
J103201.6$+$350253 & V & 10:32:01.60 & $+35$:02:53.0 & 10:31:38.97 & $+35$:00:19.7 & I & I & 0.121\tablenotemark{a} & 0.123 & $110$ & 45 & 718\\
J104104.1$+$335520 & V & 10:41:04.10 & $+33$:55:20.0 & 10:41:16.28 & $+33$:53:21.3 & I & I & 0.084\tablenotemark{a} & 0.083 & $24$ & 45 & 299\\
J105147.4$+$552309 & S & 10:51:47.46 & $+55$:23:09.4 & 10:52:56.04 & $+55$:12:17.2 & I & I & 0.074\tablenotemark{a} & 0.073 & $62$ & 85 & 358\\
J112559.8$+$252837 & V & 11:25:59.80 & $+25$:28:37.0 & 11:26:38.31 & $+25$:23:49.3 & I & I & 0.116\tablenotemark{a} & 0.113 & $53$ & 42 & 474\\
J113902.6$+$322821 & S & 11:39:02.60 & $+32$:28:21.6 & 11:38:55.83 & $+32$:26:23.4 & I & I & 0.141\tablenotemark{a} & 0.128 & $70$ & 51 & 1115\\
J114027.7$+$120308 & S & 11:40:27.76 & $+12$:03:08.0 & 11:40:10.38 & $+11$:34:18.7 & I & I & 0.081\tablenotemark{a} & 0.081 & $-17$ & 47 & 393\\
J115531.5$+$031150 & C & 11:55:31.44 & $+03$:11:50.2 & 11:55:00.14 & $+03$:27:03.1 & II & I & 0.081\tablenotemark{b} & 0.074 & $13$ & 36 & 1275\\
J120013.9$+$561502 & A & 12:00:13.91 & $+56$:15:02.0 & 12:03:21.11 & $+56$:25:25.3 & I & II & 0.061\tablenotemark{b} & 0.064 & $85$ & 83 & 610\\
J120210.3$+$274109 & S & 12:02:10.38 & $+27$:41:09.0 & 12:02:44.07 & $+27$:44:04.9 & I & I & 0.134\tablenotemark{a} & 0.137 & $52$ & 35 & 574\\
J125935.6$+$275735 & V & 12:59:35.60 & $+27$:57:35.0 & 12:59:13.83 & $+27$:53:08.4 & I & I & 0.024\tablenotemark{a} & 0.023 & $166$ & 812 & 775\\
J133425.1$+$381757 & S & 13:34:25.20 & $+38$:17:57.4 & 13:31:51.49 & $+37$:53:25.5 & II & II & 0.063\tablenotemark{a} & 0.060 & $17$ & 81 & 1127\\
J135341.7$+$331339 & A & 13:53:41.71 & $+33$:13:39.6 & 13:56:04.32 & $+33$:06:43.4 & I & I & 0.048\tablenotemark{b} & 0.052 & $56$ & 100 & 1567\\
J135449.7$+$203611 & C & 13:54:49.68 & $+20$:36:11.5 & 13:55:57.44 & $+20$:39:58.6 & I & I & 0.070\tablenotemark{a} & 0.067 & $17$ & 84 & 1201\\
J140313.2$+$061009 & A & 14:03:13.26 & $+06$:10:09.1 & 14:05:08.49 & $+06$:14:55.9 & I & I & 0.083\tablenotemark{a} & 0.084 & $25$ & 73 & 412\\
J142131.0$-$001246 & C & 14:21:30.96 & $-00$:12:46.4 & 14:18:51.37 & $+00$:07:10.0 & I & II & 0.052\tablenotemark{a} & 0.052 & $3$ & 67 & 363\\
J144818.3$+$033138 & A & 14:48:18.39 & $+03$:31:38.2 & 14:42:20.06 & $+03$:30:18.3 & II & I & 0.028\tablenotemark{b} & 0.027 & $8$ & 145 & 319\\
J145107.4$+$181417 & C & 14:51:07.44 & $+18$:14:17.2 & 14:52:46.15 & $+18$:11:09.8 & I & I & 0.062\tablenotemark{a} & 0.054 & $-18$ & 105 & 2062\\
J150957.3$+$332715 & V & 15:09:57.30 & $+33$:27:15.0 & 15:10:05.50 & $+33$:30:58.6 & I & I & 0.117\tablenotemark{a} & 0.113 & $140$ & 81 & 959\\
J151131.3$+$071506 & S & 15:11:31.38 & $+07$:15:07.0 & 15:14:23.78 & $+07$:14:29.1 & I & I & 0.045\tablenotemark{a} & 0.040 & $3$ & 220 & 1549\\
J152642.0$+$005330 & S & 15:26:42.01 & $+00$:53:30.1 & 15:27:12.18 & $+01$:10:45.0 & I & I & 0.117\tablenotemark{b} & 0.116 & $46$ & 32 & 567\\
J154624.8$+$362954 & A & 15:46:24.83 & $+36$:29:54.1 & 15:45:03.76 & $+36$:05:12.9 & II & II? & 0.051\tablenotemark{b} & 0.068 & $18$ & 111 & 654\\
J162037.0$+$252053 & V & 16:20:37.01 & $+25$:20:53.0 & 16:21:39.17 & $+25$:33:13.5 & I & I & 0.103\tablenotemark{a} & 0.101 & $36$ & 38 & 616\\
J164923.9$+$263502 & S & 16:49:24.00 & $+26$:35:02.1 & 16:48:43.42 & $+26$:51:13.8 & I & I & 0.055\tablenotemark{a} & 0.056 & $14$ & 62 & 736\\
J215423.6$+$003710 & V & 21:54:23.60 & $+00$:37:10.0 & 21:54:23.30 & $+00$:41:56.4 & II & ??? & 0.219\tablenotemark{b} & 0.217 & $156$ & 30 & 1373
\enddata
\tablecomments{Col. (1): name of the source; col. (2): identifier for the sample the source belongs to; V for the visual-bent sample, A for the auto-bent sample, S for the straight sample, and C for the single-component sample; col. (3): right ascension of the radio source; col. (4): declination of the radio source; col. (5): right ascension of the cluster center; col. (6): declination of the cluster center; col. (7) FR type based on \citet{ledlow1996}; col. (8): FR type based on visual examination; a question mark represents a dubious visual FR I/II classification; three successive question marks represent an inability to visually classify the radio source as either FR I or II; col. (9): redshift of the radio source; col. (10): redshift of the cluster center; col. (11): number of cluster galaxies based on the N$^{-19}_{1.0}$ method from \citet{wing2011}; col. (12): number of cluster galaxies with \SDSS\/ spectroscopically measured redshifts within $3.0$ Mpc and $\pm 5000$ km s$^{-1}$ of the cluster center (N$^z_{3.0}$); col. (13): the velocity dispersion of the cluster.}
\tablenotetext{a}{The redshift comes from a spectroscopic measurement.}
\tablenotetext{b}{The redshift comes from a photometric measurement.}
\label{table_sources_fixed1}
\end{deluxetable}

\begin{deluxetable}{lcrrcrcrcrc}
\tabletypesize{\tiny}
\tablewidth{0pt}
\tablecolumns{11}
\tablecaption{All Sources With Substructure Measurements (Using the Fixed Gap Method)}
\tablehead{\colhead{Source Name}&
		  \colhead{Smpl}&
		  \colhead{$\beta$}&
		  \colhead{sig}&
		  \colhead{$\Delta$}&
		  \colhead{sig}&
		  \colhead{$\alpha$}&
		  \colhead{sig}&
		  \colhead{$\epsilon$}&
		  \colhead{sig}&
		  \colhead{Abell}
		  \cr
		   & & &
		  \colhead{($\sigma$)} & & 
		  \colhead{($\sigma$)} & & 
		  \colhead{($\sigma$)} & & 
		  \colhead{($\sigma$)}
		  \cr
		  \colhead{(1)}&
		  \colhead{(2)}&
		  \colhead{(3)}&
		  \colhead{(4)}&
		  \colhead{(5)}&
		  \colhead{(6)}&
		  \colhead{(7)}&
		  \colhead{(8)}&
		  \colhead{(9)}&
		  \colhead{(10)}&
		  \colhead{(11)}}
\startdata
J005702.1$-$005231 & V & $-1.88$ & 1.594 & 1.31 & 2.727 & 3.68 & 3.615 & 0.87 & 1.987 & A0119\\
J075917.1$+$270916 & V & $-12.09$ & 0.163 & 1.15 & 1.285 & 0.94 & 0.727 & 0.49 & 2.939 & A0610\\
J080757.0$+$163716 & A & $0.50$ & 0.104 & 0.86 & 0.208 & 1.73 & 1.705 & 1.08 & 0.333 & \nodata\\
J091344.5$+$474216 & A & $2.06$ & 0.702 & 1.25 & 1.829 & 0.31 & 0.122 & 1.26 & 0.162 & A0757\\
J102236.7$+$082238 & C & $-4.49$ & 0.480 & 1.13 & 1.667 & 0.67 & 0.348 & 1.39 & 0.010 & A0989\tablenotemark{b}\\
J102757.8$+$103345 & S & $-17.14$ & 0.732 & 1.13 & 1.144 & 1.69 & 1.601 & 0.51 & 3.900 & A1020\\
J103201.6$+$350253 & V & $1.28$ & 0.152 & 1.35 & 2.172 & 2.39 & 2.036 & 0.83 & 1.234 & A1033\\
J104104.1$+$335520 & V & $5.23$ & 0.446 & 1.49 & 2.110 & 1.36 & 1.212 & 1.10 & 0.492 & \nodata\\
J105147.4$+$552309 & S & $5.10$ & 1.067 & 1.33 & 2.097 & 0.75 & 0.493 & 0.86 & 1.402 & A1112\tablenotemark{b}\\
J112559.8$+$252837 & V & $-1.74$ & 0.446 & 0.76 & 0.078 & 1.24 & 1.046 & 1.02 & 0.588 & A1258\tablenotemark{b}\\
J113902.6$+$322821 & S & $0.95$ & 0.384 & 2.12 & 3.900 & 0.17 & 0.040 & 0.78 & 2.879 & A1336\\
J114027.7$+$120308 & S & $0.23$ & 0.087 & 0.94 & 0.645 & 1.22 & 1.055 & 1.10 & 0.356 & \nodata\\
J115531.5$+$031150 & C & $0.53$ & 0.260 & 1.61 & 2.318 & 0.83 & 0.607 & 1.04 & 0.438 & \nodata\\
J120013.9$+$561502 & A & $4.32$ & 2.192 & 1.37 & 2.057 & 0.03 & 0.652 & 0.88 & 1.646 & A1436\\
J120210.3$+$274109 & S & $1.35$ & 0.641 & 1.44 & 2.262 & 2.31 & 2.717 & 0.73 & 2.034 & \nodata\\
J125935.6$+$275735 & V & $0.33$ & 0.186 & 1.33 & 3.891 & 1.66 & 1.563 & 1.00 & 0.727 & A1656\\
J133425.1$+$381757 & S & $4.54$ & 1.119 & 1.38 & 2.693 & 0.51 & 0.273 & 0.99 & 0.712 & \nodata\\
J135341.7$+$331339 & A & $16.33$ & 1.798 & 1.48 & 3.900 & 0.51 & 0.224 & 0.97 & 0.975 & \nodata\\
J135449.7$+$203611 & C & $5.31$ & 1.043 & 1.61 & 3.900 & 0.42 & 0.176 & 0.95 & 1.210 & \nodata\\
J140313.2$+$061009 & A & $4.02$ & 0.430 & 1.08 & 1.241 & 0.20 & 0.045 & 0.80 & 2.468 & \nodata\\
J142131.0$-$001246 & C & $37.35$ & 2.310 & 1.00 & 0.785 & 2.22 & 2.110 & 1.01 & 0.577 & \nodata\\
J144818.3$+$033138 & A & $-6.11$ & 2.939 & 1.14 & 1.619 & 4.95 & 3.719 & 1.13 & 0.336 & \nodata\\
J145107.4$+$181417 & C & $-6.26$ & 2.409 & 2.49 & 3.900 & 3.02 & 3.062 & 0.94 & 1.336 & \nodata\\
J150957.3$+$332715 & V & $-1.55$ & 0.244 & 1.05 & 0.925 & 1.40 & 1.256 & 0.96 & 0.867 & A2034\\
J151131.3$+$071506 & S & $-0.41$ & 0.192 & 3.43 & 3.900 & 2.22 & 2.079 & 0.59 & 3.900 & A2040\\
J152642.0$+$005330 & S & $3.53$ & 0.991 & 1.28 & 1.528 & 1.42 & 1.257 & 0.65 & 2.531 & \nodata\\
J154624.8$+$362954 & A & $-2.19$ & 0.045 & 1.13 & 1.587 & 2.43 & 2.296 & 1.02 & 0.473 & A2130\tablenotemark{b}\\
J162037.0$+$252053 & V & $1.84$ & 0.722 & 1.18 & 1.357 & 2.60 & 2.421 & 0.67 & 2.198 & A2177\tablenotemark{a}\\
J164923.9$+$263502 & S & $0.89$ & 0.531 & 1.06 & 1.072 & 1.11 & 0.921 & 0.87 & 1.795 & \nodata\\
J215423.6$+$003710 & V & $0.76$ & 0.323 & 1.35 & 2.640 & 0.52 & 0.234 & 0.91 & 1.073 & A2392\tablenotemark{b}
\enddata
\tablecomments{Col. (1): the name of the source; col. (2): the sample the source belongs to; V for the visual-bent sample, A for the auto-bent sample, S for the straight sample, and C for the single-component sample; col. (3): the normalized value for the $\beta$ test; col. (4): the confidence level in terms of $\sigma$ for the presence of optical substructure in the cluster based on the $\beta$ test; col. (5): the normalized value for the $\Delta$ test; col. (6): the confidence level in terms of $\sigma$ for the presence of optical substructure in the cluster based on the $\Delta$ test; col. (7): the normalized value for the $\alpha$ test; col. (8): the confidence level in terms of $\sigma$ for the presence of optical substructure in the cluster based on the $\alpha$ test; col. (9): the normalized value for the $\epsilon$ test; col. (10): the confidence level in terms of $\sigma$ for the presence of optical substructure in the cluster based on the $\epsilon$ test; col. (11): the Abell cluster that our source might be associated with.}
\tablenotetext{a}{The difference in redshift between the identified Abell cluster (from NED) and the radio cluster is greater than $1,500$ km s$^{-1}$.}
\tablenotetext{b}{The Abell cluster does not have a confirmed redshift in NED.}
\label{table_sources_fixed2}
\end{deluxetable}

\begin{deluxetable}{lcccccrrrrrrr}
\tabletypesize{\tiny}
\tablewidth{0pt}
\tablecolumns{13}
\tablecaption{All Sources With Substructure Measurements (Using the Shifting Gapper Method)}
\tablehead{\colhead{Source Name}&
		  \colhead{Smpl}&
		  \colhead{$\alpha$}&
		  \colhead{$\delta$}&
		  \colhead{$\alpha_{clus}$}&
		  \colhead{$\delta_{clus}$}&
		  \colhead{FR}&
		  \colhead{FR}&
		  \colhead{$z$}&
		  \colhead{$z_{clus}$}&
		  \colhead{N$^{-19}_{1.0}$}&
		  \colhead{N$^z_{3.0}$}&
		  \colhead{$\sigma$}
		  \cr
		   & & & & & & &
		   & & & & &
		  \colhead{(km s$^{-1}$)}
		  \cr
		  \colhead{(1)}&
		  \colhead{(2)}&
		  \colhead{(3)}&
		  \colhead{(4)}&
		  \colhead{(5)}&
		  \colhead{(6)}&
		  \colhead{(7)}&
		  \colhead{(8)}&
		  \colhead{(9)}&
		  \colhead{(10)}&
		  \colhead{(11)}&
		  \colhead{(12)}&
		  \colhead{(13)}}
\startdata
J005702.1$-$005231 & V & 00:57:02.10 & $-00$:52:31.0 & 00:56:39.69 & $-00$:42:59.7 & I & I & 0.044\tablenotemark{a} & 0.044 & $53$ & 237 & 508\\
J073050.5$+$445601 & S & 07:30:50.54 & $+44$:56:01.2 & 07:30:31.14 & $+44$:25:19.7 & I & II & 0.072\tablenotemark{a} & 0.074 & $20$ & 50 & 2144\\
J075917.1$+$270916 & V & 07:59:17.10 & $+27$:09:16.0 & 07:59:01.18 & $+27$:09:18.7 & I & I & 0.112\tablenotemark{b} & 0.097 & $98$ & 45 & 337\\
J080757.0$+$163716 & A & 08:07:57.05 & $+16$:37:16.2 & 08:07:58.51 & $+16$:22:43.2 & I & I? & 0.103\tablenotemark{b} & 0.100 & $0$ & 36 & 267\\
J091344.5$+$474216 & A & 09:13:44.54 & $+47$:42:16.4 & 09:13:06.12 & $+47$:39:41.9 & I & I? & 0.051\tablenotemark{a} & 0.051 & $44$ & 63 & 316\\
J092109.5$+$502802 & C & 09:21:09.36 & $+50$:28:02.3 & 09:21:46.91 & $+50$:22:20.3 & I & II & 0.087\tablenotemark{a} & 0.090 & $6$ & 42 & 2088\\
J102236.7$+$082238 & C & 10:22:36.72 & $+08$:22:37.7 & 10:24:28.95 & $+08$:17:49.1 & I & II & 0.044\tablenotemark{a} & 0.042 & $23$ & 48 & 1603\\
J102517.1$+$483213 & S & 10:25:17.13 & $+48$:32:13.7 & 10:23:29.94 & $+48$:30:25.3 & I & I & 0.149\tablenotemark{a} & 0.142 & $35$ & 37 & 1375\\
J102757.8$+$103345 & S & 10:27:57.86 & $+10$:33:45.9 & 10:27:47.12 & $+10$:33:42.7 & I & II & 0.102\tablenotemark{b} & 0.109 & $103$ & 81 & 451\\
J104104.1$+$335520 & V & 10:41:04.10 & $+33$:55:20.0 & 10:41:16.25 & $+33$:53:32.7 & I & I & 0.084\tablenotemark{a} & 0.083 & $24$ & 44 & 303\\
J105147.4$+$552309 & S & 10:51:47.46 & $+55$:23:09.4 & 10:52:54.78 & $+55$:12:43.2 & I & I & 0.074\tablenotemark{a} & 0.073 & $62$ & 78 & 345\\
J112559.8$+$252837 & V & 11:25:59.80 & $+25$:28:37.0 & 11:26:38.90 & $+25$:23:49.0 & I & I & 0.116\tablenotemark{a} & 0.113 & $53$ & 40 & 560\\
J113902.6$+$322821 & S & 11:39:02.60 & $+32$:28:21.6 & 11:39:04.86 & $+32$:26:24.1 & I & I & 0.141\tablenotemark{a} & 0.129 & $70$ & 53 & 999\\
J114027.7$+$120308 & S & 11:40:27.76 & $+12$:03:08.0 & 11:40:12.99 & $+11$:34:55.8 & I & I & 0.081\tablenotemark{a} & 0.081 & $-17$ & 40 & 363\\
J115504.6$+$230926 & S & 11:55:04.67 & $+23$:09:26.5 & 11:55:14.22 & $+23$:27:24.7 & II & II? & 0.144\tablenotemark{b} & 0.141 & $-16$ & 31 & 782\\
J115531.5$+$031150 & C & 11:55:31.44 & $+03$:11:50.2 & 11:54:55.48 & $+03$:25:14.5 & II & I & 0.081\tablenotemark{b} & 0.075 & $13$ & 47 & 1256\\
J120013.9$+$561502 & A & 12:00:13.91 & $+56$:15:02.0 & 12:03:21.19 & $+56$:25:03.1 & I & II & 0.061\tablenotemark{b} & 0.064 & $85$ & 84 & 603\\
J120210.3$+$274109 & S & 12:02:10.38 & $+27$:41:09.0 & 12:02:44.34 & $+27$:44:31.2 & I & I & 0.134\tablenotemark{a} & 0.137 & $52$ & 34 & 584\\
J121111.0$+$060743 & A & 12:11:11.02 & $+06$:07:43.5 & 12:11:34.53 & $+06$:13:41.1 & I & I & 0.139\tablenotemark{a} & 0.137 & $41$ & 30 & 351\\
J121121.1$+$141439 & S & 12:11:21.13 & $+14$:14:39.2 & 12:11:26.55 & $+14$:22:38.9 & I & I & 0.064\tablenotemark{a} & 0.075 & $-20$ & 35 & 2655\\
J125935.6$+$275735 & V & 12:59:35.60 & $+27$:57:35.0 & 12:59:12.80 & $+27$:53:10.1 & I & I & 0.024\tablenotemark{a} & 0.023 & $166$ & 807 & 773\\
J133425.1$+$381757 & S & 13:34:25.20 & $+38$:17:57.4 & 13:31:54.30 & $+37$:52:12.5 & II & II & 0.063\tablenotemark{a} & 0.060 & $17$ & 89 & 1181\\
J135341.7$+$331339 & A & 13:53:41.71 & $+33$:13:39.6 & 13:56:01.51 & $+33$:06:14.4 & I & I & 0.048\tablenotemark{b} & 0.052 & $56$ & 124 & 1664\\
J135449.7$+$203611 & C & 13:54:49.68 & $+20$:36:11.5 & 13:56:01.78 & $+20$:40:15.8 & I & I & 0.070\tablenotemark{a} & 0.068 & $17$ & 109 & 2140\\
J140148.3$+$283321 & V & 14:01:48.30 & $+28$:33:21.0 & 13:59:42.95 & $+28$:09:55.1 & I & I & 0.065\tablenotemark{a} & 0.070 & $36$ & 203 & 1880\\
J140313.2$+$061009 & A & 14:03:13.26 & $+06$:10:09.1 & 14:05:09.02 & $+06$:14:43.1 & I & I & 0.083\tablenotemark{a} & 0.084 & $25$ & 69 & 430\\
J142131.0$-$001246 & C & 14:21:30.96 & $-00$:12:46.4 & 14:18:54.00 & $+00$:05:27.0 & I & II & 0.052\tablenotemark{a} & 0.052 & $3$ & 60 & 358\\
J142921.0$+$233616 & V & 14:29:21.00 & $+23$:36:16.0 & 14:29:23.75 & $+23$:19:14.3 & I & II & 0.120\tablenotemark{b} & 0.131 & $27$ & 31 & 1874\\
J143716.9$+$245209 & V & 14:37:16.90 & $+24$:52:09.0 & 14:36:43.75 & $+24$:37:38.6 & I & I & 0.090\tablenotemark{a} & 0.088 & $68$ & 82 & 480\\
J143914.0$+$230604 & C & 14:39:13.92 & $+23$:06:04.0 & 14:38:45.59 & $+23$:05:05.0 & I & I & 0.067\tablenotemark{a} & 0.071 & $31$ & 55 & 1322\\
J144818.3$+$033138 & A & 14:48:18.39 & $+03$:31:38.2 & 14:42:15.89 & $+03$:28:48.1 & II & I & 0.028\tablenotemark{b} & 0.027 & $8$ & 126 & 326\\
J145107.4$+$181417 & C & 14:51:07.44 & $+18$:14:17.2 & 14:52:36.77 & $+18$:11:29.1 & I & I & 0.062\tablenotemark{a} & 0.054 & $-18$ & 145 & 2231\\
J150957.3$+$332715 & V & 15:09:57.30 & $+33$:27:15.0 & 15:10:03.76 & $+33$:30:58.6 & I & I & 0.117\tablenotemark{a} & 0.113 & $140$ & 79 & 976\\
J151131.3$+$071506 & S & 15:11:31.38 & $+07$:15:07.0 & 15:14:18.29 & $+07$:16:01.1 & I & I & 0.045\tablenotemark{a} & 0.041 & $3$ & 290 & 1607\\
J152642.0$+$005330 & S & 15:26:42.01 & $+00$:53:30.1 & 15:27:12.18 & $+01$:10:45.0 & I & I & 0.117\tablenotemark{b} & 0.116 & $46$ & 32 & 567\\
J154624.8$+$362954 & A & 15:46:24.83 & $+36$:29:54.1 & 15:45:06.21 & $+36$:04:52.8 & II & II? & 0.051\tablenotemark{b} & 0.068 & $18$ & 116 & 686\\
J162037.0$+$252053 & V & 16:20:37.01 & $+25$:20:53.0 & 16:21:39.17 & $+25$:33:13.5 & I & I & 0.103\tablenotemark{a} & 0.101 & $36$ & 38 & 616\\
J164923.9$+$263502 & S & 16:49:24.00 & $+26$:35:02.1 & 16:48:38.44 & $+26$:51:04.4 & I & I & 0.055\tablenotemark{a} & 0.055 & $14$ & 77 & 1774\\
J215423.6$+$003710 & V & 21:54:23.60 & $+00$:37:10.0 & 21:54:24.74 & $+00$:41:51.6 & II & ??? & 0.219\tablenotemark{b} & 0.217 & $156$ & 32 & 1485
\enddata
\tablecomments{Col. (1): name of the source; col. (2): identifier for the sample the source belongs to; V for the visual-bent sample, A for the auto-bent sample, S for the straight sample, and C for the single-component sample; col. (3): right ascension of the radio source; col. (4): declination of the radio source; col. (5): right ascension of the cluster center; col. (6): declination of the cluster center; col. (7) FR type based on \citet{ledlow1996}; col. (8): FR type based on visual examination; a question mark represents a dubious visual FR I/II classification; three successive question marks represent an inability to visually classify the radio source as either FR I or II; col. (9): redshift of the radio source; col. (10): redshift of the cluster center; col. (11): number of cluster galaxies based on the N$^{-19}_{1.0}$ method from \citet{wing2011}; col. (12): number of cluster galaxies with \SDSS\/ spectroscopically measured redshifts within $3.0$ Mpc and $\pm 5000$ km s$^{-1}$ of the cluster center (N$^z_{3.0}$); col. (13): the velocity dispersion of the cluster.}
\tablenotetext{a}{The redshift comes from a spectroscopic measurement.}
\tablenotetext{b}{The redshift comes from a photometric measurement.}
\label{table_sources_gapper1}
\end{deluxetable}

\begin{deluxetable}{lcrrcrcrcrc}
\tabletypesize{\tiny}
\tablewidth{0pt}
\tablecolumns{11}
\tablecaption{All Sources With Substructure Measurements (Using the Shifting Gapper Method)}
\tablehead{\colhead{Source Name}&
		  \colhead{Smpl}&
		  \colhead{$\beta$}&
		  \colhead{sig}&
		  \colhead{$\Delta$}&
		  \colhead{sig}&
		  \colhead{$\alpha$}&
		  \colhead{sig}&
		  \colhead{$\epsilon$}&
		  \colhead{sig}&
		  \colhead{Abell}
		  \cr
		   & & &
		  \colhead{($\sigma$)} & & 
		  \colhead{($\sigma$)} & & 
		  \colhead{($\sigma$)} & & 
		  \colhead{($\sigma$)}
		  \cr
		  \colhead{(1)}&
		  \colhead{(2)}&
		  \colhead{(3)}&
		  \colhead{(4)}&
		  \colhead{(5)}&
		  \colhead{(6)}&
		  \colhead{(7)}&
		  \colhead{(8)}&
		  \colhead{(9)}&
		  \colhead{(10)}&
		  \colhead{(11)}}
\startdata
J005702.1$-$005231 & V & $-1.97$ & 1.636 & 1.31 & 2.693 & 4.02 & 3.891 & 0.89 & 1.662 & A0119\\
J073050.5$+$445601 & S & $-5.91$ & 0.882 & 1.34 & 2.149 & 1.21 & 1.001 & 1.11 & 0.171 & \nodata\\
J075917.1$+$270916 & V & $6.20$ & 0.103 & 1.18 & 1.393 & 1.35 & 1.183 & 0.49 & 2.968 & A0610\\
J080757.0$+$163716 & A & $-0.90$ & 0.188 & 1.02 & 0.755 & 2.84 & 2.879 & 1.19 & 0.082 & \nodata\\
J091344.5$+$474216 & A & $2.10$ & 0.715 & 1.22 & 1.656 & 0.53 & 0.300 & 1.36 & 0.126 & A0757\\
J092109.5$+$502802 & C & $-2.77$ & 0.607 & 1.65 & 3.719 & 1.43 & 1.219 & 1.23 & 0.030 & \nodata\\
J102236.7$+$082238 & C & $-11.23$ & 0.365 & 1.21 & 2.282 & 0.94 & 0.635 & 1.16 & 0.095 & A0989\tablenotemark{b}\\
J102517.1$+$483213 & S & $1.75$ & 0.239 & 1.33 & 1.833 & 1.62 & 1.426 & 1.49 & 0.001 & \nodata\\
J102757.8$+$103345 & S & $-4.69$ & 0.835 & 1.25 & 1.981 & 4.15 & 3.900 & 0.72 & 3.036 & A1020\\
J104104.1$+$335520 & V & $5.82$ & 0.448 & 1.96 & 2.540 & 1.36 & 1.205 & 1.08 & 0.519 & \nodata\\
J105147.4$+$552309 & S & $5.33$ & 1.099 & 1.29 & 1.884 & 0.53 & 0.284 & 0.78 & 1.779 & A1112\tablenotemark{b}\\
J112559.8$+$252837 & V & $-1.67$ & 0.456 & 0.78 & 0.082 & 1.20 & 0.982 & 0.99 & 0.689 & A1258\tablenotemark{b}\\
J113902.6$+$322821 & S & $3.17$ & 1.210 & 1.79 & 3.719 & 1.10 & 0.881 & 0.81 & 2.383 & A1336\\
J114027.7$+$120308 & S & $-1.08$ & 0.404 & 0.88 & 0.540 & 1.25 & 1.069 & 1.02 & 0.566 & \nodata\\
J115504.6$+$230926 & S & $-0.76$ & 0.347 & 1.08 & 1.043 & 3.13 & 2.598 & 0.88 & 1.143 & A1413\\
J115531.5$+$031150 & C & $-0.81$ & 0.474 & 1.93 & 3.239 & 0.62 & 0.379 & 1.06 & 0.318 & \nodata\\
J120013.9$+$561502 & A & $4.16$ & 2.167 & 1.29 & 1.753 & 0.03 & 0.773 & 0.92 & 1.274 & A1436\\
J120210.3$+$274109 & S & $1.02$ & 0.490 & 1.44 & 2.251 & 2.24 & 2.543 & 0.74 & 1.998 & \nodata\\
J121111.0$+$060743 & A & $3.36$ & 1.397 & 0.91 & 0.521 & 0.64 & 0.372 & 0.92 & 0.953 & \nodata\\
J121121.1$+$141439 & S & $0.06$ & 0.030 & 2.45 & 3.900 & 0.69 & 0.446 & 1.08 & 0.263 & \nodata\\
J125935.6$+$275735 & V & $0.33$ & 0.183 & 1.32 & 3.900 & 1.71 & 1.616 & 1.00 & 0.680 & A1656\\
J133425.1$+$381757 & S & $5.17$ & 1.375 & 1.48 & 2.777 & 0.46 & 0.216 & 1.02 & 0.502 & \nodata\\
J135341.7$+$331339 & A & $15.37$ & 1.633 & 1.66 & 3.900 & 0.56 & 0.273 & 1.14 & 0.023 & \nodata\\
J135449.7$+$203611 & C & $15.67$ & 1.883 & 1.93 & 3.900 & 2.69 & 2.537 & 1.03 & 0.393 & \nodata\\
J140148.3$+$283321 & V & $-17.66$ & 2.450 & 2.25 & 3.900 & 1.15 & 0.962 & 0.85 & 3.900 & \nodata\\
J140313.2$+$061009 & A & $3.58$ & 0.503 & 1.16 & 1.727 & 0.30 & 0.095 & 0.73 & 3.216 & \nodata\\
J142131.0$-$001246 & C & $27.89$ & 1.763 & 0.98 & 0.741 & 2.31 & 2.266 & 0.98 & 0.669 & \nodata\\
J142921.0$+$233616 & V & $6.39$ & 1.310 & 1.80 & 2.948 & 1.50 & 1.443 & 0.83 & 1.775 & \nodata\\
J143716.9$+$245209 & V & $-12.74$ & 0.914 & 1.48 & 3.195 & 1.58 & 1.478 & 0.97 & 0.831 & A1939\\
J143914.0$+$230604 & C & $15.59$ & 0.860 & 1.10 & 1.311 & 0.95 & 0.664 & 0.80 & 2.250 & \nodata\\
J144818.3$+$033138 & A & $-6.14$ & 2.636 & 1.25 & 2.203 & 2.91 & 2.609 & 0.96 & 0.721 & \nodata\\
J145107.4$+$181417 & C & $-4.61$ & 2.661 & 2.32 & 3.900 & 0.87 & 0.600 & 0.86 & 3.195 & \nodata\\
J150957.3$+$332715 & V & $-2.69$ & 0.357 & 1.06 & 0.970 & 1.42 & 1.281 & 0.92 & 1.081 & A2034\\
J151131.3$+$071506 & S & $-0.64$ & 0.281 & 3.67 & 3.900 & 3.39 & 3.481 & 0.70 & 3.900 & A2040\\
J152642.0$+$005330 & S & $3.53$ & 0.991 & 1.28 & 1.528 & 1.42 & 1.257 & 0.65 & 2.531 & \nodata\\
J154624.8$+$362954 & A & $-3.31$ & 0.126 & 1.21 & 2.484 & 3.04 & 2.754 & 1.00 & 0.682 & A2130\tablenotemark{b}\\
J162037.0$+$252053 & V & $1.84$ & 0.722 & 1.18 & 1.357 & 2.60 & 2.421 & 0.67 & 2.198 & A2177\tablenotemark{a}\\
J164923.9$+$263502 & S & $0.97$ & 0.610 & 1.32 & 3.719 & 1.14 & 0.939 & 0.89 & 1.829 & \nodata\\
J215423.6$+$003710 & V & $0.43$ & 0.184 & 1.36 & 2.573 & 0.24 & 0.052 & 0.83 & 1.679 & A2392\tablenotemark{b}
\enddata
\tablecomments{Col. (1): the name of the source; col. (2): the sample the source belongs to; V for the visual-bent sample, A for the auto-bent sample, S for the straight sample, and C for the single-component sample; col. (3): the normalized value for the $\beta$ test; col. (4): the confidence level in terms of $\sigma$ for the presence of optical substructure in the cluster based on the $\beta$ test; col. (5): the normalized value for the $\Delta$ test; col. (6): the confidence level in terms of $\sigma$ for the presence of optical substructure in the cluster based on the $\Delta$ test; col. (7): the normalized value for the $\alpha$ test; col. (8): the confidence level in terms of $\sigma$ for the presence of optical substructure in the cluster based on the $\alpha$ test; col. (9): the normalized value for the $\epsilon$ test; col. (10): the confidence level in terms of $\sigma$ for the presence of optical substructure in the cluster based on the $\epsilon$ test; col. (11): the Abell cluster that our source might be associated with.}
\tablenotetext{a}{The difference in redshift between the identified Abell cluster (from NED) and the radio cluster is greater than $1,500$ km s$^{-1}$.}
\tablenotetext{b}{The Abell cluster does not have a confirmed redshift in NED.}
\label{table_sources_gapper2}
\end{deluxetable}

\begin{deluxetable}{llcc}
\tabletypesize{\tiny}
\tablewidth{0pt}
\tablecolumns{4}
\tablecaption{Spearman Correlations for Tests and Properties, Fixed Gap Method.}
\tablehead{\colhead{Parameter}&
		  \colhead{Parameter}&
		  \colhead{Correlation}&
		  \colhead{Significance}
		  \cr
		   & & & \colhead{($\sigma$)}
		  \cr
		  \colhead{(1)}&
		  \colhead{(2)}&
		  \colhead{(3)}&
		  \colhead{(4)}}
\startdata
$z$ & $r-i$ color & $0.907$ & $4.89$\\
$z_{clust}$ & $r-i$ color & $0.924$ & $4.98$\\
$z$ & $g-r$ color & $0.901$ & $4.85$\\
$z_{clust}$ & $g-r$ color & $0.934$ & $5.03$\\
$z$ & $g-i$ color & $0.922$ & $4.96$\\
$z_{clust}$ & $g-i$ color & $0.947$ & $5.10$\\
$z$ & $\sigma$ & $-0.076$ & $0.41$\\
$z_{clust}$ & $\sigma$ & $-0.097$ & $0.52$\\
Opening Angle & $\Delta v_{radio}$ & $0.031$ & $0.15$\\
N$_{3.0}^z$ & N$_{1.0}^{-19}$ & $-0.043$ & $0.23$\\
N$_{3.0}^z$ & M$_{r,BCG}$ & $-0.370$ & $1.92$\\
N$_{1.0}^{-19}$ & M$_{r,BCG}$ & $-0.070$ & $0.36$\\
$\alpha$ & $\alpha$ significance & $0.659$ & $3.55$\\
$\alpha$ & N$^{-19}_{1.0}$ & $-0.231$ & $1.24$\\
$\alpha$ & N$^{z}_{3.0}$ & $0.144$ & $0.78$\\
$\alpha$ & $z$ & $-0.125$ & $0.67$\\
$\alpha$ & $z_{clust}$ & $-0.078$ & $0.42$\\
$\alpha$ & $\bar{v}-v$ & $0.117$ & $0.63$\\
$\alpha$ & $\bar{v}$ & $-0.078$ & $0.42$\\
$\alpha$ & $\sigma$ & $-0.148$ & $0.80$\\
$\alpha$ & M$_{r,BCG}$ & $-0.297$ & $1.54$\\
$\alpha$ & opening angle & $-0.230$ & $1.13$\\
$\alpha$ significance & N$^{-19}_{1.0}$ & $-0.452$ & $2.44$\\
$\alpha$ significance & N$^{z}_{3.0}$ & $0.054$ & $0.29$\\
$\alpha$ significance & $z$ & $-0.282$ & $1.52$\\
$\alpha$ significance & $z_{clust}$ & $-0.262$ & $1.41$\\
$\alpha$ significance & $\bar{v}-v$ & $-0.090$ & $0.48$\\
$\alpha$ significance & $\bar{v}$ & $-0.262$ & $1.41$\\
$\alpha$ significance & $\sigma$ & $0.117$ & $0.63$\\
$\alpha$ significance & M$_{r,BCG}$ & $0.068$ & $0.35$\\
$\alpha$ significance & opening angle & $-0.172$ & $0.84$\\
$\beta$ & $\beta$ significance & $0.243$ & $1.31$\\
$\beta$ & N$^{-19}_{1.0}$ & $-0.009$ & $0.05$\\
$\beta$ & N$^{z}_{3.0}$ & $-0.072$ & $0.39$\\
$\beta$ & $z$ & $0.076$ & $0.41$\\
$\beta$ & $z_{clust}$ & $0.051$ & $0.28$\\
$\beta$ & $\bar{v}-v$ & $0.241$ & $1.30$\\
$\beta$ & $\bar{v}$ & $0.051$ & $0.28$\\
$\beta$ & $\sigma$ & $0.016$ & $0.09$\\
$\beta$ & M$_{r,BCG}$ & $0.298$ & $1.55$\\
$\beta$ & opening angle & $0.265$ & $1.30$\\
$\beta$ significance & N$^{-19}_{1.0}$ & $-0.252$ & $1.36$\\
$\beta$ significance & N$^{z}_{3.0}$ & $0.136$ & $0.73$\\
$\beta$ significance & $z$ & $-0.365$ & $1.97$\\
$\beta$ significance & $z_{clust}$ & $-0.391$ & $2.10$\\
$\beta$ significance & $\bar{v}-v$ & $-0.045$ & $0.24$\\
$\beta$ significance & $\bar{v}$ & $-0.391$ & $2.10$\\
$\beta$ significance & $\sigma$ & $0.199$ & $1.07$\\
$\beta$ significance & M$_{r,BCG}$ & $0.309$ & $1.61$\\
$\beta$ significance & opening angle & $0.213$ & $1.04$\\
$\Delta$ & $\Delta$ significance & $0.718$ & $3.87$\\
$\Delta$ & N$^{-19}_{1.0}$ & $0.055$ & $0.29$\\
$\Delta$ & N$^{z}_{3.0}$ & $0.194$ & $1.05$\\
$\Delta$ & $z$ & $-0.032$ & $0.17$\\
$\Delta$ & $z_{clust}$ & $-0.090$ & $0.49$\\
$\Delta$ & $\bar{v}-v$ & $-0.225$ & $1.21$\\
$\Delta$ & $\bar{v}$ & $-0.090$ & $0.49$\\
$\Delta$ & $\sigma$ & $0.551$ & $2.97$\\
$\Delta$ & M$_{r,BCG}$ & $0.138$ & $0.72$\\
$\Delta$ & opening angle & $0.338$ & $1.65$\\
$\Delta$ significance & N$^{-19}_{1.0}$ & $-0.132$ & $0.71$\\
$\Delta$ significance & N$^{z}_{3.0}$ & $0.249$ & $1.34$\\
$\Delta$ significance & $z$ & $-0.357$ & $1.92$\\
$\Delta$ significance & $z_{clust}$ & $-0.405$ & $2.18$\\
$\Delta$ significance & $\bar{v}-v$ & $-0.179$ & $0.96$\\
$\Delta$ significance & $\bar{v}$ & $-0.405$ & $2.18$\\
$\Delta$ significance & $\sigma$ & $0.584$ & $3.15$\\
$\Delta$ significance & M$_{r,BCG}$ & $0.216$ & $1.12$\\
$\Delta$ significance & opening angle & $0.217$ & $1.06$\\
$\epsilon$ & $\epsilon$ significance & $-0.489$ & $2.63$\\
$\epsilon$ & N$^{-19}_{1.0}$ & $-0.395$ & $2.13$\\
$\epsilon$ & N$^{z}_{3.0}$ & $0.010$ & $0.05$\\
$\epsilon$ & $z$ & $-0.402$ & $2.17$\\
$\epsilon$ & $z_{clust}$ & $-0.409$ & $2.20$\\
$\epsilon$ & $\bar{v}-v$ & $-0.001$ & $0.01$\\
$\epsilon$ & $\bar{v}$ & $-0.409$ & $2.20$\\
$\epsilon$ & $\sigma$ & $-0.135$ & $0.73$\\
$\epsilon$ & M$_{r,BCG}$ & $0.400$ & $2.08$\\
$\epsilon$ & opening angle & $-0.383$ & $1.88$\\
$\epsilon$ significance & N$^{-19}_{1.0}$ & $-0.088$ & $0.47$\\
$\epsilon$ significance & N$^{z}_{3.0}$ & $-0.056$ & $0.30$\\
$\epsilon$ significance & $z$ & $0.035$ & $0.19$\\
$\epsilon$ significance & $z_{clust}$ & $0.003$ & $0.02$\\
$\epsilon$ significance & $\bar{v}-v$ & $-0.270$ & $1.45$\\
$\epsilon$ significance & $\bar{v}$ & $0.003$ & $0.02$\\
$\epsilon$ significance & $\sigma$ & $0.400$ & $2.15$\\
$\epsilon$ significance & M$_{r,BCG}$ & $0.083$ & $0.43$\\
$\epsilon$ significance & opening angle & $0.409$ & $2.00$\\
Opening Angle & Radio-BCG Sep & $-0.182$ & $0.89$\\
Opening Angle & Radio-Cluster Sep & $0.109$ & $0.54$\\
$\frac{N^z_{3.0} - N^{-19}_{1.0}}{N^z_{3.0}}$ & Radio-BCG Sep & $0.220$ & $1.19$\\
$\frac{N^z_{3.0} - N^{-19}_{1.0}}{N^z_{3.0}}$ & Radio-Cluster Sep & $0.493$ & $2.65$\\
$\Delta$ significance & $\alpha$ significance & $-0.143$ & $0.77$\\
$\Delta$ significance & $\beta$ significance & $0.221$ & $1.19$\\
$\Delta$ significance & $\epsilon$ significance & $0.200$ & $1.08$\\
$\alpha$ significance & $\beta$ significance & $0.076$ & $0.41$\\
$\alpha$ significance & $\epsilon$ significance & $0.048$ & $0.26$\\
$\beta$ significance & $\epsilon$ significance & $0.100$ & $0.54$\\
Opening Angle\tablenotemark{a} & $\Delta v_{radio}$ & $-0.095$ & $0.45$
\enddata
\label{table_spearman_fixed}
\tablenotetext{a}{The correlation for only those sources we have identified as ``true'' bent double-lobed radio sources.}
\end{deluxetable}

\begin{deluxetable}{llcc}
\tabletypesize{\tiny}
\tablewidth{0pt}
\tablecolumns{4}
\tablecaption{Spearman Correlations for Tests and Properties, Shifting Gapper Method.}
\tablehead{\colhead{Parameter}&
		  \colhead{Parameter}&
		  \colhead{Correlation}&
		  \colhead{Significance}
		  \cr
		   & & & \colhead{($\sigma$)}
		  \cr
		  \colhead{(1)}&
		  \colhead{(2)}&
		  \colhead{(3)}&
		  \colhead{(4)}}
\startdata
$z$ & $r-i$ color & $0.895$ & $5.52$\\
$z_{clust}$ & $r-i$ color & $0.911$ & $5.61$\\
$z$ & $g-r$ color & $0.893$ & $5.50$\\
$z_{clust}$ & $g-r$ color & $0.915$ & $5.64$\\
$z$ & $g-i$ color & $0.902$ & $5.56$\\
$z_{clust}$ & $g-i$ color & $0.923$ & $5.69$\\
$z$ & $\sigma$ & $-0.098$ & $0.60$\\
$z_{clust}$ & $\sigma$ & $-0.082$ & $0.50$\\
Opening Angle & $\Delta v_{radio}$ & $0.062$ & $0.34$\\
N$_{3.0}^z$ & N$_{1.0}^{-19}$ & $0.101$ & $0.62$\\
N$_{3.0}^z$ & M$_{r,BCG}$ & $-0.232$ & $1.39$\\
N$_{1.0}^{-19}$ & M$_{r,BCG}$ & $-0.161$ & $0.97$\\
$\alpha$ & $\alpha$ significance & $0.646$ & $3.98$\\
$\alpha$ & N$^{-19}_{1.0}$ & $-0.155$ & $0.95$\\
$\alpha$ & N$^{z}_{3.0}$ & $0.085$ & $0.53$\\
$\alpha$ & $z$ & $0.007$ & $0.05$\\
$\alpha$ & $z_{clust}$ & $0.034$ & $0.21$\\
$\alpha$ & $\bar{v}-v$ & $-0.001$ & $0.01$\\
$\alpha$ & $\bar{v}$ & $0.034$ & $0.21$\\
$\alpha$ & $\sigma$ & $-0.138$ & $0.85$\\
$\alpha$ & M$_{r,BCG}$ & $-0.211$ & $1.27$\\
$\alpha$ & opening angle & $-0.000$ & $0.00$\\
$\alpha$ significance & N$^{-19}_{1.0}$ & $-0.363$ & $2.24$\\
$\alpha$ significance & N$^{z}_{3.0}$ & $0.136$ & $0.84$\\
$\alpha$ significance & $z$ & $-0.189$ & $1.17$\\
$\alpha$ significance & $z_{clust}$ & $-0.193$ & $1.19$\\
$\alpha$ significance & $\bar{v}-v$ & $-0.092$ & $0.56$\\
$\alpha$ significance & $\bar{v}$ & $-0.193$ & $1.19$\\
$\alpha$ significance & $\sigma$ & $0.119$ & $0.73$\\
$\alpha$ significance & M$_{r,BCG}$ & $0.136$ & $0.81$\\
$\alpha$ significance & opening angle & $-0.045$ & $0.25$\\
$\beta$ & $\beta$ significance & $0.060$ & $0.37$\\
$\beta$ & N$^{-19}_{1.0}$ & $0.127$ & $0.78$\\
$\beta$ & N$^{z}_{3.0}$ & $-0.196$ & $1.21$\\
$\beta$ & $z$ & $0.123$ & $0.76$\\
$\beta$ & $z_{clust}$ & $0.074$ & $0.46$\\
$\beta$ & $\bar{v}-v$ & $0.016$ & $0.10$\\
$\beta$ & $\bar{v}$ & $0.074$ & $0.46$\\
$\beta$ & $\sigma$ & $-0.139$ & $0.86$\\
$\beta$ & M$_{r,BCG}$ & $0.110$ & $0.66$\\
$\beta$ & opening angle & $0.220$ & $1.23$\\
$\beta$ significance & N$^{-19}_{1.0}$ & $-0.147$ & $0.91$\\
$\beta$ significance & N$^{z}_{3.0}$ & $0.256$ & $1.58$\\
$\beta$ significance & $z$ & $-0.292$ & $1.80$\\
$\beta$ significance & $z_{clust}$ & $-0.338$ & $2.08$\\
$\beta$ significance & $\bar{v}-v$ & $0.098$ & $0.60$\\
$\beta$ significance & $\bar{v}$ & $-0.338$ & $2.08$\\
$\beta$ significance & $\sigma$ & $0.208$ & $1.28$\\
$\beta$ significance & M$_{r,BCG}$ & $0.163$ & $0.98$\\
$\beta$ significance & opening angle & $0.012$ & $0.07$\\
$\Delta$ & $\Delta$ significance & $0.646$ & $3.98$\\
$\Delta$ & N$^{-19}_{1.0}$ & $-0.114$ & $0.71$\\
$\Delta$ & N$^{z}_{3.0}$ & $0.272$ & $1.67$\\
$\Delta$ & $z$ & $-0.165$ & $1.02$\\
$\Delta$ & $z_{clust}$ & $-0.179$ & $1.10$\\
$\Delta$ & $\bar{v}-v$ & $0.062$ & $0.38$\\
$\Delta$ & $\bar{v}$ & $-0.179$ & $1.10$\\
$\Delta$ & $\sigma$ & $0.602$ & $3.71$\\
$\Delta$ & M$_{r,BCG}$ & $0.059$ & $0.35$\\
$\Delta$ & opening angle & $0.265$ & $1.47$\\
$\Delta$ significance & N$^{-19}_{1.0}$ & $-0.255$ & $1.57$\\
$\Delta$ significance & N$^{z}_{3.0}$ & $0.408$ & $2.51$\\
$\Delta$ significance & $z$ & $-0.486$ & $2.99$\\
$\Delta$ significance & $z_{clust}$ & $-0.500$ & $3.08$\\
$\Delta$ significance & $\bar{v}-v$ & $0.174$ & $1.07$\\
$\Delta$ significance & $\bar{v}$ & $-0.500$ & $3.08$\\
$\Delta$ significance & $\sigma$ & $0.605$ & $3.73$\\
$\Delta$ significance & M$_{r,BCG}$ & $0.206$ & $1.24$\\
$\Delta$ significance & opening angle & $0.131$ & $0.73$\\
$\epsilon$ & $\epsilon$ significance & $-0.495$ & $3.05$\\
$\epsilon$ & N$^{-19}_{1.0}$ & $-0.330$ & $2.04$\\
$\epsilon$ & N$^{z}_{3.0}$ & $0.006$ & $0.04$\\
$\epsilon$ & $z$ & $-0.266$ & $1.64$\\
$\epsilon$ & $z_{clust}$ & $-0.256$ & $1.58$\\
$\epsilon$ & $\bar{v}-v$ & $0.049$ & $0.30$\\
$\epsilon$ & $\bar{v}$ & $-0.256$ & $1.58$\\
$\epsilon$ & $\sigma$ & $0.127$ & $0.78$\\
$\epsilon$ & M$_{r,BCG}$ & $0.338$ & $2.03$\\
$\epsilon$ & opening angle & $-0.092$ & $0.51$\\
$\epsilon$ significance & N$^{-19}_{1.0}$ & $-0.087$ & $0.54$\\
$\epsilon$ significance & N$^{z}_{3.0}$ & $0.110$ & $0.68$\\
$\epsilon$ significance & $z$ & $-0.028$ & $0.17$\\
$\epsilon$ significance & $z_{clust}$ & $-0.073$ & $0.45$\\
$\epsilon$ significance & $\bar{v}-v$ & $-0.085$ & $0.52$\\
$\epsilon$ significance & $\bar{v}$ & $-0.073$ & $0.45$\\
$\epsilon$ significance & $\sigma$ & $0.239$ & $1.47$\\
$\epsilon$ significance & M$_{r,BCG}$ & $0.037$ & $0.22$\\
$\epsilon$ significance & opening angle & $0.068$ & $0.38$\\
Opening Angle & Radio-BCG Sep & $-0.279$ & $1.56$\\
Opening Angle & Radio-Cluster Sep & $0.081$ & $0.45$\\
$\frac{N^z_{3.0} - N^{-19}_{1.0}}{N^z_{3.0}}$ & Radio-BCG Sep & $0.186$ & $1.15$\\
$\frac{N^z_{3.0} - N^{-19}_{1.0}}{N^z_{3.0}}$ & Radio-Cluster Sep & $0.357$ & $2.20$\\
$\Delta$ significance & $\alpha$ significance & $-0.070$ & $0.43$\\
$\Delta$ significance & $\beta$ significance & $0.142$ & $0.88$\\
$\Delta$ significance & $\epsilon$ significance & $0.007$ & $0.04$\\
$\alpha$ significance & $\beta$ significance & $-0.106$ & $0.66$\\
$\alpha$ significance & $\epsilon$ significance & $0.083$ & $0.51$\\
$\beta$ significance & $\epsilon$ significance & $0.194$ & $1.20$\\
Opening Angle\tablenotemark{a} & $\Delta v_{radio}$ & $-0.283$ & $1.36$
\enddata
\label{table_spearman_gapper}
\tablenotetext{a}{The correlation for only those sources we have identified as ``true'' bent double-lobed radio sources.}
\end{deluxetable}

\begin{deluxetable}{lccccc}
\tabletypesize{\tiny}
\tablewidth{0pt}
\tablecolumns{6}
\tablecaption{Fraction of Clusters with Substructure Detected at $>2.0\sigma$ and N$^z_{3.0}>50$, Fixed Gap Method.}
\tablehead{\colhead{Sample}&
		  \colhead{Number}&
		  \colhead{$\beta$}&
		  \colhead{$\Delta$}&
		  \colhead{$\alpha$}&
		  \colhead{$\epsilon$}
		  \cr
		   & &
		  \colhead{(\%)}&
		  \colhead{(\%)}&
		  \colhead{(\%)}&
		  \colhead{(\%)}
		  \cr
		  \colhead{(1)}&
		  \colhead{(2)}&
		  \colhead{(3)}&
		  \colhead{(4)}&
		  \colhead{(5)}&
		  \colhead{(6)}}
\startdata
True Bent Sources & 5 & 20.00\% & 60.00\% & 0.00\% & 20.00\%\\
All Bent Sources & 9 & 22.22\% & 44.44\% & 33.33\% & 11.11\%\\
All Non-Bent Sources & 9 & 22.22\% & 66.67\% & 33.33\% & 33.33\%
\enddata
\label{table_sig_thresh_fixed}
\end{deluxetable}

\begin{deluxetable}{lccccc}
\tabletypesize{\tiny}
\tablewidth{0pt}
\tablecolumns{6}
\tablecaption{Fraction of Clusters with Substructure Detected at $>2.0\sigma$ and N$^z_{3.0}>50$, Shifting Gapper Method.}
\tablehead{\colhead{Sample}&
		  \colhead{Number}&
		  \colhead{$\beta$}&
		  \colhead{$\Delta$}&
		  \colhead{$\alpha$}&
		  \colhead{$\epsilon$}
		  \cr
		   & & 
		  \colhead{(\%)}&
		  \colhead{(\%)}&
		  \colhead{(\%)}&
		  \colhead{(\%)}
		  \cr
		  \colhead{(1)}&
		  \colhead{(2)}&
		  \colhead{(3)}&
		  \colhead{(4)}&
		  \colhead{(5)}&
		  \colhead{(6)}}
\startdata
True Bent Sources & 7 & 28.57\% & 57.14\% & 0.00\% & 28.57\%\\
All Bent Sources & 11 & 27.27\% & 63.64\% & 27.27\% & 18.18\%\\
All Non-Bent Sources & 11 & 9.09\% & 63.64\% & 36.36\% & 45.45\%
\enddata
\label{table_sig_thresh_gapper}
\end{deluxetable}

\end{document}